\documentclass[11pt]{article}
\pdfoutput=1

\usepackage[normalem]{ulem}
\usepackage{amsfonts}
\usepackage{longtable}
\usepackage{makeidx}
\usepackage{fullpage}
\usepackage{amsmath}
\usepackage{amssymb}
\usepackage{comment}
\usepackage{setspace}
\usepackage{bbm}
\usepackage{dsfont}
\usepackage{graphics}
\usepackage{epsfig}
\usepackage[font=footnotesize,labelfont=bf,justification=centerlast,width=.94\textwidth]{caption}
\usepackage{cite}
 \usepackage{multirow}
\usepackage{array,booktabs}
\usepackage{color}

\usepackage{multirow}

\usepackage{changepage}

\usepackage{jheppub}
\usepackage{amsthm}
\usepackage{slashed}
\usepackage[utf8]{inputenc}
\usepackage[T1]{fontenc}
\usepackage{mathtools}
\usepackage{float}
\usepackage{empheq}
\usepackage{bm}

\def\nt{\notag}
	
\def\wt{\widetilde}

\def\R{\mathbb{R}}
\def\Z{\mathbb{Z}}

\def\cE{\mathcal{E}}

\def\cL{\mathcal{L}}

\def\cO{\mathcal{O}}

\def\cS{\mathcal{S}}
\def\cT{\mathcal{T}}

\def\dg {\dagger}
\def\p{\partial}

\def\/{\over}

\def\t{\theta}
\def\s{\sigma}
\def\e{\epsilon}
\def\ve{\varepsilon}

\def\a{\alpha}
\def\b{\beta}
\def\d{\delta}
\def\k{\kappa}
\def\g {\gamma}
\def\la {\lambda}
\def\w {\omega}
\def\z{\zeta}

\def\l{\ell}
\def\mn{{\mu\nu}}

\def\n {\nabla}
\def\L{\Lambda}
\def\D{\Delta}
\def\G {\Gamma}
\def\Om {\Omega}
\def\S{\Sigma}

\def\ra{\rightarrow}
\def\lra{\longrightarrow}

\def\r{\mathrm}
\def\hc{\text{h.c.}}

\def\_{\hspace{2cm}}
\def\-{\\\notag}
\def\={&=&}

\newcommand\be{\begin{equation}}
\newcommand\ee{\end{equation}}

\newcommand{\bea}{\begin{eqnarray}}
\newcommand{\eea}{\end{eqnarray}}

\newcommand{\bpm}{\begin{pmatrix}}
\newcommand{\epm}{\end{pmatrix}}

\newcommand{\bit}{\begin{itemize}}
\newcommand{\eit}{\end{itemize}}

\newcommand{\ben}{\begin{enumerate}}
\newcommand{\een}{\end{enumerate}}

\newcommand\bsp{\begin{split}}
\newcommand\esp{\end{split}}

\newcommand{\fsl}{\mathfrak{sl}}
\newcommand{\fu}{\mathfrak{u}}

\def\le{\left}
\def\ri{\right}

\def\ms{\medskip}

\def\l{\ell}

\def\qq{\qquad}

\def\cos{\r{cos}}
\def\sin{\r{sin}}

\def\cosh{\r{cosh}}
\def\sinh{\r{sinh}}

\renewcommand{\title}[1]{\vbox{\center\LARGE{#1}}\vspace{5mm}}
\renewcommand{\author}[1]{\vbox{\center#1}\vspace{5mm}}
\newcommand{\address}[1]{\vbox{\center\footnotesize\em#1}}
\newcommand{\email}[1]{\vbox{\center\footnotesize\tt#1}\vspace{5mm}}

\newcommand{\x}{{\bf x}}

\newcommand{\dd}{\mathrm{d}}
\newcommand{\mt}[1]{\textrm{\tiny #1}}

\newcommand{\pert}{\epsilon}

\begin{document}

{
\begin{titlepage}

\begin{center}

\hfill \\
\hfill \\
\vskip 1cm


\title{\bf Gravitational perturbations from NHEK to Kerr}


\author{Alejandra Castro$^{a}$, Victor Godet$^{a,b}$, Joan Sim\'on$^{c}$, Wei Song$^{d,\,e}$, and Boyang Yu$^{d,\,e}$
}

\address{
${}^a$ Institute for Theoretical Physics, University of Amsterdam, Science Park 904, Postbus 94485,\\
1090 GL Amsterdam, The Netherlands
\\
${}^b$ International Centre for Theoretical Sciences (ICTS-TIFR),
Tata Institute of Fundamental Research, Shivakote, Hesaraghatta, Bangalore 560089, India
\\
${}^c$ School of Mathematics and Maxwell Institute for Mathematical Sciences,
University of Edinburgh, Edinburgh EH9 3FD, United Kingdom
\\
${}^d$Department of Mathematical Sciences, Tsinghua University, Beijing 100084, China
\\
${}^e$Yau Mathematical Sciences Center, Tsinghua University, Beijing 100084, China
}

\email{ a.castro@uva.nl, victor.godet@icts.res.in, j.simon@ed.ac.uk, wsong2014@mail.tsinghua.edu.cn, yuby16@mails.tsinghua.edu.cn }

\end{center}

\vskip 1cm

\begin{center} {\bf ABSTRACT } \end{center}

\begin{adjustwidth}{10pt}{10pt}
\hspace{12pt}  We revisit the spectrum of linear axisymmetric gravitational perturbations of the (near-)extreme Kerr black hole. Our aim is to characterise those perturbations that are responsible for the deviations away from extremality, and to contrast them with the linearized perturbations treated in the Newman-Penrose formalism. For the near horizon region of the (near-)extreme Kerr solution, \emph{i.e.} the (near-)NHEK background, we provide a complete characterisation of axisymmetric modes.  This involves an infinite tower of propagating modes together with the much subtler low-lying mode sectors that contain the deformations driving the black hole away from extremality.
 Our analysis includes their effects on the line element, their contributions to Iyer-Wald charges around  the NHEK geometry, and how to reconstitute them as gravitational perturbations on Kerr.  We present in detail how regularity conditions along the angular variables modify the dynamical properties of the low-lying sector, and in particular their role in the new developments of nearly-AdS$_2$ holography. 
\end{adjustwidth}

\vfill

\end{titlepage}
}


\newpage

\tableofcontents
\newpage

\section{Introduction}
\label{sec:intro}

Gravitational perturbations of a black hole illustrate the invaluable interplay between theory and experiment (or numerical simulation) in general relativity. For example, any progress in the analytic calculations controlling the physics of extremal mass ratio inspirals (EMRI) can be of experimental relevance since it can give rise to more accurate waveforms used in the data analysis algorithms searching for gravitational wave signals.\footnote{More precisely, EMRIs are one of the most exciting sources of gravitational radiation for the space-based detector LISA \cite{Danzmann:1996da}. However, they are also very challenging to model and to extract data \cite{Amaro-Seoane:2014ela,Barack:2003fp,Chua:2017ujo}. This is because EMRIs will be observable for a large number of cycles before plunge, involving eccentric and inclined orbits up to a few cycles before the latter. This introduces a huge amount of complexity to encode and extract such information in models, but at the same time makes them suitable to test GR \cite{Gair:2012nm,Glampedakis:2005cf,Barack:2006pq}. See \cite{Barausse:2020rsu} for a broader perspective on the relevance of LISA for theoretical physics.}

This synergy between theory, experiment, and numerical simulations, has been further crossed in recent years when the primary black hole in the binary system is near-extremal. The enhancement of symmetry from $\mathbb{R}\times \fu(1)$ to $\fsl(2,\mathbb{R})\times \fu(1)$ in the near horizon region and the use of asymptotic matching techniques allows the computation of some observables either exactly, or with high analytic accuracy; see \cite{Yang:2013uba,Gralla:2016qfw,2015PhRvD..92f4029G,2018arXiv180403704C,van2015near,Porfyriadis:2014fja,Hadar:2014dpa,Hadar:2015xpa,Compere:2017hsi}  and references therein. What this body of work stresses is that the gravitational radiation from near-extremal primaries has rather unique features and can be used as a smoking gun for identifying these objects in the sky.\footnote{These features also leave a trace in the dynamics of the transition from inspiral to plunge in a circular equatorial orbit. In \cite{Compere:2019cqe,Burke:2019yek}, new potential terms responsible for different scaling behaviours were identified in the near-extremal regime, extending the original analysis by Ori and Thorne \cite{Ori:2000zn}. In fact, if near-extremal Kerr black holes exist and are observed, they are predicted to have much higher parameter estimation sensitivity, using gravitational wave probes, than regular rotating Kerr black holes and the origin for such increase can, once more, be traced to the existence of a throat in the near horizon geometry  \cite{Burke:2020vvk}.}

A further, and more recent, development in the theory side has been the identification of the relevant degrees of freedom describing the low energy physics driving a black hole away from extremality.  Based on ideas from nearly-AdS$_2$ holography \cite{Almheiri:2014cka,Maldacena:2016upp}, these degrees of freedom arise from breaking the reparametrization symmetries of the AdS$_2$ throat that appear in the near horizon region of the extremal black hole. This mechanism includes a spontaneous plus an explicit symmetry breaking pattern, leading to the construction of an effective field theory description for the resulting pseudo-Goldstone modes. This low energy sector determines important aspects of the gravitational backreaction, and several properties that are key to our microscopic (quantum) understanding of black hole physics.

However, whereas gravitational perturbations of Kerr black holes are typically formulated using the Teukolsky formalism \cite{Teukolsky:1973ha,Press:1973zz,Teukolsky:1974yv}, the description of nearly-AdS$_2$ holography physics is typically done in the context of Jackiw-Teitelboim (JT) gravity \cite{Teitelboim:1983ux,Jackiw:1984je},  or similar two-dimensional gravitational theories. The former is a gauge invariant description based on the Newman-Penrose formalism whose relation to measurable quantities in gravitational wave physics is known. The latter is based on some specific choice of gauge and is typically tied to the near horizon geometry, from which its universality comes from. It is natural to ask how the features of JT gravity appear in the Teukolsky formulation and how they are glued to the asymptotically flat physics that we observe. We will refer to the gravitational perturbations that encode these features as the JT sector, and fields that obey the same dynamics as the scalar field in JT gravity will be called JT modes. 

\begin{figure}
\begin{center}
\includegraphics[width=320pt]{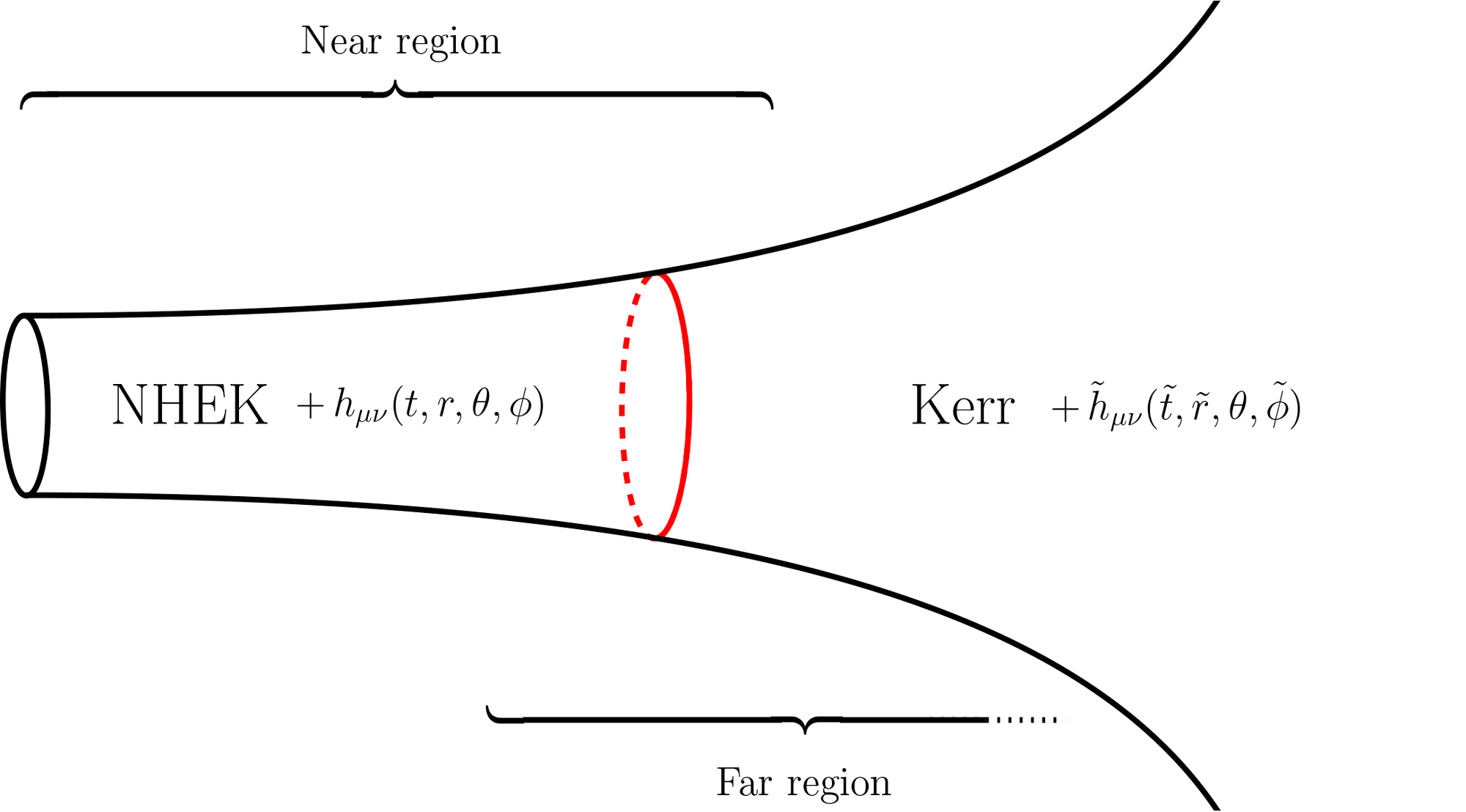}
\end{center}
\caption{The near-extremal Kerr geometry, highlighting its NHEK portion. Gravitational perturbations of the near region, described by the NHEK geometry, are glued to perturbations in the far region, which includes the asymptotically flat region of the black hole. The red circle represents the asymptotic boundary of the AdS$_2$ throat, which lies in the matching region for the perturbations.}\label{fig:nhek}
\end{figure}

Following these motivations and observations, the purpose of this work is twofold. First, we generalize the original results in \cite{Castro:2019crn} and relate axisymmetric gravitational perturbations around the near horizon geometry of the extremal Kerr black hole (NHEK)  to the gauge invariant Weyl scalars appearing in the Teukolsky formulation. This involves an infinite tower of (near-)AdS$_2$ modes together with the much subtler low-lying mode sectors that contain marginal extremal deformations and a JT sector responsible for driving the system away from extremality.
Second, we discuss how to glue the previous near horizon relations to the full asymptotically flat (near-)extremal Kerr geometry. Fig.\,\ref{fig:nhek} depicts the various regions in the geometry that are used in this gluing procedure.  For the low-lying modes this is an intricate task as we will discuss in detail: for smooth perturbations, diffeomorphisms enter in this process which become physical states on NHEK, and there are also cases when the singular nature of some of these modes adds novel features to the matching procedure.

Before presenting our general strategy and main results, and in an attempt to make this work minimally self-contained while taking into account the expertise of different readers, we have included several appendices at the end of this manuscript providing brief reviews on different topics. App.~\ref{app:Teuk} discusses some relevant aspects of the standard Teukolsky formalism of gauge invariant gravitational perturbations, applied to both the Kerr black hole and the NHEK geometry, and the content of Wald's theorem regarding gravitational perturbations on Kerr \cite{wald-theorem}. 
App.~\ref{sec:nearAdS2} reviews some of the ideas in nearly-AdS$_2$ holography.

\subsection{Summary of our strategy and results}

\label{sec:summary}

The (near-)extremal Kerr black hole is a particular example of (near)-extremal black holes where the ideas of nearly-AdS$_2$ holography, reviewed in App.\,\ref{sec:nearAdS2}, should apply. However, the explicit breaking of the spherical symmetry makes the identification of the JT sector subtle, and as we will show,  adds new intricacies to the low energy description. Prior work that  incorporates aspects of rotation in nearly-AdS$_2$ holography includes \cite{Anninos:2017cnw,Castro:2018ffi,Moitra:2019bub,Iliesiu:2020qvm,Godet:2020xpk,Heydeman:2020hhw} and see \cite{Almheiri:2016fws,Cvetic:2016eiv,Gaikwad:2018dfc,Ghosh:2019rcj,Castro:2019vog,Chaturvedi:2020jyy} for extensive work on the three-dimensional BTZ black hole. Here we follow, and generalize, the approach presented in \cite{Castro:2019crn}.

To start our summary, we first focus on our results regarding the gravitational perturbations on NHEK.  We describe axisymmetric gravitational perturbations of NHEK as follows
\bea\nt
ds^2 \=   J \le(1+\r{cos}^2\t + \pert  \chi (\x,\t)\ri)\le[g_{ab} \dd x^a\dd x^b+\dd \t^2 \ri] \label{newansatz1} \\ 
&& + 4 J\,{ {\sin}^2\t\, \over 1+{\cos}^2\t +  \pert   \chi(\x,\t)} \le(\dd\phi + A_a \dd x^a + \pert  {\cal A}_a \dd x^a\ri)^2   + O(\pert ^2)~.
\eea
Here $J=M^2$ is the maximal amount of angular momentum allowed by requiring the absence of naked singularities. Note that we are setting $G_4=1$.  At linear order in $\pert$, these perturbations are determined by a single scalar $\chi(\x,\t)$ satisfying
\be
\Box_2 \chi +{\sin^3\t\over \cos\,\theta}\partial_\theta\le({\cos^2\t\over \sin^3\t}\partial_\t\le( {\chi\over \cos\,\t}\ri)\ri)  =0~,
\ee
where $\Box_2=\nabla_a\nabla^a$ is the Laplacian on AdS$_2$ with coordinates denoted by $\x$. After separation of variables, 
\be
\chi(\x,\theta) = \r{sin}^2\t\,\sum_{\l} S_\l(\t) \chi_\l(\x)~,
\ee
the $\ell$-modes on the sphere correspond to associated Legendre polynomials with $\ell\in \Z$ and $\ell\geq 2$, while the function $\chi_\ell(\x)$ satisfies the wave equation 
\begin{equation}
  \Box_{2} \chi_\l = \l(\l+1) \chi_\l~,
\end{equation}
on AdS$_2$. These perturbations describe a tower of AdS$_2$ modes with conformal dimension $\Delta = \ell+1\geq 3$.

Since this description is not gauge invariant, its relation to the gauge invariant quantities appearing at linear order in the Teukolsky formalism  is not apparent. By computing the appropriate Weyl scalars, $\Psi_0$ and $\Psi_4$, we show that modes with $\ell\geq 2$ have a one-to-one correspondence with outgoing and incoming modes in the Teukolsky formalism. Hence, these are physical and in accordance to prior work on gravitational perturbations in NHEK \cite{Dias:2009ex,Amsel:2009ev,Hartman:2009nz}. We will refer to them as \emph{propagating} modes.

There are no associated Legendre polynomials with $\ell=0,1$. However, these modes are allowed by the AdS$_2$ Breitenlohner-Freedman bound \cite{Breitenlohner:1982jf}. We will refer to these as \emph{low-lying} modes. Our analysis shows that they give rise to non-normalizable modes on the sphere. The corresponding metric perturbations have conical singularities at either the north or the south poles.\footnote{This observation was recently made in a similar context to ours in an appendix in \cite{Hadar:2020kry}.} Furthermore, both $\Psi_0$ and $\Psi_4$ diverge at the location of such singularities. We point out that requiring the absence of such divergences, \emph{i.e.} $\Psi_0=\Psi_4=0$, which is a much milder condition that setting $\chi_\l$ to zero, gives rise to two constraints.\footnote{Since these Weyl scalars are computed on near-NHEK, it is natural to interpret these conditions as the absence of ingoing and outgoing energy flux at the horizon \cite{Hawking:1972hy,Teukolsky:1974yv,Teukolsky:1973ha,Press:1973zz}. At this early stage, it would be natural to ask why we require both $\Psi_0=\Psi_4=0$ rather than simply demanding them to be finite. As discussed in further detail in Sec.~\ref{sec:chi-2}, this is because the low-lying modes \eqref{eq:s0sol}-\eqref{eq:s1sol} are singular at either of these poles independently of the time and radial dependence of the wave function, due to the mode decomposition \eqref{eq:modek2} and as it can be explicitly seen in \eqref{eq:weyl-xp1}.} This is interesting for three reasons:
\begin{itemize}
\item   For $\Psi_0=\Psi_4=0$, the resulting perturbation due to $\chi$ is not a diffeomorphism plus a change of mass and/or angular momentum.  This does not contradict  Wald's theorem \cite{wald-theorem} since these modes produce conical singularities on the geometry. 
\item For $\ell=1$, combining these two constraints together with the AdS$_2$ wave equation, coming from the linearized Einstein's equations, can be shown to be equivalent to the JT equations of motion
\be\label{eq:JT}
  \n_a\n_b \Phi_{\mt{JT}} - g_{ab}\, \Box_{2}\Phi_{\mt{JT}} +g_{ab} \Phi_{\mt{JT}} = 0~.
\ee
This equation is the key feature of nearly-AdS$_2$ holography. As described in App.\,\ref{sec:nearAdS2}, from the dynamics of \eqref{eq:JT} one can infer the low energy sector that arises due to the symmetry breaking pattern. 
\item For $\ell=0$, the same constraint gives rise to a constant zero mode.
\end{itemize}
Thus, including the low-lying modes in the AdS$_2$ tower gives rise to an extra irrelevant perturbation with $\Delta=2$ ($\l=1$) and a marginal perturbation with $\Delta=1$ ($\l=0$). When requiring $\Psi_0=\Psi_4=0$, the former satisfies the JT equations of motion while the latter is a constant zero mode. However, both perturbations remain singular since the perturbed metric has conical singularities at either the north or south poles.

To have a complete characterization of the JT sector, we balance the conical singularity of the $\l=1$ low-lying mode using the following mechanism. First, we show that the Killing vectors $\zeta$ of AdS$_2$ backgrounds of the form
\be
d s^2=\Lambda(\theta)(g_{\mt{AdS}_2}+\dd\theta^2)+\Gamma(\theta)(\dd\phi+{A}_a \dd x^a)^2\,,
\ee
which include NHEK as a particular case, are in one--to--one correspondence with a scalar field $\Phi(\x)$ solving the JT equations of motion \eqref{eq:JT} and a constant zero mode $c_{\mt{($\phi$)}}$. More explicitly,
\be
\zeta =\varepsilon^{ba}\nabla_b\Phi_\zeta\p_a+(\Phi_\zeta+\varepsilon^{ab}{ A}_a\nabla_b\Phi_\zeta)\p_\phi~, \quad \text{with} \quad  \Phi_\zeta \equiv c_{\mt{($\phi$)}}+\Phi_{\mt{JT}}\,.
\ee
It is important to note that these Killing vectors are determined by the same differential equations that govern the AdS$_2$ low-lying modes with vanishing Weyl scalars. 
Hence, Killing vectors of NHEK naturally encode a second copy of the previously identified $\ell=1$ and $\ell=0$ AdS$_2$ modes. However, these are \emph{non-dynamical}. To make them dynamical, we observe that a \emph{non-single valued} diffeomorphism 
\be\label{eq:1234}
\xi^\mu (\x,\t,\phi) = {\pert \/2}\phi \,\z^\mu (\x,\t)~,
\ee
with $\z^\mu (\x,\t)$ a  Killing vector on NHEK, acting on $g_{\mt{NHEK}}$, gives rise to an axisymmetric perturbation, \emph{i.e.} satisfying $\p_\phi (\cL_{\xi}g_{\mt{NHEK}}) = 0$. Even though the latter is locally pure gauge, it is a \emph{physical singular} perturbation. Singular, because it gives rise to a conical defect, as one may have expected from being generated by a non-single valued diffeomorphism. Physical, because it gives rise to non-trivial Iyer-Wald charges, as we explicitly compute in Sec.~\ref{sec:charges-nhek}. 

The resulting perturbation, generated by the action of \eqref{eq:1234} on \eqref{newansatz1}, takes the form
\bea\label{NHEKfullansatz1}
ds^2 \=   J \le(1+\r{cos}^2\t + \pert  \chi (\x,\t)\ri)\le[\le(g_{ab}+ \pert h_{ab}\ri) \dd x^a \dd x^b  +\dd \t^2 \ri] \-
&& \hspace{1cm}+ 4 J\,{ {\sin}^2\t\,(1+ \pert \Phi(\x))\, \over 1+{\cos}^2\t +  \pert   \chi(\x,\t)} \le(\dd\phi +  A_a \dd x^a + \pert  {\cal A}_a\dd x^a\ri)^2  +{O}(\pert ^2)\,,
\eea
with the dynamical constraints satisfied by $h_{ab}$ and ${\cal A}_a$ given in Sec.~\ref{sec:balancing}. We show that the physics of the JT sector is driven by the NHEK perturbation with $\Phi_{\mt{JT}}=\chi$. This is the choice balancing the conical singularities associated with each $\ell=1$ mode. We confirm the physical interpretation of this near horizon perturbation as a change of mass (plus a local diffeomorphism), in agreement with Wald's theorem, by gluing this near horizon perturbation with a full Kerr perturbation in Sec.~\ref{sec:smooth}. A similar mechanism to balance conical singularities applies to the $\ell=0$ mode, albeit singularities in this sector have interesting physical interpretations discussed in Sec.\,\ref{sec:marginal}.

Our analysis of axisymmetric low-lying modes in NHEK identifies all the possibilities allowed by Wald's theorem. In the same gauge as in \eqref{newansatz1}, these perturbations are characterised by
\be
\chi(\x,\theta)=\Phi_\mt{JT}(\x)+\frac{1}{2}(1+\cos^2\theta)\,c_{\mt{($\phi$)}}~,
\label{intro:match}
\ee 
where we stressed the nature of the JT mode and we included the zero mode $c_{\mt{($\phi$)}}$. In Sec.~\ref{sec:charges-nhek}, we compute the Iyer-Wald charges carried by these near horizon perturbations. There are three $\fsl(2)$ charges
\be\label{intro:sl2}
\delta\mathcal Q_{\zeta_-}=-\pert {J}c^+~,\quad \,\delta\mathcal Q_{\zeta_{0}}= \frac{\pert}{2} J\,c^0~,\quad\,\delta\mathcal Q_{\zeta_+}=- \pert\,{J}\,c^-~,
\ee
corresponding to the three independent solutions $\Phi_{\mt{JT}}=c^i\Phi_{\zeta_{i}}$ for the JT modes, one per generator of $\fsl(2)$, and the $\fu(1)$ charge 
\be\label{intro:u1}
\delta\mathcal Q_{\p_\phi}=- \frac{\pert}{2}J\, c_{\mt{($\phi$)}}~.
\ee

The first two columns in Table \ref{t:modes} summarise our results on axisymmetric perturbations of the NHEK spacetime. Besides the set of smooth propagating modes $(\ell \geq 2)$, there are smooth low-lying modes, i.e. $\ell = 1,0$ modes with vanishing Weyl scalars and whose conical defects were compensated by a non-single valued diffeomorphism \eqref{eq:1234} giving rise to the JT modes $(\ell=1)$ and a marginal deformation of the NHEK spacetime $(\ell=0)$. Among the singular modes, we distinguish in the third column of the table among two cases: when the Weyl scalar is non-zero, and the singularity corresponds to a pole of the curvature invariant; and when the Weyl scalar is locally zero, but there is still a delta function singularity due to a conical defect on the 2-sphere. In the fourth column we give a brief characterisation of these modes on the entire Kerr background.

Having identified the dynamical mechanism that is characteristic of nearly-AdS$_2$ holography in NHEK, we undertake the second main goal in this work in Sec~\ref{sec:UV}: how to describe these near horizon perturbations as full Kerr perturbations. The matching procedure of the perturbations is as follows.  The starting point is the decoupling limit that relates the Kerr and NHEK backgrounds, a singular coordinate transformation on near-extreme Kerr of the form
\be\label{eq:near-near11}
\tilde r = \sqrt{J}+ \lambda\left(r + \frac{\tau^2}{4r}\right)\,,\qq \tilde t = 2J {{t}\/\la}~,  \qq \tilde\phi = {\phi} + \sqrt{J}{t\/\la}~, \qq \lambda\to0~.
\ee
Starting from a perturbation on Kerr, we implement this limit. Our requirement is that the metric perturbations, and associated Iyer-Wald charges, do not diverge as we take $\lambda\to0$. This allows us to match our analysis of perturbations on NHEK with those on Kerr. Our discussion here follows the same organization as above: we distinguish the propagating and low-lying modes. A summary of our results in this direction is presented in the last column in Table \ref{t:modes}.

Our reconstruction strategy for propagating modes is standard. Since the Hertz potential determines the metric perturbation in the ingoing radiation gauge (IRG), we use asymptotic matching techniques  for near-extremal Kerr \cite{Teukolsky:1974yv,Hartman:2009nz} to solve the master equation \eqref{hertzeqn} satisfied by the Hertz potential. Specifically, we follow a three-step algorithm: we first glue the Kerr Hertz potential $\tilde\Psi_{\mt{H}_0}$ to the NHEK Hertz potential $\Psi_{\mt{H}_0}$, reconstruct the NHEK perturbation in IRG using the latter, and finally use a small diffeomorphism to bring the perturbation to the gauge \eqref{newansatz1}. Our main technical result is the relation between our scalar perturbation $\chi(\x,\t)$ and $\Psi_{\mt{H}_0}$:
\be
\chi (\x,\t) =-\sin^2\theta\, \boldsymbol{l}^a\boldsymbol{l}^b\nabla_a\nabla_b\Psi_{\mt{H}_0}(\x,\t)~.
\ee
Furthermore, decomposing $\Psi_{\mt{H}_0}(\x,\t)=\sum_{\ell\geq2}U_\ell(\x)S_\ell(\theta)$, we also obtain the inverse relation
\be
U_\ell(\x)=-\frac{4}{(\ell-1)\ell(\ell+1)(\ell+2)}\boldsymbol{n}^a\boldsymbol{n}^b\nabla_a\nabla_b\chi_\ell(\x)~.
\ee
Here $\boldsymbol{l}^a$ and $\boldsymbol{n}^a$ are tetrads for AdS$_2$, given in App.\,\ref{app:proof}. These explicit maps relate, for $\l\geq 2$, our specific gauge with the more common radiation gauge used in gravitational wave physics.

\begin{table}[t]\def\arraystretch{1.4}
	\centering
	\begin{tabular}{ |m{3.5cm}  c m{2.6cm} m{7cm}|}
		\hline
		\centering {\bf Modes} & $\Delta$ & \centering{\bf Weyl scalars} &   {\bf Properties on Kerr} \\[3pt]
		\hline\hline
		&&& \\[-1.5em] 
		\centering propagating  $\ell\geq2$ & $\ell +1$ & \centering non-zero & Well-behaved axisymmetric perturbations  \\[3pt]
		\hline
		 &&& \\[-1.5em] 
		\centering smooth $\ell=1$ & 2 & \centering zero &$M$ (or $J$) deformation with fixed $J$ (or $M$), plus diffeomorphisms \\
		\centering smooth $\ell=0$ & 1 &\centering zero &$M$ and $J$ deformation with fixed $J=M^2$,  plus diffeomorphisms.		 \\  &&& \\[-1.5em]  
		\hline &&& \\[-1.5em] 
		\centering singular $\ell=0,1$ & 1,2 &\centering non-zero & Non-separable Hertz potential  \\
		&&&\\[-1.5em] 
		\centering singular $\ell=0$ & 1 & \centering zero  & Taub-NUT and/or $C$-metric deformation 		 \\
		\centering singular $\ell=1$ &2 &\centering{\footnotesize zero (NHEK) non-zero (Kerr)$^\star$} & Separable Hertz potential \\
		  \hline
	\end{tabular}	
	\caption{Summary of the different classes of perturbations considered in our work. Here $\ell$ is a nonnegative integer that controls the angular $\theta$ dependence. $\Delta=\l+1$ is the conformal dimension of the perturbations as viewed in NHEK. The third column describes the value of the Weyl scalars $\tilde \Psi_{0}$ and $\tilde\Psi_{4}$ on Kerr.  The last column briefly describes properties of the perturbations on Kerr. \\	
${}^\star$Depending on the specific configuration with vanishing Weyl scalars on NHEK, the resulting Weyl scalar on Kerr can be either zero or non-zero. } 
	\label{t:modes}
\end{table}

In the reconstruction of low-lying mode perturbations, we consider two different cases: smooth and singular perturbations. Smooth perturbations have vanishing Weyl scalars and their would-be conical singularity is compensated by the transformation \eqref{eq:1234}. Thus, their description in the full Kerr geometry is constrained by Wald's theorem, and we give a detailed analysis in Sec.\,\ref{sec:smooth}. Singular perturbations can have both vanishing or non-vanishing Weyl scalars. We treat these using similar techniques as the ones outlined for the propagating modes, and it is the focus of Sec.\,\ref{sec:singularpert}.

According to Wald's theorem, the reconstruction of smooth low-lying perturbations in Kerr, with metric $\tilde g$, must be a linear combination of mass $(\delta_M \tilde{g})$, angular momentum $(\delta_J \tilde{g})$ perturbations and a diffeomorphism $(\mathcal{L}_{\tilde\xi} \tilde{g})$, \emph{i.e.}, 
\be 
\delta \tilde{g}=\delta_M \tilde{g}+\delta_J \tilde{g}+\pert\,\mathcal{L}_{\tilde\xi} \tilde{g}~,
\ee
with $\delta M \sim \epsilon$ and $\delta J \sim \epsilon$. We show that all finite perturbations of this type as $\lambda\to 0$, correspond to smooth low-lying perturbations characterised by \eqref{intro:match}. We also quantify their near horizon charges using \eqref{intro:sl2}-\eqref{intro:u1}. These results give extra support to the physical interpretation of these near horizon perturbations. We refer the readers to the discussion in Sec.~\ref{sec:smooth} for details. The key features we would like to highlight are:
\begin{enumerate}
\item In the absence of any explicit perturbation of Kerr, we can obtain a perturbation of NHEK as the first correction to the near horizon decoupling limit \eqref{eq:near-near11}. This corresponds to the choice $\epsilon\sim \lambda$ with
\begin{align}\label{eq:decpert11}
\pert\chi(\x,\t)= \pert\chi_1(\x)=\pert\Phi_\mt{JT}(\x)
&= {2\over \sqrt{J}}{\lambda} \le(r+{\tau^2\/4 r} \ri)~.
\end{align}
The action of $\mathrm{SL}(2,\R)$ can generate the full multiplet as in \eqref{intro:sl2}.
\item There is also a sector characterised by diffeomorphisms on Kerr with $\delta_M \tilde{g}=\delta_J \tilde{g}=0$. Restricting ourselves to single-valued diffeomorphisms with support on the sphere, we show the diffeomorphisms $\tilde\xi$  that are well defined in the decoupling limit have a near horizon expansion
\be\label{eq:diffpertnhek}
\xi[\lambda]=\epsilon \, \lambda^{-1}(\xi_{\mt{(-1)}}+\lambda\xi_{\mt{(0)}}+\cdots)~,
\ee
with 
\be
\xi_{\mt{(-1)}} =a^i\zeta_i+a^\phi\zeta_{\mt{$(\phi)$}}~.
\ee
Here $\xi[\lambda]$ is the pullback of $\tilde{\xi}$ from Kerr to NHEK under the decoupling limit; $\zeta_i$ and $\zeta_{\mt{$(\phi)$}}$ are Killing vectors of NHEK. We identify this perturbation with a near horizon perturbation \eqref{NHEKfullansatz1} with $\chi=\Phi_\mt{JT}$. The Killing vector part carries $\mathfrak{sl}(2)\times \mathfrak{u}(1)$ charges with $c^0 = 0$ and $c^\pm\neq0$, \emph{i.e.}, these transformations carry neither energy nor angular momentum. 

It is very important to remark that although the procedure starts from a diffeomorphism on the Kerr geometry, as one takes the decoupling limit, the resulting perturbation is {\it not} a diffeomorphism on NHEK. Also, in our discussion, $\xi_{\mt{(0)}}$ is constructed such that the resulting perturbation matches with \eqref{NHEKfullansatz1}: 
this is a choice of boundary conditions on NHEK. And for this choice, $\xi_{\mt{(0)}}$ does not contribute to the Iyer-Wald charges.

\item Mass and angular momentum perturbations are described as follows. The first observation is that $\delta_M g[\lambda] \sim \delta M\,\lambda^{-2}$. If $\delta M \sim \lambda^2\pert$, the perturbation is finite
and corresponds to a nearby near-NHEK with Hawking temperature
\be
\tau'_H=\tau_H\le(1+\sqrt{J}\,\frac{\lambda^{-2}\delta M}{2\pi^2 \tau^2_H}\ri)~.
\ee
However the resulting perturbation carries no $\fsl(2)$ charges. When $\delta M \sim \lambda\epsilon$, we can combine this transformation with \eqref{eq:diffpertnhek} to again obtain a NHEK perturbation with $\chi=\Phi_\mt{JT}$ in \eqref{NHEKfullansatz1}. The result is
\be
\pert\, \chi(\x,\t)= \pert\, \Phi(\x)={4\lambda^{-1}\delta M\over \tau^2}\le(r+{\tau^2\/4r} \ri)~.
\ee  
The difference with \eqref{eq:decpert11} is that we don't need to identify $\pert$ with $\lambda$, and hence the Iyer-Wald charges \eqref{intro:sl2} are finite even when $\lambda =0$.
 
For an angular momentum perturbation, $\delta_J g[\lambda]$, the discussion and outcome is similar as that of the mass perturbation.

\item One interesting class of marginal perturbations corresponds to  $\delta J=2\sqrt{J}\,\delta M$, \emph{i.e.} deformations of the mass and angular momentum while keeping the black hole extremal. In this case,
\be
(\delta_M+\delta_J)\,g[\lambda]=\delta M(\lambda^{-1}h^{\mt{(-1)}}+h^{\mt{(0)}}+\cdots)~.
\ee
The most interesting scaling is when $\delta M \sim \lambda^0\pert$, which can be combined with a diffeomorphism to give one of our modes
\be
\chi(\x,\t)=\Phi(\x)={c_{(\phi)}\over2} ~,\qq \delta M={c_{(\phi)}\over4}\sqrt{J} ~.
\ee 
\end{enumerate}

Finally, we consider the reconstruction in Kerr of the singular low-lying modes, \emph{i.e.}, modes with $\ell=0,1$ for which we impose no regularity conditions on the sphere. The details are  in Sec.\,\ref{sec:singularpert}.  A very interesting feature in this case is that the Hertz potential $\Psi_{\mt{H}_0}$ on NHEK  does \emph{not} allow for a separation of variables ansatz, but has to be written as a sum of two terms. More explicitly, we have for $\ell=1$ 
\be\label{intro:hertz-low}
\Psi_{\mt{H}_0}= S_{1}(\theta)\mathcal U_{1}(\x)-4S_1^\mt{inhom}(\theta)\boldsymbol{n}^a\boldsymbol{n}^b\nabla_a\nabla_b\chi_1(\x)~.
\ee
Here $S_1^\mt{inhom}(\theta)$ is given by \eqref{Sin-sol}, and we have 
\be\label{intro:chi-u}
\boldsymbol{l}^a\nabla_a\mathcal U_1(\x)=\boldsymbol n^a\nabla_a\chi_1(\x)~.
\ee
A similar construction  also holds for $\ell=0$. Our analysis includes a discussion on  how to apply a matching procedure to $\tilde\Psi_{\mt{H}_0}$, and the challenges that a non-separable solution pose. Another striking feature concerns the special case when $\Psi_0=\Psi_4=0$ on NHEK: we show that  the corresponding Weyl scalars on Kerr are in general non-trivial, and only become zero in the decoupling limit.

This paper is organized as follows. In Sec.~\ref{sec:ExtremeKerr}, after reviewing some aspects of NHEK and near-NHEK, including how it is obtained as a near horizon limit of (near-)extremal Kerr, we describe in Sec.~\ref{sec:isometry} an explicit one--to--one correspondence between Killing vectors of NHEK-like geometries and a scalar field solving the JT equations of motion (plus a constant zero mode). In Sec.~\ref{sec:grav}, we start our study of axisymmetric perturbations in NHEK. Propagating modes are discussed in Sec.~\ref{sec:Kmodes}, while low-lying modes are presented in Sec.~\ref{sec:chi-2}. The balancing mechanism giving rise to the smooth JT mode is discussed in Sec.~\ref{sec:balancing}, while the discussion of marginal deformations of NHEK corresponding to the ones allowed by Wald's theorem is given in Sec.~\ref{sec:marginal}, though the technical derivations are left to App.~\ref{App:Delta1}. Iyer-Wald charges of our NHEK perturbations are computed in Sec.~\ref{sec:IW}. In Sec.~\ref{sec:UV}, we explain our techniques to match/glue our NHEK perturbations with Kerr perturbations using Wald's theorem and the reconstruction of gauge invariant perturbations based on the Hertz potential. Finally, we discuss in Sec.\,\ref{sec:singularpert} some properties of singular low-lying perturbations. Our appendices include various complementary material related to the main sections. 

\section{(Near-)extreme Kerr}\label{sec:ExtremeKerr}

The Kerr black hole metric in Boyer-Lindquist coordinates is
\begin{align}\label{eq:kerr}
ds^2&=-{\Sigma\,\Delta\over (\tilde r^2+a^2)^2-\Delta\, a^2\,\sin^2\theta}\dd\tilde t^2+{\Sigma}\left({\dd\tilde r^2\over\Delta}+\dd\theta^2\right) \cr
&\quad +{\sin^2\theta\over\Sigma}((\tilde r^2+a^2)^2-\Delta \,a^2\,\sin^2\theta)\left(\dd\tilde \phi-{2a M \tilde r\over (\tilde r^2+a^2)^2-\Delta\, a^2\,\sin^2\theta}\dd\tilde t\right)^2~,
\end{align}
with
\begin{equation}
\Delta=(\tilde r-r_-)(\tilde r-r_+)~,\quad \Sigma=\tilde r^2+a^2\cos^2\theta~.
\end{equation}
The outer $(r_+)$ and inner $(r_-)$ horizons are $r_\pm = M\pm \sqrt{M^2-a^2}$. We set Newton's constant $G_4=1$, so that $M$ is the mass and $J=a M$ is the angular momentum of the black hole. Tilde coordinates $(\tilde t,\tilde r,\tilde \phi)$ refer to the asymptotically flat black hole to distinguish them from $(t,r,\phi)$ in the near horizon geometry below; $\theta$ is unchanged.

We are interested in extreme Kerr corresponding to $J=M^2$, \emph{i.e.} $a=M$, when both horizons coalesce $(r_+=r_-)$. These black holes develop an AdS$_2$ throat in the region close to the horizon that can be decoupled from the asymptotically flat description in \eqref{eq:kerr} by taking the limit $\lambda\to 0$ in the change of coordinates
\be\label{eq:near-horizon}
\tilde r = r_+ +{\la} r \qq \tilde t = 2r_+^2 {{t}\/\la}~,  \qq \tilde\phi = {\phi} + r_+{t\/\la}~,
\ee
while keeping all other parameters fixed. This near horizon limit leads to the line element
\begin{align}\label{NHEK}
d s^2 &=  g^{\mt{NHEK}}_{\mu\nu} \dd x^\mu  \dd x^\nu  \cr
&=J (1+\r{cos}^2\t ) \le[ -r^2{\dd t^2} + {\dd r^2\/r^2 } + \dd\t^2\ri] + J {4\,\r{sin}^2\t\/1+\r{cos}^2\t}\le[\dd\phi + r\,\dd t\ri]^2  ~.
\end{align}
This is the Near Horizon geometry of Extreme Kerr (NHEK) \cite{Bardeen:1999px,Guica:2008mu}. 

The isometries for the full Kerr geometry \eqref{eq:kerr}, given by $\mathbb{R}\times \fu(1)$, are enhanced  to $\fsl(2,\mathbb{R})\times \fu(1)$ in NHEK \eqref{NHEK}.  The four Killing vectors generating the latter are 
\be\label{NHEK:iso}
\zeta_{-} = \p_t ~, \quad \zeta_{0} = t\p_t -r\p_r~, \quad \zeta_+ = \le({1\/r^2 }+ t^2\ri) \p_t -2  r t \p_r -{2\/r }\p_\phi ~, 
\ee
and
\be
 \zeta_{\mt{$(\phi)$}}= \p_\phi~.
\ee
The decoupling limit \eqref{eq:near-horizon} introduces some arbitrariness on how we relate the AdS$_2$ time $t$ with the asymptotically flat time $\tilde t$. This freedom gives rise to a set of diffeomorphisms preserving the asymptotic structure of the NHEK metric. More explicitly, we take  \cite{Castro:2009jf,Kapec:2019hro}
\begin{align}\label{Sdiffeo}
t &\lra f(t) + {2 f''(t) f'(t)^2 \/ 4 r^2 f'(t)^2-f''(t)^2}~, \cr r&\lra {4 r^2 f'(t)^2 - f''(t)^2 \/ 4r \,f'(t)^3}~,\cr \phi &\lra \phi + \log \le({2r f'(t)-f''(t)\/2r f'(t)+f''(t)} \ri)~.
\end{align}
The arbitrariness is reflected on the arbitrary function $f(t)$ that redefines the time in the near horizon region. Acting on \eqref{NHEK}, this diffeomorphism gives  
\bea\label{eq:nhek2}
 ds^2 \=  J(1+\r{cos}^2\t)\le[ -r^2\le(1+ {\{f(t),t\}\/2 r^2}\ri)^2 \dd t^2 + {\dd r^2\/r^2}+\dd \t^2 \ri] \-
&&+ {4 J\,\r{sin}^2\t\/1+\r{cos}^2\t}\le[ \dd \phi + r \le(1- {\{f(t),t\}\/2 r^2}\ri) \dd t\ri]^2~,
\eea
where 
\be
\{f(t),t\} = \le({f''\over f'}\ri)' -{1\over 2}\le({f''\over f'}\ri)^2~.
\ee
We can see here that leading terms as $r\to \infty$ in \eqref{eq:nhek2} approach \eqref{NHEK}, \emph{i.e.} the diffeomorphism \eqref{Sdiffeo} only affects subleading components of the line elements in an expansion in $r$.  It is important to note that these are not the same boundary conditions used in Kerr/CFT \cite{Guica:2008mu}: the set of allowed diffeomorphisms there does not overlap with those here (with the exemption of the $\fu(1)$ Killing vector). 

 In the analysis of the subsequent sections, a particular choice of $f(t)$ selects a background for which we will quantify the gravitational perturbations. For example, $f(t)=t$ returns us to \eqref{NHEK}. Another choice which we will use frequently is $f(t)=e^{\tau t}$, with $\tau$ constant. It follows that $\{f(t),t\} =-\frac{\tau^2}{2}$, leading to the background
\bea\label{eq:near-nhek}
 ds^2 \=  J(1+\r{cos}^2\t)\le[ -r^2\le(1- \frac{\tau^2}{4r^2}\ri)^2 \dd t^2 + {\dd r^2\/r^2}+\dd \t^2 \ri] \-
&&+ {4 J\,\r{sin}^2\t\/1+\r{cos}^2\t}\le[ \dd \phi + r \le(1 + \frac{\tau^2}{4 r^2}\ri) \dd t\ri]^2~.
\eea
This corresponds to the so-called near-NHEK geometry \cite{Amsel:2009ev}.  It can be also obtained from a near-extremal Kerr black hole \eqref{eq:kerr}, where the decoupling limit \eqref{eq:near-horizon} allows for a small increase of the mass while keeping the angular momentum of the black hole fixed to $J$. More concretely, we introduce a small deviation away from extremality  of the form 
\be\label{eq:deviation}
r_\pm = \sqrt{J}\pm \lambda \tau + \frac{\tau^2\lambda^2}{4\sqrt{J}} + {O}(\lambda^3)~,
\ee 
with $\tau$ finite and positive, and the decoupling limit $\lambda\to 0$ becomes 
\be\label{eq:near-near}
\tilde r = \sqrt{J}+ \lambda\left(r + \frac{\tau^2}{4r}\right)\,,\qq \tilde t = 2J {{t}\/\la}~,  \qq \tilde\phi = {\phi} + \sqrt{J}{t\/\la}~,
\ee
where the choice of coordinate $r$ is a choice of gauge to keep the radial metric component independent of $\tau$. These steps lead to the near-NHEK geometry \eqref{eq:near-nhek}.

\subsection{Isometry-scalar duality}
\label{sec:isometry}

In this subsection we revisit the isometries of (near-)NHEK using a more covariant formalism. We will extend the original discussion in \cite{Mann:1992yv} for AdS$_2$  to backgrounds of the form
\be
d s^2=\Lambda(\theta)(g_{ab} \dd x^a\dd x^b+\dd\theta^2)+\Gamma(\theta)(\dd\phi+{ A}_a \dd x^a)^2\,.
\label{background}
\ee
$\Lambda(\theta)$ and $\Gamma(\theta)$ are two functions of $\theta$.\footnote{The subsequent discussion does not depend on the explicit form of $\Lambda(\theta)$ and $\Gamma(\theta)$. For NHEK and near-NHEK, they can be read from \eqref{NHEK} and \eqref{eq:near-nhek}, respectively.} Here $x^a$ are coordinates in 2D, and the metric $g_{ab}$ corresponds to a locally AdS$_2$ spacetime. Tensors and other covariant objects below are defined relative to this spacetime, e.g., covariant derivatives $(\nabla_a)$ or the Laplacian $(\Box_2=\nabla_a\nabla^a)$. ${A}_a$ is a gauge field supported on this 2D spacetime, and the field strength associated to it is
\be
  \dd A= -\frac{1}{2}\varepsilon_{ab} \dd x^a\wedge \dd x^b\,,
\label{field-strength}
\ee
with $\varepsilon_{ab}$ the Levi-Civita tensor.  Our expressions will turn out to the covariant with respect to the AdS$_2$ metric, and hence hold as well for the nearly-AdS$_2$ geometries such as the one in \eqref{eq:nhek2}. Still, it will be convenient to write some expressions explicitly; a choice of background we will commonly use for the AdS$_2$ metric and gauge field are
\be\label{eq:2dmetric}
g_{ab} \dd x^a\dd x^b=   -r^2 \dd t^2 + {\dd r^2\/r^2} ~, \qquad  A_a \dd x^a={r}\, \dd t~,
\ee
with Levi-Civita tensor $\varepsilon_{tr}=1$.

In the following we will build a one-to-one map between the isometries of \eqref{background} and a scalar field satisfying some suitable equations of motion. We will show this scalar field corresponds to the JT field in parallel with \cite{Mann:1992yv}, with an additional term due to the axisymmetry of the background \eqref{background}.

Let $\zeta$ be a Killing vector field of \eqref{background}. It follows from $\nabla_{\theta}\zeta_{\theta}=0$ that $\zeta^\theta=0$.  We can then split the Killing vector $\zeta$ into a 2D vector field $\zeta^a\p_a$ and the $\phi$-component $\zeta^\phi$ according to
\be
  \zeta=\zeta^a(x^a,\phi)\p_a+\zeta^\phi(x^a,\phi)\p_\phi\,.
\label{g-isometry}
\ee
The variation of the line element under the diffeomorphism generated by \eqref{g-isometry} equals 
\bea
\delta_\zeta (d s^2)&=&2\Lambda(\theta)(\nabla_{(a}\zeta_{b)}\dd x^a\dd x^b+g_{ab}\p_\phi\zeta^b\dd x^a\dd\phi)\nonumber
\\&&+2\Gamma(\theta)((\p_\phi\zeta^\phi+A_a\p_\phi\zeta^a)\dd\phi+(\p_a\zeta^\phi+\mathcal L_{\zeta} A_a)\dd x^a)(\dd\phi+ A_a\dd x^a)~,
\eea
where all indices are raised and lowered by $g_{ab}$. Since $\Lambda(\theta)$ and $\Gamma(\theta)$ are two independent functions, and $\zeta,g_{ab}, A_a$ are independent of $\theta$, the Killing equations guaranteeing the vanishing of $\delta_\zeta (d s^2)$ reduce to
\begin{align}\label{killingeqn}
\p_\phi\zeta^a=\p_\phi\zeta^\phi=0~,\\  
\nabla_{(a}\zeta_{b)}=0~,\label{kill-eqn} \\ 
\p_a\zeta^\phi+\mathcal L_\zeta A_a=0~.\label{killingeqn3}
\end{align}
The first implies that $\zeta$ only depends on 2D coordinates, \emph{i.e.} $\zeta^a=\zeta^a(x^a),\zeta^\phi=\zeta^\phi(x^a)$, whereas \eqref{kill-eqn} implies the 2D vector $\zeta^a\p_a$ satisfies a 2D Killing equation.
Contracting \eqref{kill-eqn} with $g^{ab}$, we get $\zeta^a\p_a$ is divergence free, \emph{i.e.} $\nabla\cdot\zeta=0$, 
enabling us to write it as the curl of an scalar $\Phi_\zeta$ 
\be
  \zeta^a=\varepsilon^{ba}\nabla_b\Phi_\zeta\,.
\label{zeta2d}
\ee
Integrating \eqref{killingeqn3} determines the full Killing vector $\zeta$ to be 
\be\label{vec-scalar}
\zeta =\varepsilon^{ba}\nabla_b\Phi_\zeta\p_a+(\Phi_\zeta+\varepsilon^{ab}{ A}_a\nabla_b\Phi_\zeta)\p_\phi~,
\ee
where we used \eqref{field-strength}, \eqref{kill-eqn}-\eqref{zeta2d} and absorbed the integral constant into $\Phi_\zeta$. Notice that given a Killing vector $\zeta$, the scalar $\Phi_\zeta$ can be reconstructed by 
 \be
  \Phi_\zeta=\zeta^\phi+{ A}_a\zeta^a\,.
\label{scalarfromvector}
\ee
Finally, substituting \eqref{zeta2d} into the 2D Killing equations \eqref{kill-eqn}, we get 
\be
\nabla_{(a}\zeta_{b)}=\varepsilon_a\,^c(\nabla_c\nabla_b\Phi_\zeta-\frac{1}{2}g_{cb}\Box_2\Phi_\zeta)=0\,.
\ee
This is equivalent to a set of differential equations 
\be
T_{ab}[\Phi_\zeta]\equiv \nabla_a\nabla_b\Phi_\zeta-\frac{1}{2}g_{ab}\Box_2\Phi_\zeta=0\,.
\label{scalareq}
\ee
Thus, the existence of  Killing vectors $\zeta$ solving \eqref{killingeqn}-\eqref{killingeqn3} is equivalent to \eqref{scalareq} evaluated on the background \eqref{background}. 

We now show that the solutions to \eqref{scalareq} are equivalent to the JT modes \eqref{eq:JT} plus the addition of a zero mode. First, we note that \eqref{scalareq} is the traceless portion of  \eqref{eq:JT}, and hence any solution to the JT equation will comply with $T_{ab}[\Phi_\zeta]=0$.  However \eqref{scalareq} has one additional solution. To see this, evaluate the divergence of $T^{ab}[\Phi_\zeta]$: this gives
\be
\nabla_aT^{ab}[\Phi_\zeta]=\frac{1}{2}\nabla^b(\Box_2-2)\Phi_\zeta=0\,.
\ee
Its general solution is a linear combination of a constant mode, which we will denote as $c_{\mt{($\phi$)}}$, and the solution to 
\be
(\Box_2-2)\Phi_\zeta=0~.\label{JTmass}
\ee
It is then clear that  \eqref{JTmass} together with the Killing equation \eqref{scalareq} is equivalent to the JT equations \eqref{eq:JT}. 
Hence, the general solution of \eqref{scalareq} consists of JT modes and a zero mode which we cast as
\be
  \Phi_\zeta \equiv c_{\mt{($\phi$)}}+\Phi_{\mt{JT}}\,.
\label{eq:temp}
\ee
To sum up, given a Killing vector $\zeta$, we can construct an scalar field $\Phi_\zeta$ via \eqref{scalarfromvector} satisfying the equations of motion \eqref{scalareq}. Conversely, given a scalar $\Phi_\zeta$ satisfying \eqref{scalareq}, the vector field \eqref{vec-scalar} is an isometry. This establishes the sought equivalence between the isometries of the background \eqref{background} and a linear combination of JT modes and a zero mode $c_{\mt{($\phi$)}}$ as reflected in \eqref{eq:temp}.

We close this general discussion by connecting the above conclusion with the explicit Killing vectors in \eqref{NHEK:iso} for the NHEK geometry \eqref{NHEK}. Since $\zeta_\Phi$ is a Killing vector, it is a linear combination of the $\fsl(2,\mathbb{R})\times \fu(1)$ generators
\be\label{eq:kvnhek}
\zeta_\Phi=c^-\zeta_-+c^{0}\zeta_0+c^+\zeta_++c_{\mt{($\phi$)}}\p_\phi\,,
\ee
with $c^{\pm,0}$ constants. The corresponding scalars, dual to the $\fu(1)$ and the $\fsl(2,\mathbb{R})$ isometries, are the zero mode $c_{\mt{($\phi$)}}$ and the JT modes $\Phi_{\mt{JT}}$, respectively
\be
  \Phi_{\zeta} = c_{\mt{($\phi$)}}+\Phi_{\mt{JT}}~, \qquad  \Phi_{\mt{JT}}=c^i\Phi_{\zeta_{i}},\quad\quad i=-,\,0,\,+~.
\label{phidecomposition}
\ee
 where the components of the JT field for \eqref{eq:2dmetric} read
 \be
 c^i\Phi_{\zeta_{i}} = c^- \, r  + c^0\, r\,t + c^+\le(r\,t^2-{1\over r}\ri)~.
 \ee
See App.\,\ref{app:isometries} for a construction of the Killing vectors and $\Phi_{\mt{JT}}$ for near-NHEK.

Given the one-to-one map between $\zeta_i$ and $\Phi_{\zeta_i}$, the set $\Phi_{\zeta_i}$ forms a representation of $\fsl(2,\mathbb{R})$. Let us define the bilinear 
\be
\eta_{ij}\equiv \eta(\Phi_{\zeta_i},\Phi_{\zeta_j})= -2\Big(\nabla_a\Phi_{\zeta_i}\nabla^a\Phi_{\zeta_j}-\Phi_{\zeta_i}\Phi_{\zeta_j}\Big)~.
\ee
This is invariant under the adjoint action, \emph{i.e.} $\eta([\zeta_i,\zeta_j],\zeta_k)+\eta(\zeta_j,[\zeta_i,\zeta_k])=0$. Since the $\fsl(2)$ algebra is simple, the invariant bilinear form is unique up to a constant factor. By explicit computation, one can verify that $\eta_{ij}$ is just the Killing form of the $\fsl(2)$ algebra, whose nonzero entries are given by
\be\label{kill-form}
\eta_{-+}\equiv \eta(\Phi_{\zeta_-},\Phi_{\zeta_+})=\eta(\Phi_{\zeta_+},\Phi_{\zeta_-})\equiv \eta_{+-} =-4~,\quad \eta_{00}\equiv \eta(\Phi_{\zeta_{0}},\Phi_{\zeta_0})=2~.
\ee
This bilinear form can be used to write the coefficients $c^i$ in \eqref{phidecomposition} in terms of $\Phi_{\zeta_i}$ and $\Phi_{\mt{JT}}$ using the inverse
matrix $\eta^{ij}$ to $\eta_{ij}$
\be
c^{i}=\eta^{ij}\eta(\Phi_{\zeta_j},\Phi_{\mt{JT}})\,.
\ee
More explicitly, 
\be
  c^-=-\frac{1}{4}\eta(\Phi_{\zeta_+},\Phi_{\mt{JT}})\,, \quad c^0=\frac{1}{2}\eta(\Phi_{\zeta_{0}},\Phi_{\mt{JT}})\,, \quad
  c^+=-\frac{1}{4} \eta(\Phi_{\zeta_-},\Phi_{\mt{JT}})\,.
\ee
It is also useful to record the identity
\be\label{eq:slLie}
\Phi_{[\zeta_i,\zeta_j]}=\varepsilon^{ba}\nabla_b\Phi_{\zeta_i}\nabla_a\Phi_{\zeta_j}=\zeta_i^a\nabla_a\Phi_{\zeta_j}~,
\ee
for any $\zeta_i,\zeta_j\in \fsl(2)$, which simply follows from the algebra.

\section{Gravitational perturbations on NHEK}\label{sec:grav}

In this section we will characterise axisymmetric ($\phi$-independent) gravitational perturbations around the NHEK background \eqref{NHEK}, generalizing the results in \cite{Castro:2019crn}. In particular, we will relate our description of these perturbations to the more familiar Teukolsky formalism for gravitational perturbations \cite{teukolsky1972rotating,Teukolsky:1973ha}: this will allow us to distinguish among excitations that correspond to normalizable propagating degrees of freedom and modes that affect the global properties of the  black hole. 

Let us cast the gravitational fluctuations around the generalized NHEK background as
\bea\nt
ds^2 \=   J \le(1+\r{cos}^2\t + \pert  \chi (\x,\t)\ri)\le[g_{ab} \dd x^a\dd x^b+\dd \t^2 \ri] \\ \label{newansatz}
&& + 4 J\,{ {\sin}^2\t\, \over 1+{\cos}^2\t +  \pert   \chi(\x,\t)} \le(\dd\phi +  A_a \dd x^a + \pert  {\cal A}\ri)^2   + O(\pert ^2)~.
\eea
The background, corresponding to $\pert =0$, is described in \eqref{background}. The axisymmetric deformations from NHEK we have introduced here involve an scalar field $\chi (\x,\t)$ and a one-form ${\cal A}$ supported in the $x^a=(t,r)$ subspace\footnote{Here and in subsequent expressions we are using the shorthand notation $f(\x):= f(x^a)=f(t,r)$.}
\be
{\cal A}= {\cal A}_a(\x,\theta) \dd x^a~.
\ee
Consider the metric \eqref{newansatz} at linear order in $\pert $. To study the dynamics of the perturbations, we impose
that \eqref{newansatz} satisfies the linearized vacuum Einstein equations, \emph{i.e.} 
\be
R_{\mu\nu}= R^{(0)}_{\mu\nu} + \pert R^{(1)}_{\mu\nu} + O(\pert ^2)\stackrel{!}{=} 0~.
\ee
Setting $R_{\mu\nu}^{(1)}=0$, gives 
\begin{align}\label{eq:cond1}
\partial_\theta \le( \nabla_b {\cal A}^b\ri)&=0~,\cr
 \sin^2\t\, \partial_\t\le( \varepsilon_{ab} {\cal A}^b\ri) +\cot\t \, \nabla_a \chi &=0~,\cr
 \varepsilon^{ab}\nabla_a {\cal A}_b-{\cos^2\t\over \sin^3\t}\partial_\t\le( {\chi\over \cos\,\t}\ri)&=0~,
\end{align}
and 
\be
\Box_2 \chi +{\sin^3\t\over \cos\,\theta}\partial_\theta\le({\cos^2\t\over \sin^3\t}\partial_\t\le( {\chi\over \cos\,\t}\ri)\ri)  =0~,
\label{eq:chi-motion}
\ee
where $\Box_2$ is the Laplacian on AdS$_2$. 
 Note that once $\chi$ is specified, it is straightforward to solve for ${\cal A}_a$ from \eqref{eq:cond1}. For this reason, from now on we will treat $\chi$ as the independent variable for the metric perturbation, whose equation of motion is given by \eqref{eq:chi-motion}. 
 
It is not common to cast gravitational perturbations for the Kerr black hole, or its near horizon NHEK geometry, as explicitly as in \eqref{newansatz}. 
The drawbacks of starting from such an ansatz are at least two-fold: we have not identified gauge redundancies in our parametrization, and it is unclear if we have a complete basis for the perturbations.\footnote{To emphasize, the advantages of using \eqref{newansatz} are that it was simple to solve the linearized Einstein equations, and we can easily quantify their effect in the spacetime as we will see in subsequent analysis.} In the following, we explain how to systematically overcome both of them.

The perturbations of the Kerr metric are commonly characterised by the Weyl scalars in the Teukolsky formalism, since these are gauge invariant quantities at linear order. In App.\,\ref{app:Teuk} we review the general strategy of this approach, and provide the relevant definitions. In particular we will focus on 
 \begin{align}\label{eq:weyl04}
  \Psi_0 &= C_{\mu\nu\alpha\beta}\,l^\mu\,m^\nu\,l^\alpha\,m^\beta\,, \cr
  \Psi_4 &= C_{\mu\nu\alpha\beta}\,n^\mu\,\bar{m}^\nu\,n^\alpha\,\bar{m}^\beta\,,
\end{align}
where $C_{\mu\nu\alpha\beta}$ is the Weyl tensor and the vectors $l^\mu$, $n^\mu$ and $m^\mu$ are introduced in \eqref{NPrelations} and identified in \eqref{eq:NHEK-tetrad} for NHEK. $ \Psi_0$ and $ \Psi_4$ are the Weyl scalars that characterise in a diffeomorphism invariant way a massless spin $s=\pm 2$ perturbation. To relate the Teukolsky formalism with our ansatz, we evaluate \eqref{eq:weyl04} using our perturbations \eqref{newansatz}. To linear order in $\pert $, this gives
 \begin{align}\label{eq:weyl-xp}
  \Psi_0 = &- {\pert \over 2\,\sin^2\theta} \boldsymbol{l}^a\boldsymbol{l}^b\nabla_a\nabla_b \chi + O(\pert ^2)~,  \cr
   \Psi_4 =& - {\pert \over 2 J^2\,\sin^2\theta\, (1-i\,\cos\,\theta)^4}  \boldsymbol{n}^a\boldsymbol{n}^b\nabla_a\nabla_b \chi + O(\pert ^2)~,
\end{align}
where $\boldsymbol{n}$ and $ \boldsymbol{l}$ are the AdS$_2$ counterparts of \eqref{eq:NHEK-tetrad}, defined in \eqref{eq:nldef}.
Note that in deriving \eqref{eq:weyl-xp} we have not used the equations of motion for $\chi$, \eqref{eq:chi-motion}, but we did use \eqref{eq:cond1}.

The direct relation between $\chi(\x,\theta)$ and $\Psi_{0,4}$ captured by \eqref{eq:weyl-xp} establishes the physical content of our scalar perturbations $\chi(\x,\theta)$. In the next subsection we will make this connection more explicit by analyzing the solution to \eqref{eq:chi-motion} and placing it in the context of the solutions to the Teukolsky equation \eqref{eq:axi-master-NHEK}.

\subsection{Propagating modes}\label{sec:Kmodes}

In this subsection we describe the physical content encoded in the scalar perturbation $\chi(\x,\theta)$ in \eqref{newansatz}. Its equation of motion \eqref{eq:chi-motion} allows to use separation of variables
\begin{equation}
   \chi (\x,\theta) = \r{sin}^2\t\,S(\theta)\,\chi (\x)\,.
\end{equation}
It follows $S(\theta)$ satisfies  
\begin{equation}
  S^{\prime\prime} + \cot\theta\,S^\prime + \left(K- \frac{4}{\sin^2\theta}\right)\,S = 0~.
\label{eq:stheta}
\end{equation} 
This is the same linear operator appearing in the separation of variables of the Teukolsky's master equation in NHEK (see \eqref{eq:theta-NHEK}). Furthermore, $\chi (\x)$ satisfies
\begin{equation}\label{eq:delta-chi}
  \Box_2\chi(\x) = K\chi(\x)\,.
\end{equation}
At this stage $K$ is a real eigenvalue relating the angular equation \eqref{eq:stheta} to the Laplacian on AdS$_2$ in \eqref{eq:delta-chi}. Using the terminology of the AdS/CFT correspondence, the AdS$_2$ conformal dimension equals
\be
\D_\pm = {1\over 2}\pm\sqrt{{1\over 4}+K}~.
\ee
We observe that for $4K>-1$, the field $\chi(\x)$ is above the Breitenlohner-Freedman stability bound in AdS$_2$ \cite{Breitenlohner:1982bm}.

The allowed values of $K$ can be assessed from properties of the solutions to \eqref{eq:stheta}. Changing its independent variable to $x=\r{cos}\,\t$, the latter becomes
\be
\p_x \le((1-x^2) \p_x S\ri)+ \le(K- {4\/1-x^2} \ri) S = 0~.
\label{eq:int-1}
\ee
We recognise this as a particular case of the \emph{spin-weighted spheroidal harmonics} \cite{Teukolsky:1973ha}
\be\label{SWSH}
\p_x \le((1-x^2) \p_x S\ri)+ \le(\la + s +c^2 x^2 - 2 c s x - {(m+s x)^2\/1-x^2} \ri) S = 0~,
\ee
corresponding to the specific values
\be
m=c=0,\qq \la = K - s(s+1),\qq s = \pm 2~.
\ee
To identify the space of normalizable solutions, with respect to the inner product inherited from the Sturm-Liouville theory, notice
that \eqref{eq:int-1}, or \eqref{eq:stheta}, is mathematically equivalent to the standard spherical harmonics equation. Hence,
the general solution to \eqref{eq:int-1} is
\be\label{angSol}
S(\t) = c_1 \,P_\l^{(2)} (\r{cos}\,\t) + c_2\, Q_\l^{(2)}(\r{cos}\,\t)~,\qq \l ={1\/2}(-1+\sqrt{1+4 K})~,
\ee
where $P_\l^{(m)}$ and $Q_\l^{(m)}$ are the associated Legendre functions.  Requiring the solutions to be {\it smooth} and {\it normalizable} functions of $x=\cos\,\theta$ restricts $\l$ to be
\be
\l\in \Z,\qq \l \geq 2~,
\ee
and discards the $Q$-branch of solutions in \eqref{angSol}.
To sum up, regularity of the solutions to \eqref{eq:chi-motion} in the angular $\t$-direction gives rise to the mode expansion 
\be
\chi(\x,\theta) = \r{sin}^2\t\,\sum_{\l \geq 2} S_\l(\t) \chi_\l(\x)~.
\label{eq:higher-K}
\ee
The spin-weighted spherical harmonic $S_\l(\t)$ equals an associated Legendre polynomial
\be\label{eq:regSl}
S_\l(\t) = P_\l^{(2)}(\cos\,\t)~, \qq \l=2,3,\ldots ~,
\ee
and $\chi_\l(\x)$ satisfies the AdS$_2$ wave equation
\begin{equation}
  \Box_{2} \chi_\l = \l(\l+1) \chi_\l~,
\label{eq:ads2box}
\end{equation}
where we used that the separation constant is $K=\l(\l+1)$ in \eqref{eq:delta-chi} . In the AdS$_2$ terminology, these regular modes are interpreted as fields of conformal dimension $\Delta=\l+1\geq 3$. Hence, they correspond to irrelevant operators in the context of AdS/CFT. 

Requiring smoothness and normalizability is common in the discussion of gravitational perturbations when determining a basis of angular eigenfunctions.
To relate this discussion further to the traditional literature, we return to the Weyl scalars in the Teukolsky formalism: for each single mode $\ell$ in \eqref{eq:higher-K} inserted in \eqref{eq:weyl-xp}, we find 
\begin{align}\label{eq:weyl-xp1}
  \Psi_0 = &- {\pert \over 2}S_\l(\t) \, \boldsymbol{l}^a\boldsymbol{l}^b\nabla_a\nabla_b \chi_\l(\x) + O(\pert ^2)~,  \cr
   \Psi_4 =& - {\pert \over 2 J^2 }{S_\l(\t)\over(1-i\,\cos\,\theta)^4}   \boldsymbol{n}^a\boldsymbol{n}^b\nabla_a\nabla_b \chi_\l(\x)  + O(\pert ^2)~.
  \end{align}
It is evident that the same  special function $S_\ell(\t)$ controls both $\chi(\x,\t)$ and $\Psi_{0,4}$. In particular, they satisfy the same ODE \eqref{eq:theta-NHEK}. Furthermore, it is also straightforward to verify that
\be\label{eq:derivchi}
 \boldsymbol{l}^a\boldsymbol{l}^b\nabla_a\nabla_b \chi_\l(\x) ~,\qquad \boldsymbol{n}^a\boldsymbol{n}^b\nabla_a\nabla_b \chi_\l(\x)~,
\ee
correspond to $U_{s}(\x)$ in \eqref{eq:decomppsi1} with $s=\pm2$ respectively, and the radial equation \eqref{eq:radial11} is compatible with the wave equation \eqref{eq:ads2box}. All these features identify $\chi(\x,\t)$ in terms of Teukolsky modes and show that our NHEK ansatz captures all the gravitational modes in the $m=0$ sector.\footnote{Here $m$ is the Fourier mode for the azimuthal direction as defined in \eqref{eq:psi17}.} These axisymmetric normalizable  propagating modes appear as a particular case of the characterisation of general spin-2 perturbations in NHEK spacetime presented in \cite{Dias:2009ex,Amsel:2009ev,Hartman:2009nz}. As we will further discuss in Sec.~\ref{sec:UV}, this is correct for $\ell \geq 2$ because \eqref{eq:derivchi} is non-zero for these propagating modes. The discussion is subtler for the $\ell=0,1$ sectors, as we shall start discussing in Sec.~\ref{sec:chi-2}.

The two derivative combinations in \eqref{eq:derivchi} have a natural interpretation which is manifest when working in Eddington-Finkelstein coordinates, either $(u,r)$ or $(v,r)$,  
\be
u = t+{1\/r},\qq v = t- {1\/r}~.
\ee
These are smooth coordinates across the horizons allowing to write the AdS$_2$ Poincar\'e metric \eqref{eq:2dmetric} as
\be
ds^2 = -r^2 \dd v^2 + 2 \dd v \dd r = - r^2\dd u^2 -2 \dd u\dd r~.
\ee
The Weyl scalars \eqref{eq:weyl-xp1} in these coordinates simplify to
\begin{align}
&\Psi_0 = -{\pert \/2} S_\l(\t)\, \p_r^2 \chi_\l(u,r) + O(\pert ^2)\,: \quad \text{outgoing mode}\,, \cr
&\Psi_4 = -{\pert  r^4 \/8J^2} { S_\l(\t)\over(1- i\,\r{cos}\,\t)^4}\, \p_r^2 \chi_\l(v,r)+ O(\pert ^2)\,: \quad \text{ingoing mode}\,.
\end{align}
This matches the physical interpretation of $\Psi_0$ and $\Psi_4$ as describing outgoing and ingoing flux for the AdS$_2$ scalar perturbations, as customary in the Teukolsky formalism \cite{Teukolsky:1973ha}.

\subsection{Low-lying modes: $\l=0$ and $\l=1$}\label{sec:chi-2}

Based on the regularity conditions around \eqref{eq:regSl} satisfied by the functions $S_\l(\t)$, it would seem natural to end the discussion of the spectrum of axisymmetric perturbations there. However, it is worth exploring whether there is any physics in the solutions that are not regular on the sphere. We will see that these modes tamper with the global properties of the geometry. Furthermore, they do it in an interesting way that will allow us to identify the JT mode responsible for making the extremal Kerr black hole non-extremal, as we will discuss in subsequent sections.

Let us relax the smoothness and normalizable restrictions on $S_\l(\t)$ in \eqref{angSol}, by allowing meromorphic solutions on the sphere while respecting the Breitenlohner-Freedman bound in \eqref{eq:delta-chi}.  This permits two more values of $K$:
\begin{align}
K=0~: \quad \Delta =1~, ~~\ell=0~,\cr
K =2 ~:\quad \Delta= 2~, ~~\ell = 1~.
\end{align}
We have the decomposition 
\be\label{eq:modek2}
\chi(\x,\t)= \sin^2\t\sum_{\l=0,1}S_\l(\t)\, \chi_\l (\x)~,
\ee
where $S_\l(\t)$ should solve \eqref{eq:stheta}. For $\l=0$ and $\l=1$, there is no associated Legendre polynomials of the first kind but we find in each case the two linearly independent solutions\footnote{Indeed we have $P^{(2)}_\l(\r{cos}\,\t) = 0$ for $\l=0,1$. Note that one of the two solutions with $\l=0,1$ is an associated Legendre function of the second kind: $Q_0^{(2)}(\r{cos}\,\t) = {2\, \r{cos}\,\t/\r{sin}^2\t} $ and $Q_1^{(2)}(\r{cos}\,\t) = {2/\r{sin}^2\t}$.} 
\bea\label{eq:s0sol}
\l=0 &: &\qq S_0 (\t) = {1\/\r{sin}^2\t}\le(s_0^+ (1+\r{cos}^2\t) +s_0^- \,\r{cos}\,\t\ri)~,\\\label{eq:s1sol}
\l=1 &: &\qq S_1(\t) ={1\over \sin^2 \theta} \le(  s_1^+ + s_1^-\,\r{cos}\,\t\,\le(\cos^2\theta-3\ri)\ri)~,
\eea
with $s_0^\pm$ and $s_1^\pm$ constants, differing in their parity properties under $\theta\to \pi - \theta$. Both $S_0(\t)$ an $S_1(\t)$ are singular at the north and/or the south pole. A suitable choice of $s_\l^\pm$ can cancel one of the two singularities, but never both. As a consequence, they are non-normalizable with the inner product inherited from the Sturm-Liouville theory associated with the linear operator in \eqref{eq:stheta}. Alternatively, the gauge invariant Weyl scalars \eqref{eq:weyl-xp1} diverge at either the north and south pole or both in the NHEK region.

\paragraph{$\l=1$: the JT mode.}
 
Instead of disregarding the $\l=1$ sector by setting $\chi_1 (\x)=0$, we can impose a softer condition on $\chi_1 (\x)$ by requiring the Weyl scalars \eqref{eq:weyl-xp1} to vanish:\footnote{This requirement can be interpreted as demanding finiteness of the ingoing and outgoing energy fluxes associated with the perturbation, as measured by integrating $\Psi_0$ and $\Psi_4$ on the sphere \cite{Hawking:1972hy,Teukolsky:1974yv,Teukolsky:1973ha,Press:1973zz}.}
\be\label{eq:vanishingWeylchi}
\Psi_0=\Psi_4=0~.
\ee 
This leads to two constraints 
\begin{align}\label{eq:JTchiconstraints}
  \boldsymbol{l}^a\boldsymbol{l}^b\nabla_a\nabla_b \chi_1&=0~,\cr
  \boldsymbol{n}^a\boldsymbol{n}^b\nabla_a\nabla_b \chi_1&=0~,
\end{align} 
on top of the equation of motion \eqref{eq:ads2box} which reads
\be
  \Box_{2} \chi_1 = 2 \chi_1~.
\ee
A simple computation shows these three conditions are equivalent to
\be\label{eq:jteom1}
 \n_a\n_b \chi_1 - g_{ab}\, \Box_{2}\chi_1+g_{ab}\, \chi_1 = 0~,
\ee
which we recognize as the equation of motion \eqref{eq:JT} for the dilaton field in Jackiw-Teitelboim gravity \cite{Teitelboim:1983ux,Jackiw:1984je}. Hence $\chi_1$ behaves like a JT mode.

At this stage there is an important remark about the properties of $\chi_1$. Based on Wald's theorem for the Kerr geometry \cite{wald-theorem}, it is tempting to conclude that imposing \eqref{eq:vanishingWeylchi} leads to a trivial perturbation, \emph{i.e.} a diffeomorphism possibly combined with a change of mass and/or angular momentum.\footnote{Wald's theorem also implies that for Kerr perturbations, imposing $\Psi_0=0$ is equivalent to imposing $\Psi_4=0$. Our analysis shows that this conclusion is incorrect on NHEK since the first and second lines of \eqref{eq:JTchiconstraints} are independent.} However, this is the wrong conclusion since one can explicitly verify that it's impossible to cast the line element \eqref{newansatz}, under the restriction \eqref{eq:JTchiconstraints}, as a diffeomorphism. Hence, $\chi_1$ carries additional information besides its potential interpretation as a change of the constant parameters in NHEK. We will return to this point in Sec.\,\ref{sec:singularpert} as we discuss the matching conditions of perturbations that have vanishing Weyl scalar contributions and its interplay with Wald's theorem.

\paragraph{$\l=0$: marginal deformations.} The $\l=0$ mode corresponds to a marginal operator with conformal dimension $\D=1$. Hence, this should correspond to perturbations preserving extremality. As above, despite the singularities of $S_0(\t)$, we will not set $\chi_0(\x)=0$ but, consider the milder condition  $\Psi_0=\Psi_4=0$. This leads to
\begin{align}
  \boldsymbol{l}^a\boldsymbol{l}^b\nabla_a\nabla_b \chi_{0}&=0~,\cr
  \boldsymbol{n}^a\boldsymbol{n}^b\nabla_a\nabla_b \chi_{0}&=0~,
\end{align} 
which together with $\Box_{2} \chi_{0}=0$  is equivalent to the equation
\be\label{eq:chi0}
\n_a\n_b\chi_0=0~.
\ee
Its unique solution is 
\be
\chi_0 = \r{const}~.
\ee

\paragraph{Conical singularities.} The pathologies associated to the poles in \eqref{eq:s1sol} leave an imprint on the geometry, even after imposing \eqref{eq:jteom1}. Indeed, the metric of the sphere at fixed 2D coordinates $x^a$ in \eqref{newansatz} is given by
\bea\label{eq:cone}
ds^2\Big|_{\x } =   J \le(1+\r{cos}^2\t + \pert  \chi (\x,\t)\ri)\dd \t^2   + {4 J\, {\sin}^2\t\, \over 1+{\cos}^2\t } \le(1-{\pert  \chi(\x,\t)\/1+\r{cos}^2\t}\ri)\dd\phi ^2 \cr 
+ O(\pert ^2)  \,.
\eea
The troublesome points are the poles $\t=0,\pi$ where the one-form $\dd\phi$ is ill-defined. Near these points, the term linear in $\pert  $ takes the form
\begin{align}\label{eq:cone1}
ds^2\Big|_{\x}~ \underset{\t\to 0,\pi}{\sim} 2 J  \le( \dd \t^2 + \r{sin}^2\t \,\dd\phi^2\ri) +  \pert   J \,\chi (\x,\t) \le( \dd \t^2 -  \r{sin}^2\t \,\dd\phi^2\ri) \cr 
+ O(\r{sin}^4\t)+ O(\pert ^2) ~.
\end{align}
This makes manifest the presence of conical singularities at both poles.
\footnote{Note that for $\ell\geq2$, the modes are described by associated Legendre Polynomial which vanish at $\t=0,\pi$. And hence, these perturbations are well supported on the sphere as expected.} Note that such conical singularities appear at any fixed 2D coordinates $x^a$, and are therefore string-like singularities extended along the radial direction in four dimensions. Equivalently, the metric now satisfies the Einstein equation with delta function sources proportional to $\delta (\theta)$ and $\delta(\theta-\pi)$. 
We will show next that when the conditions \eqref{eq:jteom1} and \eqref{eq:chi0} are satisfied, and for the even modes $s_0^+$ and $s_1^+$, these conical singularities can be cancelled by an appropriate diffeomorphism.

\subsubsection{Balancing singularities: the other JT mode.}\label{sec:balancing}

It is interesting to identify a perturbation within the propagating sector leading to the JT mode. However, it is disappointing the latter has a conical singularity. In this section we will show that this singularity can be removed by acting with a non-single valued diffeomorphism.

In order to potentially remove the conical singularity in \eqref{eq:cone1} we will tamper with the topology of the sphere as follows.  Consider a non-single valued diffeomorphism of the form
\be\label{eq:ldfif}
\xi^\mu (\x,\t,\phi) = {\pert \/2}\phi \,\z^\mu (\x,\t)~.
\ee
This transformation clearly changes the size of $g_{\phi\phi}$, among other effects. It is also not well defined on the sphere for an arbitrary $\z^\mu (\x,\t)$. However, we will tolerate this provided the resulting Lie derivative is single-valued on the sphere, so we impose
\be\label{eq:lcond}
\p_\phi (\cL_{\xi}g_{\mu\nu}) = 0~,
\ee
where $g$ is the NHEK metric \eqref{NHEK}. For \eqref{eq:ldfif} we have 
$$\p_\phi (\cL_{\xi}g_{\mu\nu})  ={\pert\/2} \cL_\z g_{\mu\nu}~,$$ 
and hence we can comply with \eqref{eq:lcond}  provided that $\z$ is one of the NHEK Killing vectors. As in \eqref{vec-scalar}, we will write it in the basis
\be\label{eq:ddkerr}
\zeta =\varepsilon^{ba}\nabla_b\Phi \p_a+(\Phi+\varepsilon^{ab}{ A}_a\nabla_b\Phi)\p_\phi~,
\ee
where
\be\label{eq:decompphi}
\Phi (\x) = c_{\mt{($\phi$)}}+\Phi_{\mt{JT}}~, \qquad \Phi_{\mt{JT}}= c^i \Phi_{\zeta_{i}}(\x) ~.
\ee
Recall that $c_{\mt{($\phi$)}}$ parametrizes the $\fu(1)$ isometries, and $c^i$ the $\fsl(2)$ symmetry of NHEK;  $\Phi_{\mt{JT}}$ obeys the JT equations \eqref{eq:JT}. The transformation now has the desired properties: For instance, applying \eqref{eq:ldfif} with \eqref{eq:ddkerr}  to the background \eqref{background}, one finds that the fiber changes as
\be
\dd\phi+{ A}_a \dd x^a ~\rightarrow  \le(1+{\pert \/2} \Phi(\x)\ri)\dd\phi+{ A}_a \dd x^a~,
\ee
which has the effect of modifying the size of the sphere. 
Using the jargon of AdS/CFT,  this transformation can be interpreted as turning on an irrelevant deformation with $\Delta=2$ due to $\Phi_{\mt{JT}}$, and a marginal deformation, with $\Delta =1$, due to $c_{\mt{($\phi$)}}$.

 Applying \eqref{eq:ldfif}, with \eqref{eq:ddkerr},  to the perturbation \eqref{newansatz} leads to the metric\footnote{To place the result in the same gauge as in \cite{Castro:2019crn}, we add a correction term which doesn't affect the sphere, with the total diffeomorphism being
\be
\xi = {1\/2}\phi\, \z+\xi_\r{corr}~,\qq \xi_\r{corr} \equiv {1\/2} \ve^{ab} (\n^c {A}_b) \p_c\Phi \,\p_a~.
\ee} 
\bea\label{NHEKfullansatz}
ds^2 \=   J \le(1+\r{cos}^2\t + \pert  \chi (\x,\t)\ri)\le[\le(g_{ab}+ \pert h_{ab}\ri) \dd x^a \dd x^b  +\dd \t^2 \ri] \-
&& \hspace{1cm}+ 4 J\,{ {\sin}^2\t\,(1+ \pert \Phi(\x))\, \over 1+{\cos}^2\t +  \pert   \chi(\x,\t)} \le(\dd\phi +  A_a \dd x^a + \pert  {\cal A}\ri)^2  +O(\pert ^2)\,.
\eea
The fields $h_{ab}(\x)$ and $\cal A$ are determined by $\Phi(\x)$ and $\chi (\x,\t)$. The contribution of $\chi (\x,\t)$ to these modes is given by \eqref{eq:cond1} and $h_{ab}=0$; the dependence on $\Phi(\x)$ is given by
\be\label{abb}
{\cal A} =-{1\over 2}\Phi(x){A}_a \dd x^a {-}   {\le(1+{\cos^2\theta}\ri)^2\over 8\,\sin^2\theta} \ve_{a b} \nabla^b \Phi \,\dd x^a~,
\ee
and
\begin{align}\label{aaa}
 h_{ab}=  A_{(b}\ve_{a) c}  \nabla^c \Phi~.
\end{align}
 
Next, we will show how  $\Phi(\x)$ can be used to cancel the conical singularity described in \eqref{eq:cone1}. The metric of the sphere, obtained by considering a slice at fixed $\x$ in \eqref{NHEKfullansatz}, takes the form
\begin{align}
ds^2\Big|_{\x} = &J \le((1+\r{cos}^2\t)\,\dd\t^2 + {4\,\r{sin}^2\t\/1+\r{cos}^2\t} \,\dd\phi^2 \ri) \-
&+ \pert \le[ \chi(\x,\t) \,\dd\t^2 + {4\,\r{sin}^2\t\/1+\r{cos}^2\t}\le(\Phi(\x)-{\chi(\x,\t)\/1+\r{cos}^2\t} \ri)\dd\phi^2 \ri] + O(\pert ^2)~.
\end{align}
Expanding the above metric near the poles $\t=0,\pi$, we obtain the condition for the absence of conical singularity
\begin{align}
\Phi(\x) & = \chi(\x,0) & \hspace{-2cm}&\Longleftrightarrow & & \text{regular at }\t=0~,\\
\Phi(\x)  & = \chi(\x,\pi) & &\Longleftrightarrow & &\text{regular at }\t=\pi~. 
\end{align}
This condition selects the parity even solutions in \eqref{eq:s0sol}, \emph{i.e.} we need to set $s^-_{0}=0=s^{-}_1$. For the $\l=1$ ($\Delta =2$) mode, we have to equate the JT component of  $\Phi(\x)$ in \eqref{eq:decompphi}  to $\chi_1$, \emph{i.e.}
\be\label{eq:conical-erasure}
\Phi_\mt{JT}(\x) = \chi_1(\x)~,
\ee
to obtain a regular perturbation. For $\l=0$ ($\Delta =1$), demanding regularity gives
\be\label{eq:regl0}
c_{\mt{($\phi$)}}={1\over 2}\chi_0~.
\ee
Still for $\ell=0$ one could allow singular behaviours, which we will explain in the next subsection. 

Let us summarize our findings on the $\l=1$ sector of NHEK perturbations. There are two fields with conformal dimension $\Delta =2$ whose origin and main features are the following :
\begin{enumerate}
\item $\chi_1(\x)$ arises as part of the tower of AdS$_2$ modes contained within the Weyl scalars $\Psi_{0,4}$. Due to the poles in the angular eigenfunction $S_1(\t)$ \eqref{eq:s1sol}, the Weyl scalars would typically diverge at both poles \eqref{eq:weyl-xp1}. Demanding the vanishing of the Weyl scalars, the perturbation $\chi_1(\x)$ becomes equivalent to a JT mode solving the JT equations \eqref{eq:JT}. However, even after imposing the latter, this mode remains physical
and the perturbed geometry contains conical singularities.
\item $\Phi_\mt{JT}$ is generated by a non-single valued diffeomorphism modifying the size and shape of the 2-sphere, while adding a further conical singularity. Preservation of the axial symmetry again leads to the JT equations \eqref{eq:JT}.
\end{enumerate}
The combination of both modes, together with  \eqref{eq:conical-erasure}, gives rise to a smooth perturbation driving the extreme Kerr black hole away from extremality. It is this combination that we will colloquially refer to as the JT sector. We will confirm this interpretation by computing the contribution of these modes to the Iyer-Wald charges in Sec. \ref{sec:IW}. Furthermore, in Sec. \ref{sec:UV}, we will show these smooth modes can be glued to asymptotically flat modes corresponding to a change in the mass of the extreme Kerr black hole, in agreement with Wald's theorem \cite{wald-theorem}.

\subsubsection{Marginal deformations of NHEK}\label{sec:marginal}

The complete family of type D spacetimes in 4D Einstein gravity with $\L=0$ is contained in the  Pleba\'{n}ski–Demia\'{n}ski  family of solutions. In addition to the mass and angular momentum of the Kerr geometry, the metric also has a NUT parameter $ n $ and an acceleration parameter $\a$. The accelerating Kerr black hole is also known as the spinning $C$-metric \cite{Kinnersley:1970zw,Pravda:2000vh}. The upshot is that the $\l=0$ mode described above captures deformations of the NHEK corresponding to changing these parameters while preserving extremality. We only present the results here and refer to App.\,\ref{App:Delta1} for the derivations.\footnote{The parameters $n$ and $\a$ in this subsection and App.\,\ref{App:Delta1} to describe the NUT parameter and acceleration should not be confused with the Newman-Penrose variables defined in App.\,\ref{app:Teuk}. The context of the discussion should make clear the distinction.}

\paragraph{Change of extremal entropy.}
We consider the perturbation of the NHEK metric \eqref{NHEK} corresponding to a change of extremal mass
\be\label{eq:changeMmarginal}
J \ra J+ \pert\, \d J+ O(\pert ^2)~.
\ee
After taking the decoupling limit, this leads to the perturbation \eqref{NHEKfullansatz} with
\be
\chi(\x,\t) = {\d  J\/J} (1+\r{cos}^2\t)~,\qq\Phi(\x) = { 2\d J\/J}~.\label{marginal}
\ee
We recognize this as the even ($s_0^-=0$) $\D=1$ mode together with the $\Phi=c_{\mt{($\phi$)}}$ mode in \eqref{eq:regl0}, which cancels the canonical singularities at $\t=0$ and $\t=\pi$. According to the previous section, $\Phi=c_{\mt{($\phi$)}}$ is simply generated by a rescaling of the angle $\phi$:
\be\label{rescalephimarginal}
\phi \ra \le(1+{\pert \/2}c_{(\phi)} \ri)\phi~.
\ee
We will see this perturbation again in Sec.\,\ref{sec:charges-nhek} as a contribution of the marginal deformation to angular momentum.  

\paragraph{Towards the $C$-metric.} The perturbation towards the spinning $C$-metric  is obtained by taking the NUT parameter $n=0$ and the acceleration parameter
\be
\a =  \pert\, \d\a + O(\pert ^2)~.
\ee
In the extremal case $J=M^2$, the decoupling limit leads to the perturbation \eqref{NHEKfullansatz}  with
\be
\chi(\x,\t) = 4 J \d\a\,\r{cos}\,\t~,\qq\Phi(\x) = 0~.
\ee
We recognize the odd ($s^+_0=0$) $\D=1$ mode. Since $\Phi(\x)=0$, we have conical singularities at both the south and north poles of the sphere. These  singularities follow from the corresponding singularities of the $C$-metric.  We note that one of the two conical singularities can be cancelled by rescaling the angle $\phi$ according to \eqref{rescalephimarginal}. A possible physical interpretation of these singularities in terms of  meromorphic superrotations was proposed in \cite{Strominger:2016wns}.

\paragraph{Towards Kerr-NUT.} The addition of NUT charge corresponds to the perturbation  with NUT parameter
\be
 n  = \pert \,\d n  + O(\pert ^2)~.
\ee 
and with acceleration parameter $\a=0$.  In the extremal case $M^2=a^2 -n^2\equiv M^2_0 $, the decoupling limit leads to the perturbation \eqref{NHEKfullansatz}  with
\be
\chi(\x,\t) = {2 \d n \/M_0} \,\r{cos}\,\t~,\qq\Phi(\x) = {2\d n \/ M_0}~.
\ee
We see here that the $\Phi(\x) = c_{(\phi)}$ mode cancels the conical singularity at $\t=0$ but not at $\t=\pi$. This leads to a conical singularity at the south pole which is interpreted as coming from the corresponding singularity in the Kerr-NUT geometry. The constant value of $\Phi(\x)$ can be changed by rescaling the angle $\phi$ according to \eqref{rescalephimarginal}. For example, this can be used to move the conical singularity to the north pole by changing the sign of $\Phi(\x)$.

\section{Gravitational charges on NHEK}\label{sec:IW}

After decoding the spectrum of axisymmetric perturbations on NHEK,  in this section we quantify the Iyer-Wald Noether charges associated to them. The emphasis will be mostly placed on the low-lying modes with $\ell=1,0$ ($\Delta=2,1$) which affect the global properties of the NHEK background. As we will see, conservation and finiteness of the gravitational charges of these modes is tied to the global regularity requirements discussed in the prior section. 

\subsection{Review of the covariant formalism}

We would like to collect a pair of facts of the covariant formalism for gravitational charges \`{a} la Iyer-Wald \cite{Iyer:1994ys} for the Einstein-Hilbert action in four dimensions : the existence of Noether charges and their relevance to reproduce the first law of black hole thermodynamics. We recommend \cite{Compere:2018aar} for a more extensive and pedagogical review. Our main goal is to highlight that the presence of singularities and sources,
such as the conical singularities carried by some of the (near-)NHEK perturbations can still lead to finite quantities affecting the conservation properties of the ought-to-be (near-)NHEK charges.

Let $\textbf{L}[g]$ be the Lagrangian 4-form, its variation defines the 3-form pre-symplectic potential $\Theta[\delta g,g]$,  \footnote{Note that $\Theta[\delta g,g]$ is ambiguous up to terms of the form, $\Theta\to \Theta + \delta \mu + \dd Y$, which are important for defining a classical phase space in the presence of boundaries \cite{Wald:1999wa,Harlow:2019yfa}. In the following will be ignoring those contributions since they seem to not affect our final results; still it might be worth to investigate them more closely.}
 \be
\delta\textbf{L}[g]=E[g]\delta g+\dd\Theta[\delta g,g]~,
\ee
where $E[g]$ are the Einstein's equations, $g$ is the on-shell background metric and $\delta g$ is an arbitrary variation near $g$. Given a pair $\delta_i g$ $(i=1,2)$ of such variations, the Lee-Wald symplectic current is defined as
\be  
\omega[\delta_1 g, \delta_2 g]=\delta_1 \Theta[\delta_2 g,g]-\delta_2 \Theta[\delta_1 g,g]~.
\ee
Notice this is a 2-form in phase space, \emph{i.e.} with arguments $\delta_i g$, and a 3-form in spacetime. When one of the perturbations is a diffeomorphism generated by a vector field $\xi$, \emph{i.e.} $\delta_1 g = \delta_\xi g$, the integral of the symplectic current over a co-dimension one spatial region $\Sigma$ equals the infinitesimal variation of the Hamiltonian 
\be  
  \delta H_\xi [g;\Sigma]=  \int_\Sigma \omega [\delta_\xi g, \delta g]~.
\label{Hamiltonian}
\ee
Furthermore, since $\omega [\delta_\xi g, \delta g]$ is closed, up to the linearised equations of motion $\delta E_{\mu\nu}^g$, it can locally be written as an exact form, \emph{i.e.} there exists a 2-form $k_\xi$ satisfying
\be\label{eq:simk}
\omega [\delta_\xi g, \delta g]-2\delta E^g_{\mu\nu}\, \xi^\mu \ve^\nu=\dd k_\xi[\delta g,  g]~,
\ee
where $\ve_\nu$ is defined as the particular case of the volume form of a $(d+1-k)$-dimensional surface in a $(d+1)$-dimensional spacetime
\begin{equation}\label{eq:volform}
\ve_{\mu_{1} \cdots \mu_{k}}=\frac{1}{k!(d+1-k) !} \sqrt{-g} \,\ve_{\mu_{1} \cdots \mu_{k} \mu_{k+1} \cdots \mu_{d+1}} \dd x^{\mu_{k+1}} \wedge \cdots \wedge \dd x^{\mu_{d+1}}~.
\end{equation}
Using Stokes' theorem, the Hamiltonian \eqref{Hamiltonian} can be written as a surface integral 
\be  
  \delta H_\xi [g;\Sigma]-2\int_\Sigma \delta E_{\mu\nu}^g \xi^\mu \ve^\nu=  \int_{\p\Sigma} k_\xi[\delta g,  g]~.
\label{Hamiltonian2}
\ee
Notice that whenever $\xi$ is a Killing vector of the background $g$ and $\delta g$ satisfies the linearised Einstein's equations everywhere in the region $\Sigma$, the left hand side of \eqref{Hamiltonian2} vanishes. This allows to define the Noether charge for a spatial region bounded by a closed co-dimension two surface $\mathcal C$ as
\be\label{surfacecharge}
\delta \mathcal{Q}_\xi[g;\mathcal C]\equiv \int_{\mathcal C} k_\xi[\delta g,  g]~.
\ee
The latter is invariant if the surface $\mathcal C$ is deformed to another surface $\mathcal C'$ which is homologous to $\mathcal C$. This conclusion does not hold if there are sources or singularities between both surfaces $\mathcal C$ and $\mathcal C'$. This fact will play an important role in our specific discussions for linearised perturbations of near-NHEK.

Next, we review how this formalism captures the first law of black hole thermodynamics. Consider an stationary background solution containing a bifurcate Killing horizon $\mathcal H$. For an appropriate choice of constant angular velocity $\Omega_H$, the generator of this null surface is the Killing vector
\be 
\xi=\p_{t}+\Omega_H \p_{\phi}~,
\ee
with the defining properties that it vanishes on this surface, $\xi|_{\mathcal H}=0$, and the Hawking temperature ($T_H$) is given by 
\be\label{killing}
\nabla_{[\mu}\xi_{\nu ]}\Big|_{\cal H}=2\pi T_H \ve_{\mu \nu}~,
\ee
where $\ve_{\mu\nu}$ is the binormal vector on the bifurcation surface, \emph{i.e.} the entries of \eqref{eq:volform} with $k=2$ at ${\cal H}$.
The infinitesimal entropy can be defined as the Noether charge associated  with the horizon generator \cite{Wald:1993nt}
\be
  \delta S={1\over T_H} \delta \mathcal Q^{}_\xi[g,\mathcal H]~.
\label{deltaS}
\ee
For the Einstein-Hilbert action the expression for $k_\xi[\delta g,  g]$ in \eqref{surfacecharge} is 
\be
  k_\xi[\delta g,  g]=\delta \Big({1\over 16\pi} \nabla^\mu\xi^\nu \ve_{\mu\nu} \Big)-\xi\cdot \Theta[\delta g,g]~.
\label{khorizon}
\ee
Since the horizon generator vanishes at the bifurcation surface $\mathcal H$, the second term of \eqref{khorizon} vanishes, and 
the first term is a total variation. This allows to integrate the infinitesimal entropy in \eqref{deltaS} in the solution (tangent) space, recovering
the well known result that the entropy is given by the horizon area
\be 
S={{\rm Area}[\mathcal H]\over 4}~.
\ee
On the other hand, the surface charge could have also been evaluated at infinity,
\be
\delta \mathcal{Q}^{}_{\xi}[g;\mathcal C_\infty]\equiv \delta M-\Omega_H \delta J~, \ee 
where we used
\be
\delta M=\delta Q_{\p_{t}}^{}[g;\mathcal C_\infty]~,\qquad \delta J=-\delta Q_{\p_\phi}^{}[g;\mathcal C_\infty]~.
\ee
On the space of all Kerr black hole solutions parameterised by the mass and angular momentum, we can choose $\Sigma $ to be the spatial region between the event horizon and asymptotic infinity, where the variation $\delta g$ satisfies the linearised Einstein's equations with no matter sources. Charge conservation gives rise to the first law of black hole thermodynamics 
\be
T_H\delta S=\delta \mathcal{Q}^{}_{\xi}[g;\mathcal H]=\delta \mathcal{Q}^{}_{\xi}[g;\mathcal C_\infty] =\delta M-\Omega_H\delta J~.
\ee
As stressed earlier, this argument would fail if the perturbation $\delta g$ encodes a singularity. This is the case we will discuss in the next subsection for near-NHEK perturbations.

\subsection{$\fsl(2)$ charges and angular momentum}\label{sec:charges-nhek}

Let us compute the Iyer-Wald charge differences for the gravitational perturbations of Sec.\,\ref{sec:grav} associated to the isometries of near-NHEK. As discussed in Sec.\,\ref{sec:isometry}, the latter consist of three $\fsl(2)$ generators $\zeta_{\pm, 0}$ and an additional $\fu(1)$ generator $\p_\phi$. These preserve the near-NHEK (background) metric $g$, whereas $\delta g$ will be one of the axisymmetric near-NHEK perturbations, including both the propagating and the low-lying modes.  The analysis will incorporate the singular modes in the low-lying sector, paying careful attention on how the singular behaviour affects the Iyer-Wald charges; our aim is to show that one obtains finite, and non-zero, conserved charges when the perturbations are balanced appropriately to form a smooth configuration as done in Sec.\,\ref{sec:chi-2}.

Consider the near-NHEK background metric \eqref{eq:near-nhek}. The explicit expression for $k_\xi$ in \eqref{khorizon} is given by
\be \label{eq:k2f}
k_\xi[\delta g, g]=\frac{1}{8\pi }\ve_{\mu\nu}\Big(\xi^\nu\nabla^\mu\delta g^{~\sigma}_\sigma-\xi^\nu\nabla_\sigma\delta g^{\mu\sigma}+\xi_\sigma\nabla^\nu \delta g^{\mu\sigma}+\frac{\delta g^{~\sigma}_\sigma}{2}\nabla^\nu\xi^\mu-\delta g^{\sigma\nu}\nabla_\sigma\xi^\mu\Big)~.
\ee 
Since near-NHEK has a bifurcation horizon, the discussion leading to \eqref{Hamiltonian2} and \eqref{surfacecharge} applies, modulo the presence of singularities and/or sources. In particular, it is natural to define the total gravitational charge of near-NHEK perturbations as the surface charge evaluated on a closed co-dimensional two sphere $\mathcal C_\infty$ at the asymptotic boundary
$
\delta \mathcal{Q}_\xi[{\mathcal C_\infty}], 
$
while the charge of the black hole $\delta \mathcal{Q}_\xi[{\mathcal H}]$ can be evaluated at the event horizon. However, explicit calculations show these charges depend on the choice of surface whenever the perturbations $\delta g$ of near-NHEK have conical singularities at the poles located at $\theta=0,\,\pi$. This is because at these locations, Einstein's equations are \emph{not} satisfied. Notice the left hand side of \eqref{Hamiltonian2} would still vanish if we chose a source free region $\Sigma^\prime$. For instance, we can choose  $\Sigma^\prime$ to be the constant time slice $t=t_0$ between the horizon and asymptotic infinity, excluding the strings from the north and south poles. The boundary $\p\Sigma^\prime$ now contains $\mathcal H$, $ {\cal C}_\infty$, and the strings from the north and south poles. This means the following relation must be satisfied 
\be
\delta \mathcal{Q}_\xi[\mathcal C_\infty]=\delta \mathcal{Q}_\xi[\mathcal H]+\delta \mathcal{Q}_\xi[N]+\delta \mathcal{Q}_\xi[S]~,
\ee
where $\delta \mathcal{Q}_\xi[N]$ and $\delta \mathcal{Q}_\xi[S]$ stand for the extra boundary contribution at each pole. An equivalent interpretation of the above equation is to choose $\Sigma$ as the constant time slice $t=t_0$ between the horizon and asymptotic infinity. The conical singularities at the poles can be understood as adding a source localised along the spin axis. Using Einstein's equations in the presence of matter, the additional term on the left hand side of  \eqref{Hamiltonian2} is the matter stress tensor. 

We can write a more general charge conservation equation by considering 2-sphere shells ${\cal C}_r$ at a constant radius $r$, rather than at infinity and/or at the event horizon. The relation between the charge at radius $r_0$ and $r$ is given by
\begin{align}
\delta \mathcal{Q}_\xi[{\cal C}_{r}]&\equiv\oint_{{\cal C}_{r}}(k_\xi)_{\theta\phi}\cr&
=\delta \mathcal{Q}_\xi[{\cal C}_{r_0}]+\int_{r_0}^r \dd r\int_0^{2\pi}\dd\phi\Big((k_\xi)_{r\phi}\Big|_{\theta=\pi}-(k_\xi)_{r\phi}\Big|_{\theta=0}\Big)~.
\label{eq:charge-shell}
\end{align}
The latter can be derived by choosing $\Sigma^\prime$ as a region bounded by the shells at $r_0$, $r$ and the strings between both shells. What we learn is that if $k_\xi$ is a regular 2-form, the last two terms vanish. However, both the low-lying $\chi$ modes with $\ell=0,1$ and the $\Phi$ perturbation, produce conical singularities. Hence, $k_\xi$ is no longer regular and we must keep the second and third terms to comply with \eqref{eq:simk}.\footnote{The non-vanishing of these extra terms, i.e. fluxes, suggests a different physical interpretation from the more standard conical defects in 3d corresponding to  masses of bulk particles. As was mentioned below \eqref{eq:cone1}, these are string-like singularities and
the non-vanishing of these fluxes is due to delta function sources to the Einstein equation at the poles.} In the following, we present the values of these charges for the modes discussed in Sec.\,\ref{sec:grav}.

\paragraph{$\fsl(2)\times \fu(1)$ charges  for axisymmetric modes.}

Consider the metric variation $\delta g_{\mu\nu}$ due to the terms linear in $\pert $ in \eqref{newansatz}. These are metric perturbations induced by the axisymmetric modes $\chi$. We will compute the charges \eqref{eq:charge-shell} associated to the Killing vector $\zeta$, the $\fsl(2)\times \fu(1)$ generators in \eqref{vec-scalar}, evaluated on the surface at
$$\x_0=(t=t_0,r=r_0)~.$$
The result can be expressed in terms of the dual scalars $\Phi_\zeta$ in \eqref{phidecomposition} as
\begin{align}\label{chicharge}
\delta_\chi\mathcal Q_\zeta[{\cal C}_{r_0}]&=\oint_{{\cal C}_{r_0}}(k_\zeta)_{\theta\phi} \cr
&=
- \pert \, J\, \Phi_\zeta(\x_0)  \le(\frac{\cos \,\theta }{(1+\cos^2\theta)^2}\chi(\x_0,\theta) \ri)\Bigg|^{\theta=\pi}_{\theta=0}~.
\end{align}

For modes with $\ell\geq2$,  $\chi(\x,\theta)$ vanishes at the poles. Hence their contribution to the charges \eqref{chicharge} is zero. 
However, for the low-lying modes the contribution is non-trivial. For $\ell=1$ ($\Delta =2$) we select $s_1^+=1$ and $s_1^-=0$ in \eqref{eq:s1sol}, \emph{i.e.} $\chi(\x,\theta)=\chi_1(\x)$; the corresponding  $\fsl(2)$ charges are
\bea\label{chicharge1}
&&\delta_{\chi_1}\mathcal Q_\zeta[{\cal C}_{r_0}]= \frac{\pert }{2} J\, \Phi_\zeta(\x_0) \, \chi_1(\x_0) ~.
\eea
For the $\ell=0$ ($\Delta =1$) mode, we will set $\chi(\x,\theta)=(1+\cos^2\theta)\,\chi_0(\x)$ for simplicity. The corresponding charges are
 \bea\label{chicharge2}
&&\delta_{\chi_0}\mathcal Q_\zeta[{\cal C}_{r_0}]= \pert\, J\,\Phi_\zeta(\x_0) \,\chi_0 (\x_0)~.
\eea
Both $\ell=1$ and $\ell=0$ charges depend on both $t_0$ and $r_0$ for generic solutions.  The dependence on $t_0$ implies that these charges are not conserved and the dependence on $r_0$ implies the existence of sources between the two sphere shells with different radius. This is due to the fact that the $\chi (\x,\t)$ perturbation will create a conical singularity unless compensated  by a $\Phi$ mode, as discussed in Sec.\,\ref{sec:grav}.  

\paragraph{$\fsl(2)\times \fu(1)$ charges for  the global mode $\Phi$.}
Let us compute the same charges evaluated on the same surfaces due to metric variations induced by the global mode $\Phi$, as described by the metric perturbations in \eqref{NHEKfullansatz}, where 
\be
\Phi(\x) = \Phi_\mt{JT} + c_{\mt{($\phi$)}} = c^i \Phi_{\zeta_i} + c_{\mt{($\phi$)}}~,
\ee
and we set $\chi(\x,\theta)=0$. The resulting expression at ${\cal C}_{r_0}$ reads
\bea\label{phicharge}
\delta_\Phi\mathcal Q_\zeta[{\cal C}_{r_0}]&=& \frac{\pert}{4}J\Big(\Phi\Box_2 \Phi_\zeta+\Phi_\zeta\Box_2\Phi-2\nabla_a\Phi_\zeta\nabla^a\Phi-4\Phi_\zeta\Phi\Big)\Big|_{\x=\x_0}~.
\eea
For the $\fsl(2)$ charges, we can use the relation \eqref{JTmass} to reduce both $\Phi$ and $\Phi_{\zeta_i}$, leading to
\begin{align}\label{phicharge1}
\delta_\Phi\mathcal Q_{\zeta_i}[{\cal C}_{r_0}]&= \frac{\pert}{4} J\Big(-2\nabla_a\Phi_{\zeta_i}\nabla^a\Phi-2\Phi_\zeta\Phi\Big)\Big|_{\x=\x_0}\cr
&= \frac{\pert}{4} J\Big(\eta_{ij}c^j-2\Phi_{\zeta_i}\Phi\Big)\Big|_{\x=\x_0}~.
\end{align}
In the last equality we used the definition of the $\fsl(2)$ Killing form $\eta_{ij}$ defined in \eqref{kill-form}.
For the $\fu(1)$ charge we simply obtain
\be\delta_\Phi\mathcal Q_{\p_\phi}[r_0]= \frac{\pert}{4} J \le( \Box_2\Phi (\x_0) - 4 \Phi(\x_0)\ri)~.
\ee
As for the low-lying axisymmetric modes, the charges are again neither conserved nor position independent due to the presence of conical singularities. 

\paragraph{$\fsl(2)\times \fu(1)$ charges for smooth low-lying perturbations.}
As discussed in Sec.\,\ref{sec:grav},  regular perturbations require the low-lying modes of $\chi(\x,\theta)$ to be accompanied by a $\Phi$ mode satisfying
\be\label{eq:lowmodesmooth}
\chi(\x,\theta)=\Phi_\mt{JT}(\x)+\frac{1}{2}(1+\cos^2\theta)\,c_{\mt{($\phi$)}}~.
\ee 
Adding the charge formula \eqref{chicharge} for the $\chi$ mode and  \eqref{phicharge} for the $\Phi$ mode, we find four charges for the regular perturbation: the three $\fsl(2)$ charges
\be\label{sl2u1charge}
\delta\mathcal Q_{\zeta_-}=-\pert \, J\,c^+~,\quad \,\delta\mathcal Q_{\zeta_{0}}= \frac{\pert}{2} \, J\,c^0~,\quad\,\delta\mathcal Q_{\zeta_+}=- \pert \,J\,c^-~,\ee
and the $\fu(1)$ charge 
\be\label{sl2u1charge1}
\delta\mathcal Q_{\p_\phi}=- \frac{\pert}{2} J\, c_{\mt{($\phi$)}}~.
\ee
All these charges are both conserved and independent of the radius due to the regularity of these perturbations, in agreement with the general discussion based on the covariant formalism. Note that in \eqref{sl2u1charge} we wrote $\Phi_\mt{JT}(\x) = c^i\Phi_{\zeta_{i}}$, with $c^i$ constant, and used \eqref{kill-form}. The value obtained in \eqref{sl2u1charge1} is in accordance to the change in $\delta J$ discussed around \eqref{eq:changeMmarginal}-\eqref{rescalephimarginal}.

\paragraph{Thermodynamics for near-NHEK.}
Now, we aim to place the $\fsl(2)\times\fu(1)$ charges in a thermodynamic context, for which the appropriate geometry is the near-NHEK solution \eqref{eq:near-nhek}. 
This background  is by itself a black hole, with a horizon generator
\be
\xi=\p_t+\Omega_H\p_\phi~,\quad \,  \Omega_H=-\tau ~,
\ee
for which we can read the horizon and the effective temperature,
\be
r_h=\frac{\tau}{2}~,\qquad \tau_H=\frac{\tau}{2\pi}~.
\ee
The three local $\fsl(2)$ generators, up to automorphism, are given by \eqref{vec-scalar}, and the explicit profile for the  three dual scalars is
\be\label{eq:phi241}
\Phi_{\zeta_0}={1\/\tau} \le(r+{\tau^2\/4 r}\ri),\qq \Phi_{\zeta_\pm}={1\/\tau} \le(r-{\tau^2\/4 r}\ri)e^{\pm\tau t}~,
\ee
which corresponds to the Killing vectors
\be\label{SL2expbasis}
\zeta_0=\frac{1}{\tau}\p_t,\qq \zeta_\pm=\le(\frac{4 r^2+\tau^2}{\tau(4r^2-\tau^2)}\p_t\mp r\p_r-\frac{4\tau r}{4 r^2-\tau^2}\p_\phi\ri)e^{\pm\tau t}~.
\ee
The smooth perturbation for the low-lying modes, with $\ell=1,0$, is described by \eqref{eq:lowmodesmooth} where in particular
\be\label{eq:jt-nk}
\Phi_\mt{JT} = \tau \le(  c^+\Phi_{\zeta_{+}} +  c^{0}\Phi_{\zeta_{0}}  +c^-\Phi_{\zeta_{-}}\ri) ~,
\ee
with $\Phi_{\zeta_{i}}$ as in \eqref{eq:phi241} and $c^i$ are constants. The overall factor of $\tau$ is introduced for the large $r$ behavior of $\Phi_\mt{JT}$ to depend only on the constants $c^i$. 

The $\fsl(2)\times \fu(1)$ charges for near-NHEK are given by \eqref{sl2u1charge}-\eqref{sl2u1charge1}. And in particular, the energy associated to $\p_t$ and the angular momentum are
\begin{align}\label{jt-energy-ang}
\delta E&\equiv \delta\mathcal Q_{\p_t}= {\pert \over 2 } J\,\tau^2  c^0~,\cr
\delta J&\equiv-\delta\mathcal Q_{\p_\phi}= {\pert\over 2 }J\,c_{\mt{($\phi$)}}~.
\end{align}
Note that the increase of energy depends on $\tau^2$, \emph{i.e.} a quadratic response on temperature $\tau_H$ modulated by the JT field as expected from the 2D gravity arguments in \cite{Maldacena:2016upp,Almheiri:2014cka}. 

As was argued before, the variation of the entropy due to the perturbation is given by the Iyer-Wald charge associated with horizon generator
\be
\delta S\equiv {1\over \tau_H}\delta\mathcal Q_{\xi} =\frac{\delta E-\Omega_H\delta J}{\tau_H}= \pi \pert\, J\,( c^0 \tau+ c_{\mt{($\phi$)}})~.
\ee
One can easily verify that this is just the variation of the area of the bifurcation surface $\mathcal H$
\be
\delta S={\delta \text{Area}(\mathcal H)\over 4}=\pi \pert \, J\, \Phi\Big|_{r=r_h}~,
\ee
since $\Phi_{\z_\pm }|_{r=r_h}=0$ and $\Phi_{\z_0}|_{r=r_h}=1 $. Therefore, the variation of the entropy satisfies the area law as it should be.

\section{Matching near and far region perturbations}
\label{sec:UV}

In this section, we match the near and far region perturbations in Kerr. Following the near horizon perturbations presented in Sec.\,\ref{sec:grav}, we will divide the discussion into propagating modes for which the Weyl scalars are non-vanishing, and regular (smooth) low-lying modes.

Since this section focuses heavily on interpolating between Kerr and NHEK, let us reinforce the notation used in Sec.\,\ref{sec:ExtremeKerr}: variables with a tilde, such as, $\tilde g_{\mu\nu}$ or $\tilde x^\mu$, correspond to quantities in the Kerr geometry \eqref{eq:kerr}; and variables without a tilde correspond to quantities defined on near-NHEK \eqref{eq:near-nhek}.\footnote{Several of our results are valid more generally for \eqref{background}, \emph{i.e.} for any locally AdS$_2$ background. But for concreteness we will write our results for the near-NHEK geometry.} They are related by the decoupling limit  \eqref{eq:near-nhek} with the deformation   \eqref{eq:deviation}. Fig.\,\ref{fig:nhek} depicts the relevant regions of the near-extremal Kerr geometry.

\subsection{Propagating modes}\label{sec:UVpropagating}

Our aim is to extend the propagating $\chi$-modes with $\ell\geq 2$ discussed in Sec.\,\ref{sec:Kmodes} into the far region of Kerr. Since our treatment of gravitational perturbations, via the introduction of $\chi(\x,\t)$, is unconventional, we will briefly discuss how to carry out the matching procedure and place them in the context of well known results in the literature, in particular \cite{Hartman:2009nz}. This will also serve to contrast against the subtleties that arise in the low-lying sector.  

The $\chi$-modes constitute a complete set of normalizable modes with non-vanishing Weyl scalars on NHEK. There are multiple ways to perform the matching. Here, we use the Hertz map because it gives a concise relation between the perturbed metric $h_{\mu\nu}$ and a scalar potential $\Psi_{\mt{H}_0}$, called the Hertz potential. On a vacuum type-D spacetime, which encompasses Kerr, this map is given by
\bea\nt
h^\mt{IRG}_{\mu\nu} \={\epsilon}\Big\{l_{(\mu}m_{\nu)}[(D-\rho+\bar\rho)(\delta+4\beta+3\hat\tau)+(\delta+3\beta-\bar\alpha-\hat\tau-\bar\pi)(D+3\rho)] \\ 
&&-l_\mu l_\nu(\delta+3\beta+\bar\alpha-{\hat\tau})(\delta+4\beta+3\hat\tau)-m_\mu m_\nu(D-\rho)(D+3\rho)\Big\}\Psi_{\mt{H}_0}+\text{c.c.}\,
\label{hertzmap}
\eea
where $h^\mt{IRG}_{\mu\nu}$  denotes the perturbation in the ingoing radiation gauge
\be
l^\mu h_{\mu\nu}^\mt{IRG}=g^{\mu\nu}h_{\mu\nu}^\mt{IRG}=0~.
\ee
In these definitions $l^\mu$ and $m^\mu$ are the complex tetrads defined in \eqref{NPrelations}. The Einstein equation for the perturbation $h_\mn^\mt{IRG}$ translates into  the Teukolsky equation for the Hertz potential $\Psi_{\mt{H}_0}$:
\be
\Big[(\hat\Delta+3\gamma-\bar\gamma+\bar\mu)(D+3\rho)-(\bar\delta+\bar\beta+3\alpha-\bar\tau)(\delta+4\beta+3\hat\tau)-3\psi_2\Big]\Psi_{\mt{H}_0}=0~.\label{hertzeqn}
\ee
The operators and quantities that enter in \eqref{hertzmap} and \eqref{hertzeqn} are defined in App.\,\ref{app:Teuk}.

In the following we describe a three-step algorithm to relate the Hertz potential in Kerr to the $\chi$-modes. These steps involve first solving for the Hertz potential in the far and near region (which we will define below),  then reconstructing the metric in terms of $\Psi_{\mt{H}_0}$, and finally establishing the relation to $\chi(\x,\t)$. 

\paragraph{Step 1: $\tilde\Psi_{\mt{H}_0}\to\Psi_{\mt{H}_0}$.} Let $\tilde\Psi_{\mt{H}_0}$ and $\Psi_{\mt{H}_0}$ denote the Hertz potentials in Kerr and NHEK, respectively. The first task is to solve the Hertz potential on the Kerr geometry in a low frequency regime by a matching procedure to the Hertz potential in the near horizon region. 

Consider axisymmetric Hertz potentials $\tilde\Psi_{\mt{H}_0}(\tilde{\x},\theta)$ on the Kerr background, where we use $\tilde{\x}$ to collectively denote the 2D coordinates $(\tilde t,\tilde r)$. The master equation \eqref{hertzeqn} gives
\bea\label{Hertzeq2} 
\Big( {\cal L}_{\tilde\x}-2\Big)\tilde\Psi_{\mt{H}_0}(\tilde{\x},\theta)
=
{\cal L}_{\theta}\tilde\Psi_{\mt{H}_0}(\tilde{\x},\theta)+ \Big( 4i\,a\,\r{cos}\,\theta\,\p_{\tilde t}-a^2\cos^2\theta \,\p_{\tilde t}^2\Big) \tilde\Psi_{\mt{H}_0}  (\tilde{\x},\theta)~,
\eea
where the differential operator $\mathcal L_\theta$ acts on the angular coordinate $\theta$,  and $\mathcal L_{\tilde{\x}}$ acts on the 2D coordinates $\tilde{\x}=(\tilde t,\tilde r)$, 
\bea\label{defLtheta}
\mathcal L_\theta&\equiv& \p_\theta^2+\cot\theta\,\p_\theta-\frac{4}{\sin^2\theta}~,\\
\label{Ltr}
\mathcal L_{\tilde{\x}}&\equiv&\frac{(\tilde r^2+a^2)^2-a^2\Delta}{\Delta}\p_{\tilde t}^2-\Delta^2\p_{\tilde r}\Big(\Delta^{-1}\p_{\tilde r}\Big)+4\Big(\frac{M(\tilde r^2-a^2)}{\Delta}-\tilde r\Big)\p_{\tilde t}~.
\eea
Due to the terms in the parenthesis on the right hand side of  \eqref{Hertzeq2}, it is in general  not possible to have
 separation of variables in the form $\tilde\Psi_{\mt{H}_0}(\tilde{\x},\theta)=\tilde{\mathcal{U}}(\tilde{\x})S(\theta)$. One possible exception is to consider a regime of parameters where the contribution from the non-separable part is negligible compared to the other terms. While this possibility is potentially interesting, we will not explore it in this paper. 
A more obvious choice is to perform a Fourier expansion:
\be\label{tildePsiFourier}
\tilde\Psi_{\mt{H}_0}(\tilde{\x},\theta)=\int d\tilde \omega\sum_{\ell}e^{-i\tilde \omega\tilde t}\tilde R_{\ell\tilde \omega}(\tilde r)\tilde S_{\ell\tilde \omega}(\theta)~.
\ee
Plugging it into \eqref{Hertzeq2}, we get two decoupled differential equations: one for the angular dependence 
\bea
 \Big({\cal L}_{\theta} +K_{\ell\tilde \omega}\Big) \tilde S_{\ell\tilde \omega}+\Big(4a\tilde\omega\,\cos\,\theta+a^2\tilde\omega^2\cos^2\theta\Big)\tilde S_{\ell\tilde \omega}=0~,\label{ang-eqn}
\eea
and one for the radial dependence
\bea
\Big(\mathcal L_{\tilde \omega,\tilde r}-2-K_{\ell\tilde \omega}  \Big)\tilde R_{\ell\tilde \omega}=0~,\label{rad-eqn}
\eea
where $\mathcal L_{\tilde \omega,\tilde r}$ is the operator $\mathcal L_{\tilde{\x}}$ in frequency basis,  obtained from \eqref{Ltr} with the replacement $\p_{\tilde t}\to -i\tilde \omega$. 
The eigenvalue $K_{\ell\tilde \omega}$ in \eqref{ang-eqn} also serves as a separation of variable constant with the radial equation \eqref{rad-eqn}. 

Both the angular and radial equations are within the class of Heun's differential equations, for which it is not known how to construct an explicit solution for generic parameters. However, there are simplifications for the low energy excitations $a\tilde\omega\ll1$, and modes near the superradiant bound in the near-extremal Kerr background \cite{Press:1973zz,Bardeen:1999px,Bredberg:2009pv}. Since the latter permits a well defined decoupling limit, it can be analyzed in NHEK, and it is relevant for our purposes. In addition, requiring axisymmetric modes to be near the superradiant bound also implies that they have low frequencies. 
Putting all this together, we will solve for $\tilde\Psi_{\mt{H}_0}$ in a near extremal Kerr background in the regime
\be\label{eq:lowfregime}
a\tilde\omega\ll1~,\quad \tilde\tau_H\equiv \frac{r_+-r_-}{r_+}\ll1 ~, \quad \hbox{with} \quad \mathfrak n\equiv \frac{4 M}{\tilde\tau_H} \tilde \omega\quad \hbox{fixed}~. 
\ee
First, the solutions to the angular equation \eqref{ang-eqn} greatly simplify. Imposing regularity of $S_{\ell\tilde \omega}$ at $\theta=0,\pi$ determines $K_{\ell\tilde \omega}$ \cite{Press:1973zz,Abramowitz}. The corresponding solutions $S_{\ell\tilde \omega}$ are called spin-weighted spheroidal harmonics which form an orthonormal basis of functions of $\t\in[0,\pi]$ with the inner product
\be
\int^\pi_0S_{\ell\tilde \omega}(\theta)S^*_{\ell'\tilde \omega}(\theta)\,\sin\,\theta \,d\theta=2\pi\delta _{\ell\ell'}~.
\ee
For small $\tilde \omega$, we have
\be
K_{\ell\tilde \omega}\sim \ell(\ell+1) + O(\tilde \omega^2)~,
\ee
with $\ell\geq 2$ a positive integer. Therefore, in the low frequency regime $a\tilde \omega\to0$, we have $\tilde S_{\ell\tilde \omega}(\t)\to S_\ell(\t)$, \emph{i.e.} the same spin-2 spherical harmonics in \eqref{eq:theta-NHEK} and \eqref{eq:regSl} describing the angular dependence of the gravitational perturbations in NHEK given in Sec.\,\ref{sec:Kmodes}.

Second, the radial equation \eqref{rad-eqn} can be solved in both the near region ${\tilde r-r_+\over r_+} \ll1$  and the far region $\tilde\tau_H\ll{\tilde r-r_+\over r_+}$. These solutions can then be glued together by matching their asymptotic expansions in the region of overlap where $\tilde\tau_H\ll {\tilde r-r_+\over r_+}   \ll1$. This procedure has been systematically studied in the literature; see for example \cite{Hartman:2009nz}. 
This matching condition gives us an expression for $\tilde\Psi_{\mt{H}_0}$ everywhere in the regime \eqref{eq:lowfregime}, and reconstructs the metric $\tilde h^\mt{IRG}_{\mu\nu}$ through the map \eqref{hertzmap}. 

Our last and most important portion of this first step is to relate the Hertz potential in Kerr $\tilde\Psi_{\mt{H}_0}$  to the Hertz potential in NHEK $\Psi_{\mt{H}_0}$. 
The corresponding NHEK perturbation $h^\mt{IRG}_{\mu\nu}$ in IRG gauge is also related to $\Psi_{\mt{H}_0}$ by \eqref{hertzmap} and is expected to match the Kerr perturbation  in the decoupling limit \eqref{eq:near-horizon}, \emph{i.e.}  
\be\label{eq:limith}
\lim_{\lambda\to0}\tilde h^\mt{IRG}_{\mu\nu} \dd \tilde x^\mu \dd \tilde x^\nu=h^\mt{IRG}_{\mu\nu} \dd  x^\mu \dd  x^\nu~.
\ee
In order to get a well defined limit for the line element as $\lambda\to0$, we require the Hertz potential to satisfy the matching condition
\be\label{hertz-glue}
\lim_{\lambda\to0}\lambda^{-2}\,\tilde\Psi_{\mt{H}_0}\le(\tilde\x,\theta\ri) =\Psi_{\mt{H}_0}(\x,\theta)~,
\ee
where $\tilde\x$ is given in terms of $\x$ and $\la$ by the decoupling limit \eqref{eq:near-near}. 

The master equation \eqref{hertzeqn} for the Hertz potential in the NHEK  can be written as 
\begin{align}\label{hertzNHEK}
&({\mathcal L}_\x-{\mathcal L}_\theta-2)\Psi_{\mt{H}_0}(\x,\theta)=0~,
\end{align}
where ${\mathcal L}_\theta$ was defined in \eqref{defLtheta} and the differential operator acting on the 2D part is given by
\bea
\label{hertz-2d}
 {\mathcal L}_\x&\equiv& 2(\boldsymbol n^a\nabla_a )(\boldsymbol{l}^b\nabla_b) +4\boldsymbol{n}^b  A_b\, (\boldsymbol{l}^a\nabla_a)~,
\eea
One can explicitly check that the 2D differential operator in Kerr \eqref{Ltr} directly reduces to \eqref{hertz-2d}  in NHEK  
\be 
\lim_{\lambda\to0}\mathcal{L}_{\tilde \x}=\mathcal L_{\x}~.\label{ltrlimit}
\ee
This means that the master equation \eqref{Hertzeq2} reduces to the master equation \eqref{hertzNHEK} in NHEK in the decoupling limit, provided 
the terms in the parenthesis on the right hand of \eqref{Hertzeq2} are subleading. Notice the latter is indeed satisfied in our low frequency regime \eqref{eq:lowfregime}.

To make the matching more explicit, let us decompose the Hertz potential $\Psi_{\mt{H}_0}(\x,\theta)$ in Fourier modes as
\be\label{hertznhek}
\Psi_{\mt{H}_0}(\x,\theta)=\int d\omega\sum_{\ell}e^{-i\omega t}R_{\ell \omega}(r)S_{\ell}(\theta)~,\quad \mathcal L_\theta S_\ell=-\ell(\ell+1)S_\ell~.
\ee
Notice that $\mathfrak n={4M\tilde\omega\over\tilde\tau_H}={\omega\over \tau}$, where $\omega $ is the frequency used in the near horizon variables \eqref{hertznhek}. This makes the implementation of this limit compatible with \eqref{eq:lowfregime}.
Therefore solutions to the master equation \eqref{Hertzeq2} and \eqref{hertzNHEK}  will comply with 
the matching condition \eqref{hertz-glue} if we require
\begin{align}\label{Rnear2R}
R_{\ell \omega}=\lim _{\lambda \rightarrow 0} \lambda^{-2} \tilde{R}_{\ell \tilde{\omega}}\,, \quad S_{\ell}=\lim _{\lambda \rightarrow 0} S_{\ell \tilde{\omega}}~,
\end{align}
in the Fourier basis. 

Finally, let us comment on the Weyl scalars in this context.
The relation between the Weyl scalars and the Hertz potential is given by \eqref{weylhertz}. 
In NHEK, \eqref{weylhertz} simplifies significantly, and for axisymmetric modes we have 
\be\label{weylhertz-nhek}
\Psi_{0}(\x,\theta)=\frac{\pert}{2}(\boldsymbol{l}^a\nabla_a)^4\Psi_{\mt{H}_0}(\x,\theta)\,,\quad\Psi_4(\x,\theta)=\frac{\pert}{8J^2(1-i\,\cos\,\theta)^4}\mathcal L_\theta(\mathcal L_\theta+2)\Psi_{\mt{H}_0}(\x,\theta)\,.
\ee
In Kerr, the Weyl scalars can be expanded in the decoupling limit \eqref{eq:near-horizon} as
\begin{align}\label{weylglue}
&\tilde\Psi_0=\pert\frac{\lambda^{-2}}{2}((\boldsymbol{l}^a\nabla_a)^4\Psi_{\mt{H}_0}+O(\lambda))=\pert\lambda^{-2}(\Psi_{0}+O(\lambda))~,\nonumber
\\&\tilde\Psi_4=\pert\frac{\lambda^{2}}{8J^2(1-i\,\cos\,\theta)^4}(\mathcal L_\theta(\mathcal L_\theta+2)\Psi_{\mt{H}_0}+O(\lambda))=\pert\lambda^{2}(\Psi_4+O(\lambda))~,
 \end{align}
 which is indeed consistent with \eqref{hertz-glue}.

To recapitulate, the Hertz potential $\tilde\Psi_{\mt{H}_0}$ in Kerr and the Hertz potential $\Psi_{\mt{H}_0}$ in NHEK can be related by the gluing condition \eqref{hertz-glue} in the regime of parameters \eqref{eq:lowfregime}.
More explicitly, in the Fourier basis, solutions in Kerr and in NHEK are related through \eqref{Rnear2R}.

\paragraph{Step 2: Reconstruction in NHEK, $\Psi_{\mt{H}_0}\to h^\mt{IRG}$.} 
The decoupling limit of $\tilde\Psi_{\mt{H}_0}$ in \eqref{hertz-glue} leads to the Hertz potential in NHEK, which takes the form \eqref{hertznhek}. In this limit, the angular dependence reduces to $S_\ell$, which is independent of $\omega$. Hence it is also consistent to cast \eqref{hertznhek} into the form
\be\label{hertz-regular}
\Psi_{\mt{H}_0}(\x,\t)=\sum_{\ell\geq2}U_\ell(\x)S_\ell(\theta)~,
\ee
with no Fourier decomposition in time. This will allow our expressions in NHEK to be covariant with respect to the 2D coordinates $\x=(t,r)$.

Assuming $\Psi_{\mt{H}_0}(\x,\theta)$ to be real, we use \eqref{hertzmap} to reconstruct $h^\mt{IRG}_{\mu\nu}$. In the tetrad basis, we have
\be
h_\mn^\mt{IRG} ={1\/2} h_{nn}^\mt{IRG} \l_\mu \l_\nu +  h_{m m}^\mt{IRG} \bar{m}_\mu \bar{m}_\nu-{1\/2} h_{n m}^\mt{IRG} (\l_\mu \bar{m}_\nu+\l_\nu \bar{m}_\mu)+ \hc~,
\ee
with components 
\begin{equation}
\begin{aligned}
&h^{\mt{IRG}}_{nn}=-{ \epsilon}\frac{\sin^2\theta}{J(1+\cos^2\theta)^2}(\mathcal L_\theta+2)\Psi_{\mt{H}_0}~,\\
&h^\mt{IRG}_{mm}=h^\mt{IRG}_{\bar m\bar m}=-{\epsilon}\,\boldsymbol l^a\boldsymbol l^b\nabla_a\nabla_b\Psi_{\mt{H}_0}~,\\
&h^\mt{IRG}_{nm}=-{\epsilon}{(1+i\,\r{cos}\,\t)\/\sqrt{2J}\,\r{sin}^2\t} \,\boldsymbol l^a\nabla_a \p_\theta\le( {\r{sin}^2\t\/ 1+\r{cos}^2\t} \Psi_{\mt{H}_0}\ri) ~,\\
&h^\mt{IRG}_{n\bar m}=-{ \epsilon}\,{(1-i\,\r{cos}\,\t)\/\sqrt{2J}\,\r{sin}^2\t} \,\boldsymbol l^a\nabla_a \p_\theta\le( {\r{sin}^2\t\/ 1+\r{cos}^2\t} \Psi_{\mt{H}_0}\ri) ~.
\end{aligned}
\end{equation}
We are particularly interested in the angular components of the metric perturbation
\begin{align}\label{hIRG}
h^\mt{IRG}_{\theta\theta}=-\frac{(1+\cos^2\theta)^2}{4\,\sin^2\theta}h^\mt{IRG}_{\phi\phi}&=J\,\sin^2\theta\, h_{mm}^\mt{IRG}
\cr
&=-{\epsilon}J\,\sin^2\theta\, \boldsymbol{l}^a\boldsymbol{l}^b\nabla_a\nabla_b\Psi_{\mt{H}_0}~,
\end{align}
since they will suffice to illustrate how to relate $\Psi_{\mt{H}_0}$ to the $\chi$-modes. 

\paragraph{Step 3: Gauge transformation from $h^\mt{IRG}$ to $h^\chi$.}  To relate $\Psi_{\mt{H}_0}$ to the $\chi$-modes, we need to bring $h^\mt{IRG}_{\mu\nu}$ to the same gauge as in \eqref{newansatz} via a diffeomorphism $\hat\xi$
\be\label{eq:chiirg}
h^\chi=h^\mt{IRG}+ \epsilon\,\mathcal L_{\hat\xi} g_\mt{NHEK}~.
\ee
Here $h^\chi_{\mu\nu}$ denotes the metric perturbation that we used in \eqref{newansatz}.
To construct $\hat \xi$, notice that both \eqref{hIRG} and \eqref{newansatz} preserve the determinant of the sphere, \emph{i.e.} $h_{\t\t}$ and $h_{\phi\phi}$ are related according to the first line in \eqref{hIRG}. To preserve this feature, we set $\hat\xi^\theta =0$. This ensures
\begin{align}
h^\mt{IRG}_{\theta\theta}=h^\chi_{\theta\theta}~,\qquad
 h_{\phi\phi}^\mt{IRG}=h^\chi_{\phi\phi}~.
\end{align}
Choosing the diffeomorphism
\be\label{eq:difffixH}
\hat\xi=-\frac{\sin^2\theta}{1+\cos^2\theta}\,\boldsymbol{l}^a\nabla_a\Psi_{\mt{H}_0}\,l^\mu\p_\mu+\int \dd\theta\cot\theta\, \boldsymbol{l}^a\boldsymbol{l}^b\nabla_a\nabla_b\Psi_{\mt{H}_0}\p_\phi~,
\ee
the resulting components $h^\chi_{\mu\nu}$ from \eqref{eq:chiirg} satisfy all the gauge conditions in \eqref{newansatz}. Matching components, we find 
\be\label{chi-hertz}
\chi (\x,\t)=-\sin^2\theta\, \boldsymbol{l}^a\boldsymbol{l}^b\nabla_a\nabla_b\Psi_{\mt{H}_0}(\x,\t)~.
\ee
Thus, given a Hertz potential $\Psi_{\mt{H}_0}$ characterising a NHEK perturbation, this relation determines the corresponding $\chi(\x,\t)$ propagating modes. Note that \eqref{eq:weyl-xp}, \eqref{weylhertz-nhek}, and \eqref{chi-hertz} are compatible relations among the $\chi$-modes, Weyl scalars, and Hertz potential.

Following the steps $\tilde{\Psi}_{{\mt H}_0}\to \Psi_{{\mt H}_0}\to h^{\mt{IRG}}\to h^\chi$, we established a map from the UV to the IR. Namely, given a Kerr perturbation in the ingoing radiation gauge and reconstructed from a Hertz potential $\tilde\Psi _{H_0}$, we can take the decoupling limit and get a Hertz potential  $\Psi_{{\mt H}_0}$ in the near horizon region by \eqref{hertz-glue}, from which we can read the $\chi$ mode using \eqref{chi-hertz}.

Conversely, given a $\chi$ mode in the NHEK region, 
the Hertz potential $\Psi_{{\mt H}_0}$ can be determined by solving \eqref{chi-hertz}  together with the master equation \eqref{hertzNHEK}.
Then $\Psi_{{\mt H}_0}$ can further be glued to a Hertz potential $\tilde{\Psi}_{{\mt H}_0}$ in Kerr. This process can be regarded as being from the IR to the UV. In fact, \eqref{chi-hertz} can be inverted for $\ell\geq2$ allowing us to express $\Psi_{\mt{H}_0}$ in terms of $\chi$ in a compact form.
Indeed, acting with the operator $\boldsymbol{n}^a\boldsymbol{n}^b\nabla_a\nabla_b$ on both sides of \eqref{chi-hertz}, we get
\be\label{eqn-chi-hertz}
\boldsymbol{n}^a\boldsymbol{n}^b\nabla_a\nabla_b\chi
=-\frac{1}{4}\,\sin^2\theta\,\mathcal L_\theta(\mathcal L_\theta+2)\Psi_{\mt{H}_0}~,
\ee
where on the right hand side we have first used the relation \eqref{nnll}  and then the master equation \eqref{hertzNHEK}.
Using the separation of variables as in \eqref{eq:higher-K} and \eqref{hertz-regular},  the relation \eqref{eqn-chi-hertz} leads to the compact relation between the 2D parts of $\chi$ and $\Psi_{\mt{H}_0}$, 
\be\label{hertz-chi}
U_\ell(\x)=-\frac{4}{(\ell-1)\ell(\ell+1)(\ell+2)}\boldsymbol{n}^a\boldsymbol{n}^b\nabla_a\nabla_b\chi_\ell(\x)~.
\ee
Therefore, given a solution  of $\chi(\x,\t)$, we can use \eqref{hertz-chi} to obtain the corresponding Hertz potential. Altogether, relations
\eqref{chi-hertz} and \eqref{hertz-chi} give a one-to-one correspondence between the solution space of $\chi$ and $\Psi_{\mt{H}_0}$. 

This conclusion is the central result in our reconstruction of propagating modes since it provides an explicit relation between the $\chi (\x,\t)$ modes and the Hertz potential $\Psi_{\mt{H}_0}$ in NHEK, allowing us to extend the $\chi (\x,\t)$ modes to the full Kerr geometry using the asymptotic matching of the Hertz potentials in \eqref{hertz-glue}.

  \subsection{Low-lying modes: $\ell =0$ and $\ell=1$}
\label{sec:smooth}

  Having understood the reconstruction of the regular perturbations with $\ell\geq 2$, we would like to extend this result for the low-lying modes, including the smooth JT modes driving the system out of extremality. While specific parts of the propagating modes analysis are still applicable to $\ell=0,1$, such as the gauge fixing diffeomorphism \eqref{eq:difffixH} and the relation \eqref{chi-hertz}, it is also evident that  \eqref{hertz-chi} breaks down for the low-lying modes. 
All of this boils down to the delicate nature of the modes we have  discussed in Sec.\,\ref{sec:grav} and Sec.\,\ref{sec:IW}. They are intrinsically problematic since their angular dependence is supported by meromorphic functions on the sphere.  It is only after balancing these singularities that we can discuss them as well-behaved perturbations. 

In this subsection we focus on how to reconstruct smooth low-lying modes, \emph{i.e.}, those complying with the regularity conditions in Sec.\,\ref{sec:balancing}. These modes have therefore vanishing Weyl scalars and are well-behaved on the sphere. As a result, they fall into the class of Kerr perturbations considered in \cite{wald-theorem}. Our task is then to match them: starting from a smooth perturbation in Kerr with vanishing Weyl scalars, we will  identify those that are smooth and finite as we take $\lambda\to 0$ in the decoupling limit \eqref{eq:near-nhek}. This will allow us to interpret them as perturbations of the NHEK geometry, and relate them to our analysis in Sec.\,\ref{sec:balancing}.

In \cite{wald-theorem},  it was proved that smooth perturbations on Kerr black holes with vanishing Weyl scalar can only be a linear combination of changing the mass, angular momentum, and a diffeomorphism,  namely
\be 
\delta \tilde{g}=\delta_M \tilde{g}+\delta_J \tilde{g}+\pert\,\mathcal{L}_{\tilde\xi} \tilde{g}~.
\label{Kerr-pert}
\ee
Here $\delta M =O(\pert)$ and $\delta J = O(\pert)$. 
Since the NHEK perturbations we consider are axisymmetric, we also focus on axisymmetric perturbations in Kerr. This requires
\be
\p_{\tilde\phi}\delta\tilde g=\p_{\tilde\phi}\mathcal{L}_{\tilde\xi} \tilde{g}=\mathcal L_{\p_{\tilde\phi}\tilde\xi}\tilde g=0~.
\ee
This implies the vector field $\p_{\tilde\phi}\tilde\xi$ must be a Kerr isometry, \emph{i.e.} a linear combination of $\p_{\tilde t}$ and $\p_{\tilde\phi}$.  
However, when integrating for the vector field $\tilde\xi$, the non-vanishing part of $\p_{\tilde\phi}\tilde\xi$ leads to a linear piece in $\tilde\phi$. As discussed in Sec.\,\ref{sec:grav}, such non-single valued diffeomorphisms create conical singularities and  we will not be considered in this section.
Hence, from now on, we restrict to the case $\p_{\tilde\phi}\tilde\xi=0$. We will now show how the different perturbations allowed by Wald's theorem in \eqref{Kerr-pert} are related to perturbations in the near horizon limit as presented in \eqref{NHEKfullansatz}.

Before starting this analysis, let us introduce some convenient notation. Given a tensor $\tilde T$ in Kerr, the near horizon expansion is given by performing the coordinate transformation \eqref{eq:near-near} and expanding in $\lambda$. This procedure defines a tensor $T[\lambda]$ in NHEK given by
\be
T[\lambda]^{\mu_1\cdots\mu_n}_{\hspace{0.90cm}\nu_1\cdots\nu_m}=\frac{\p x^{\mu_1}}{\p\tilde x^{\alpha_1}}\cdots \frac{\p x^{\mu_n}}{\p\tilde x^{\alpha_n}}\frac{\p \tilde x^{\beta_1}}{\p x^{\nu_1}}\cdots \frac{\p\tilde x^{\beta_m}}{\p x^{\nu_m}}\tilde T^{\alpha_1\cdots\alpha_n}_{\hspace{1cm}\beta_1\cdots\beta_m}~.
\ee
Note that $T[\la]$ is just the pullback of $\tilde{T}$  from Kerr to NHEK  under the map $x\mapsto \tilde{x}$  given in \eqref{eq:near-near}. We then study the tensor $T[\la]$ in an expansion in $\la$. Now we discuss Kerr perturbations given by \eqref{Kerr-pert} surviving the near horizon limit \eqref{eq:near-near}.

\paragraph{Decoupling limit with no additional perturbations.} Before turning on any perturbation, let us consider the $\lambda$ expansion of the near extremal Kerr metric $\tilde g$ itself
\be\label{eq:g111}
g[\lambda]=g_\mt{NHEK}+\lambda\, g_{\mt{(1)}}+\cdots~.
\ee
The leading order term $g_\mt{NHEK}$ gives the near-NHEK metric \eqref{eq:near-nhek}. The $O(\lambda)$ term $g_{(1)}$ satisfies the near-NHEK linearized Einstein's equations and can be viewed as a near-NHEK perturbation with perturbative parameter $\pert \sim \lambda$. It is this
agreement that motivated the ansatz in \eqref{NHEKfullansatz} as originally introduced in \cite{Castro:2019crn}. 

Let us be more precise. We can show that $\lambda g_{\mt{(1)}}$, together with the radial redefinition
\be
r\to r+\lambda\frac{\tau^2r^2}{\sqrt{J}(4r^2-\tau^2)} ~,
\ee
gives a time independent perturbation of the type described in \eqref{NHEKfullansatz} with
\begin{align}\label{flow-mode}
\pert\,\chi(\x,\t)= \pert\,\chi_1(\x)=\pert\,\Phi_\mt{JT}(\x)
&= {2\la\/ \sqrt{J}}\le(r+{\tau^2\/4 r} \ri).
\end{align}
This means the coefficients in \eqref{eq:jt-nk} are given by
\be\label{anabasis} 
c^0 = {2\la\/\pert \sqrt{J}},\qq c^\pm = 0~.
\ee
We learn the leading near horizon expansion mode can be interpreted as a particular smooth perturbation in near-NHEK
\be
\lambda\,  g_{\mt{(1)}}=\big(\delta g\big)_{\chi=\Phi_\mt{JT}}~,
\ee 
where $\d g$ refers to \eqref{NHEKfullansatz}. In particular, we note that the angular components of $g_{(1)}$ can be written as
\be\label{g1com}
(g_{\mt{(1)}})_{\mt{$\theta\theta$}}=2\sqrt{J}\,\tau\,\Phi_{\p_t}~,  \quad   (g_{\mt{(1)}})_{\mt{$\phi\phi$}}=\frac{4\sqrt{J}\,\sin^2\theta\,\r{cos}^2\t}{(1+\cos^2\theta)^2}\,\tau\Phi_{\p_t}~.
\ee
These are the same perturbations recently discussed in \cite{Hadar:2020kry}, and also discussed in \cite{Castro:2018ffi} for five-dimensional black holes. Notice that by performing a finite $\mathrm{SL}(2,\R)$ coordinate transformation
on \eqref{eq:g111}, the near-NHEK background metric $g_\mt{NHEK}$ remains invariant, but one can get more general JT modes \eqref{eq:jt-nk} than the stationary one in \eqref{flow-mode}. Of course, this is consistent with the symmetry breaking discussion reviewed in App.\,\ref{sec:nearAdS2}. $g_\mt{NHEK}$ is invariant under the full $\mathrm{SL}(2,\R)$, but NHEK perturbations captured by JT modes \eqref{eq:jt-nk} break this isometry group. The specific perturbation coming from Kerr is time translationally invariant, hence it must correspond to \eqref{anabasis}, \emph{i.e.} it must be associated with $\Phi_{\p_t}$. However, from a purely IR perspective, the action of $\mathrm{SL}(2,\R)$ on the latter can still generate the full multiplet of NHEK perturbations. 

Since the interpretation of the near-NHEK perturbation requires $\pert \sim \lambda$, this mode disappears when $\lambda=0$. In the following, we will consider near-NHEK perturbations surviving the $\lambda\to0$,  \emph{i.e.} keeping $\pert $ fixed as  $\lambda\to0$. 

\paragraph{Diffeomorphism sector.} 
Let us start our analysis of Kerr perturbations consistent with vanishing Weyl scalars, see \eqref{Kerr-pert}, by focusing on those generated by a
diffeomorphism parameterized by the vector $\tilde\xi$, \emph{i.e.} we set  $\delta_M \tilde{g}=\delta_J \tilde{g}=0$ in \eqref{Kerr-pert}. This vector can be expanded in $\lambda$ as 
\be\label{expand-xi}
\xi[\lambda]=\lambda^{n}(\xi_{\mt{(n)}}+\lambda\xi_{\mt{(n+1)}}+\cdots)~,
\ee
where $n$ is an integer which can be negative. The corresponding diffeomorphism acting on Kerr admits a $\lambda$ expansion according to
\be\label{eq-purediff}
\lambda^{n}\Big(\mathcal L_{\xi_{\mt{(n)}}}g_\mt{NHEK}+\lambda(\mathcal L_{\xi_{\mt{(n+1)}}}g_\mt{NHEK}+\mathcal L_{\xi_{\mt{(n)}}}g_{\mt{(1)}})+\cdots\Big)~.
\ee
Our goal is to restrict the form of $\xi[\lambda]$ by demanding regularity of \eqref{eq-purediff}. We claim $n\geq -1$ in order to have a finite perturbation as $\lambda\to0$. To prove this, suppose $n\leq -2$. Regularity requires the first two terms in the expansion to
vanish 
\be
\mathcal L_{\xi_{\mt{(n)}}}g_\mt{NHEK}=0~,\quad \mathcal L_{\xi_{\mt{(n+1)}}}g_\mt{NHEK}=-\mathcal L_{\xi_{\mt{(n)}}}g_{\mt{(1)}}~.
\ee
The first equation implies that $\xi_{\mt{(n)}}$ has to be a NHEK isometry, so that
\be\label{xi-1}
\xi_{\mt{(n)}}=a^-\zeta_-+a^0\zeta_0+a^+\zeta_++a^\phi\zeta_{\mt{$(\phi)$}}~.
\ee
We already showed $g_{\mt{(1)}}$ is a particular mode with $\chi=\Phi_\mt{JT}$. Hence, $g_{\mt{(1)}}$ is not pure gauge. Applying the isometry \eqref{xi-1} to $g_{\mt{(1)}}$ simply rewrites such modes in a different coordinate system keeping the same metric. Thus $\mathcal L_{\xi_{\mt{(n)}}}g_{\mt{(1)}}$ can still not be written as a diffeomorphism acting on NHEK, and the second equation in \eqref{eq-purediff} has no solution for $\xi_{\mt{(n+1)}}$. Therefore, we conclude that $n\geq -1$ . 

Let us consider $n=-1$. The expansion \eqref{expand-xi} starts from ${\lambda^{-1}}\xi_{\mt{(-1)}}$, where $\xi_{\mt{(-1)}}$ is an isometry of NHEK as in \eqref{xi-1} in order to make \eqref{eq-purediff} finite. It follows that the most general diffeomorphism surviving the decoupling limit is given by
\be\label{hdiff}
\hat h\equiv \mathcal L_{\xi_{\mt{(-1)}}}g_{\mt{(1)}}+\mathcal L_{\xi_{\mt{(0)}}}g_\mt{NHEK}~.
\ee   
Next we adjust the subleading term  $\xi_{\mt{(0)}}$ such that $\hat h$ takes the form \eqref{NHEKfullansatz} of our NHEK perturbations. This gives the following conditions
\bea\label{gauge1}
&&\p_\theta \Big(\delta \log \det g_{S^2}\Big)=\p_\theta\Big(g_{\theta\theta}^{-1}\hat h_{\theta\theta}+g_{\phi\phi}^{-1}\hat h_{\phi\phi}\Big)=0~,\\
&&g_{\theta\theta}^{-1}\hat h_{\theta\theta}+g_{\phi\phi}^{-1}\hat h_{\phi\phi}\equiv  \,\Phi(\x)~,\label{gauge2}\\
&&\hat h_{\theta\theta}\equiv  J\,\chi(\x,\t)~,\label{gauge3}\\
&&\hat h_{t\theta}=\hat h_{r\theta}=\hat h_{\phi\theta}=0~.\label{gauge4}
\eea
In addition we have to adjust the AdS$_2$ components of \eqref{NHEKfullansatz}, on which we will comment briefly.  Using  \eqref{g1com}, and the isometry-scalar relation \eqref{vec-scalar}, we get
\begin{align}\label{lg1}
\Big[\mathcal L_{\xi_{\mt{(-1)}}}g_{\mt{(1)}}\Big]_{\theta\theta}&=2\tau\sqrt{J}\,\xi_{\mt{(-1)}}^a\nabla_a\Phi_{\p_t}=2\tau \sqrt{J}\,\Phi_{[\xi_{\mt{(-1)}},\p_t]}~,\\
\Big[\mathcal L_{\xi_{\mt{(-1)}}}g_{\mt{(1)}}\Big]_{\phi\phi}&=\frac{2\tau\sqrt{J}\,\sin^2(2\theta)}{(1+\cos^2\theta)^2}\Phi_{[\xi_{\mt{(-1)}},\p_t]}~,
\end{align}
where \eqref{eq:slLie} was used to write the right hand side as the dual scalar of a commutator. While $\mathcal L_{\xi_{\mt{(-1)}}}g_{\mt{(1)}}$ already satisfies the condition \eqref{gauge1}, requiring the latter for $\mathcal L_{\xi_{(0)}}g_\mt{NHEK}$ amounts to 
\be
\p_\theta^2\xi^\theta_{\mt{(0)}}+\cot\theta\,\p_\theta\xi^{\theta}_{\mt{(0)}}-\frac{1}{\sin^2\theta}\xi^\theta_{\mt{(0)}}=0~.
\ee
This is the spin-weighted spherical harmonics equation \eqref{eq:theta-NHEK}, with $K=0$ and spin 1, whose non-vanishing solutions are 
always divergent at the poles. Therefore for smooth, axisymmetric perturbations, we conclude that $\xi_{\mt{(0)}}^\theta=0$ and $\xi_{\mt{(0)}}$ will not contribute to $\hat h_{\theta\theta}$ and $\hat h_{\phi\phi}$.  
The remaining gauge conditions \eqref{gauge2}-\eqref{gauge4}, when combined with \eqref{lg1}, lead to a consistent matching of the $\hat h_{\mu\nu}$ components provided we identify 
\be
\chi(\x)=\chi_1(\x)~, \qquad \Phi =\Phi_\mt{JT}~,
\ee
together with the constraint
\be 
2 \Phi_{[\xi_{\mt{(-1)}},\p_t]}=\sqrt{J} \,\chi_1 =\sqrt{J}\,\Phi_\mt{JT}= \sqrt{J}\,\Phi_{c^i\zeta_i}~.
\ee
The latter reduces to a relation between Killing vectors 
\be 
  2 [\xi_{\mt{(-1)}},\p_t]=\sqrt{J}\, c^i\zeta_i~.
\label{eq:-1matching}
\ee
Hence, given a diffeomorphism with near horizon expansion \eqref{expand-xi} starting at $n=-1$, the isometry $\xi_{\mt{(-1)}}$ with arbitrary parameters as in \eqref{xi-1}, determines a near-NHEK perturbation of the form \eqref{NHEKfullansatz} that is fully determined by \eqref{eq:-1matching} using the $\mathfrak{sl}(2)$ commutation relations. Solving for the constants $c^i$ gives
\be
c^0 = 0,\qq c^\pm = \mp {2\tau \/\sqrt{J}}a^\pm\,.
\label{eq:cs-diff}
\ee
We can read the corresponding $\mathfrak{sl}(2)\times \mathfrak{u}(1)$ charges of the near-NHEK perturbation in \eqref{sl2u1charge}-\eqref{sl2u1charge1}. However, we see from  \eqref{jt-energy-ang} that both the energy and the angular momentum are zero in the near horizon region. This is expected since these perturbations originate from the decoupling limit of a diffeomorphism in the full Kerr geometry.

We have determined $\xi_{\mt{(-1)}}$ up to the isometries of Kerr $\p_{\tilde{t}}$ and $\p_{\tilde\phi}$ which act trivially. Now we need to solve the remaining gauge fixing conditions that $\hat h_{\mu\nu}$ should satisfy
to determine $\xi_{\mt{(0)}}$. This is straightforward, and the details just depend on the choice of AdS$_2$ coordinates in \eqref{NHEKfullansatz} and residual transformations that affect $h_{ab}$ and ${\cal A}_a$ in \eqref{abb}-\eqref{aaa}. In general it takes the form 
\be\label{xi-0}
\xi_{\mt{(0)}}= \xi^\mt{res} + \xi^\mt{sub}
\ee
where $\xi^{\mt{sub}}$ is determined by requiring $h_{ab}$ to satisfy \eqref{aaa} and $\xi^{\mt{res}}$ is the residual gauge transformation originating from the ambiguity of $\mathcal A$ in solving \eqref{eq:cond1}, 
\bea
\xi^{\mt{sub}}\=\frac{4\tau r(r^4-4\tau^2r^2+16\tau^4)(e^{\tau t}a^++e^{-\tau t}a^-)}{\sqrt{J}(4r^2-\tau^2)^3}\p_t+\frac{\tau^2 r^2(4r^2+\tau^2)(e^{\tau t}a^+-e^{-\tau t}a^-)}{\sqrt{J}(4r^2-\tau^2)^2}\p_r\nonumber
\\\xi^{\mt{res}}\=f(t,r)\,\p_\phi\label{xires}
\eea
where $f(t,r)$ is an arbitrary function.  One can explicitly check that the perturbation generated by $\xi_{\mt{(0)}}$ in \eqref{hdiff} carries no $\fsl(2)\times\fu(1)$ charges. 

Finally if the transformation in \eqref{expand-xi} starts from the zeroth order in $\lambda$, then the diffeomorphism surviving the decoupling limit and preserving the gauge choices \eqref{gauge1}-\eqref{gauge4} is also given by $\xi^\mt{res}$,  which is a pure diffeomorphism in NHEK that is smooth and has trivial Iyer-Wald charges associated to it. 

To summarize, a general diffeomorphism in Kerr that reduces to a finite metric perturbation in NHEK has an expansion \eqref{expand-xi} starting from $n=-1$, whose first two leading orders are 
\be\label{purediffeo}
\xi[\lambda]=\lambda^{-1}(a^i\zeta_i+a^\phi\zeta_{\mt{$(\phi)$}}+\lambda\xi_{\mt{(0)}}+\cdots)~.
\ee
The near horizon limit of the perturbation takes the form \eqref{hdiff}, and it can be identified with a near horizon perturbation \eqref{NHEKfullansatz} with $\chi=\Phi_\mt{JT}$ carrying $\mathfrak{sl}(2)\times \mathfrak{u}(1)$ charges \eqref{sl2u1charge}-\eqref{sl2u1charge1} evaluated for \eqref{eq:cs-diff}.

\paragraph{Mass perturbation.} 
Let us continue with the analysis of the individual Kerr perturbations in \eqref{Kerr-pert} by turning on a mass perturbation while keeping the angular momentum fixed. Expanding the mass variation $\delta_M g$ of the Kerr metric in the decoupling limit \eqref{eq:near-near} gives
\be\label{eq:dM11}
\delta_M g[\lambda]=\delta M\Big(\lambda^{-2}  h_\mt{M}^{\mt{(-2)}}+ \lambda^{-1}  h_\mt{M}^{\mt{(-1)}}+h_\mt{M}^{\mt{(0)}}+\cdots\Big)~.
\ee 
Hence, to have a finite limit as $\lambda\to0$, the mass perturbation $\d M$ should be of order $\lambda^2$. The surviving perturbation is given by
\be\label{eq:pertMass2}
h_\mt{M}^{\mt{(-2)}} = 4 J^{3/2}(1+\r{cos}^2\t)\le( \dd t^2 + {16\/(4 r^2-\tau^2)^2} \dd r^2\ri)~.
\ee
From the near horizon point of view, the near-NHEK background with temperature $\tau_H=\frac{\tau}{2\pi}$ is perturbed to a nearby near-NHEK with temperature
\be
\tau'_H=\tau_H\le(1+\sqrt{J}\,\frac{\lambda^{-2}\delta M}{2\pi^2 \tau^2_H}\ri)~.
\ee
as can be directly seen from the near extremal limit \eqref{eq:deviation}.  In AdS$_2$ language, mass perturbation of order $\lambda^2$ change the AdS$_2$ temperature without turning on any dynamics. Indeed, one can explicitly check that this mode carries no $\mathfrak{sl}(2)\times \mathfrak{u}(1)$ charges.

In order to have dynamical fields in NHEK emerging in the $\lambda\to 0$ limit, we turn to a scenario with  $\delta M \sim\lambda $ or $\delta M \sim\lambda^{0}$. In this case, the mass perturbation cannot survive the limit by itself. However, \eqref{Kerr-pert} allows us to combine the $\delta M$ perturbation with a diffeomorphism, that we shall specifically denote by $\tilde\xi^\mt{M}$, so that the combined perturbation is finite in the decoupling limit.

Given the conditions we found around \eqref{eq-purediff} and the order of the divergences appearing in \eqref{eq:dM11} when $\delta M$ is order $\lambda^1$ or $\lambda^{0}$, the expansion of the diffeomorphism must be of the form
\be  
\label{xiMexp}
  \xi^\mt{M}[\lambda]=\delta M \Big(\lambda^{-2}\xi^\mt{M}_{\mt{(-2)}} + \lambda^{-1} \xi^\mt{M}_{\mt{(-1)}} +\xi^\mt{M}_{\mt{(0)}}+\cdots\Big)~,
\ee
The resulting metric perturbation from the combined effect of \eqref{xiMexp} and \eqref{eq:dM11} is
\be\label{mass-dif-expand}
\delta g[\lambda]=\delta M \Big(\lambda^{-2}\Big(h_\mt{M}^{\mt{(-2)}}+\mathcal L_{\xi^{\mt{M}}_{\mt{(-2)}}}g_\mt{NHEK}\Big)+\lambda^{-1}\Big(h_\mt{M}^{\mt{(-1)}}+\mathcal L_{\xi^\mt{M}_{\mt{(-2)}}}g_{\mt{(1)}}+\mathcal L_{\xi^\mt{M}_{\mt{(-1)}}}g_\mt{NHEK}\Big)+\cdots\Big)~.
\ee
The leading divergent term requires 
\be
h_\mt{M}^{\mt{(-2)}}+\mathcal L_{\xi^\mt{M}_{\mt{(-2)}}}g_\mt{NHEK}=0~.
\ee
This determines 
\be\label{xiM-2}
\xi^\mt{M}_{\mt{(-2)}}=\frac{2\sqrt{J}}{\tau^2}\le(-t \,\p_t + \frac{4r^2+\tau^2}{4r^2-\tau^2} \,r \,\p_r\ri)+\zeta~,
\ee
up to a Killing vector $\zeta$ of the near-NHEK background. Since we already discussed the contribution of $\zeta$ to the NHEK perturbation in the preceding paragraph on the diffeomorphism sector, we will set $\zeta=0$ in the following. 

The vanishing of the leading order $\lambda^{-1}$ coefficient in \eqref{mass-dif-expand} requires
\be
{-}\mathcal L_{\xi^\mt{M}_{\mt{(-1)}}}g_\mt{NHEK} = h_\mt{M}^{\mt{(-1)}}+\mathcal L_{\xi^\mt{M}_{\mt{(-2)}}}g_{\mt{(1)}}~.
\ee
However, since the source in the right hand side is not pure gauge, there exists no vector field $\xi^\mt{M}_{\mt{(-1)}}$ mapping the near-NHEK metric to this physical perturbation. Thus, for this combined metric and diffeomorphism perturbation to have a finite decoupling limit, $\delta M$ must be at least of order $\lambda$. The corresponding surviving perturbation in NHEK is given by 
\be
\lim_{\lambda\to0}\delta g[\lambda]=\delta M\,\lambda^{-1} \le(h_{\mt{M}}^{\mt{(-1)}}+\mathcal L_{\xi^\mt{M}_{\mt{(-2)}}}g_{\mt{(1)}}+\mathcal L_{\xi^\mt{M}_{\mt{(-1)}}}g_\mt{NHEK}\ri)\equiv h^\mt{M}~.\label{dgM}
\ee
To compare with the low-lying modes in Sec.\,\ref{sec:grav}, we transform $h^\mt{M}_{\mu\nu}$ to the appropriate gauge as described in \eqref{gauge1}-\eqref{gauge4}. This imposes $(\xi^\mt{M}_{\mt{(-1)}})^\theta=0$. Furthermore, comparing $h^\mt{M}_{\theta\theta}$ and $h^\mt{M}_{\phi\phi}$ with \eqref{gauge2} and \eqref{gauge3}, we learn  
\be\label{eqn-mass-jt}
\pert\,\chi(\x,\t)=\pert\,\Phi(\x)={4\delta M\over\lambda\tau^2}\le(r+{\tau^2\/4r} \ri)~.
\ee
 This allows us to identify the constants $c^i$, defined in \eqref{eq:jt-nk}, that determine the low-lying mode perturbation in \eqref{NHEKfullansatz}. We learn that
\be
c^\pm = 0,\qq c^0=\frac{4\delta M}{\pert\lambda\tau^2} ~.
\label{m2c}
\ee
Finally, the remaining vector field $\xi^\mt{M}_{\mt{(-1)}}$ can be determined by requiring $h^\mt{M}$ to satisfy \eqref{gauge3} and \eqref{gauge4}. This gives
\begin{align}\label{xiM-1}
\xi^\mt{M}_{\mt{(-1)}}=&
\frac{2r^2}{4r^2-\tau^2}\p_r+\xi^\mt{res}~,
\end{align}
where $\xi^\mt{res}$ is given by \eqref{xires}. 

It is reassuring to check the value of $\delta M$ in \eqref{m2c} can also be determined using the relation between charges in Kerr and in NHEK. According to the coordinate transformation \eqref{eq:near-horizon}, we have
\be
\p_t=\frac{2J}{\lambda}\p_{\tilde t}+\frac{\sqrt{J}}{\lambda}\p_{\tilde\phi}\,,\qq \p_\phi=\p_{\tilde\phi}~.
\ee
Therefore, the charges in Kerr and near-NHEK are related by
\bea\label{charge-kerr-nhek}
\delta\mathcal Q_{\p_t}=\frac{2J}{\lambda}\delta\mathcal Q_{\p_{\tilde t}}+\frac{\sqrt{J}}{\lambda}\delta\mathcal Q_{\p_{\tilde\phi}}~,\quad \delta\mathcal Q_{\p_\phi}=\delta\mathcal Q_{\p_{\tilde\phi}}~.
\eea
A mass perturbation of Kerr corresponds to $\delta\mathcal Q_{\p_{\tilde t}}=\delta M$ and $\delta\mathcal Q_{\p_{\tilde \phi}}=0$, giving the NHEK charges
\be
\delta\mathcal Q_{\p_{t}}=\frac{2J}{\lambda} \delta M\,,\qq \delta\mathcal Q_{\p_{\phi}}=0~.
\ee
Comparing with \eqref{jt-energy-ang}, we recover
\be
\delta M=\pert \lambda\frac{c^0\tau^2}{4}~.
\ee
 
To summarize, Kerr mass perturbations $\delta M\sim \lambda^2$ can be interpreted as changing the temperature in the near-NHEK metric. When $\delta M\sim \lambda$, they can be combined with an additional diffeomorphism, as in \eqref{xiMexp}, and explicitly given by \eqref{xiM-2} and \eqref{xiM-1}, to give a $\chi=\Phi_\mt{JT}$ mode in near NHEK with coefficients given by \eqref{m2c}. Note that the mass+diffeo perturbation \eqref{m2c} has the same structure as the anabasis mode \eqref{anabasis}. However, while the latter has $\pert \sim \la$, the mass perturbation has $\pert \sim \la^0$. Hence, they belong to different parameter regions. One intuition for the similarity is that adding finite energy to AdS$_2$ corresponds to an RG flow from the conformal fixed point at the IR towards UV, and hence corresponds to a deviation from the near horizon throat.

As a final remark, we restricted ourselves to the {\it minimal } diffeomorphism necessary to cancel the divergent part of \eqref{eq:dM11}. Note that adding a pure diffeomorphism of the form \eqref{purediffeo} still leads to a perturbation surviving the decoupling limit as $\lambda\to0$. The resulting matching would be modified accordingly and in general will be different from \eqref{eqn-mass-jt}. 

\paragraph{Angular momentum perturbation.} The discussion of angular momentum perturbations is technically similar to the mass perturbation one. Let us just summarize the conclusion. To make $\delta_J\tilde g$ finite in the decoupling limit, we need $\delta J\sim \lambda^2$. This perturbation changes the IR temperature to $\delta \tau_H=\tau_H(1-\frac{\delta J}{\lambda^2\tau^2})$.

When $\delta J\sim \lambda$, $\delta_J\tilde g$ is divergent in the decoupling limit. However, we can combine the latter with a diffeomorphism $\xi^J$ to cancel this divergence. Choosing $\xi^J$ appropriately, we can also match the JT mode which nonzero energy, $\delta\mathcal Q_{\p_t}$, in near-NHEK. In particular, we have 
\be
\delta J=\pert \la \frac{ c^0\tau^2}{4\sqrt{J}}~,
\ee 
which agrees with \eqref{charge-kerr-nhek}. Note that $\delta\mathcal Q_{\p_\phi}$ is zero as we take $\lambda\to0$. When $\delta J\sim \la^0$, the divergence can not be removed by a diffeomorphism, as in the discussion of mass perturbations.

\paragraph{Marginal deformation.}
A general Kerr perturbation with vanishing Weyl scalars \eqref{Kerr-pert} can be glued to a linear combination of perturbations in near-NHEK.  
In the following, we discuss the particular case 
\be
\delta J=2\sqrt{J}\,\delta M
\label{eq:marg}
\ee
preserving the extremality condition $J=M^2$. This property is responsible for the cancellation of the leading order term in the $\lambda$-expansion, leading to a perturbation of the form
\be
(\delta_M+\delta_J)\,g[\lambda]=\delta M(\lambda^{-1}h^{\mt{(-1)}}+h^{\mt{(0)}}+\cdots)~.
\ee
When $\delta M\sim \lambda$, the perturbation surviving the limit is given by the leading term
\be
h^{\mt{(-1)}} = J{(1+\cos^2\theta)(4r^2+\tau^2)\/2r} \le(  \dd t^2 + \frac{16}{(4r^2-\tau^2)^2} \dd r^2\ri)~.
\ee
which can be written as a diffeomorphism with 
\bea\label{ximinus1}
h^{\mt{(-1)}}+\mathcal L_{\xi_{\mt{(-1)}}}g_\mt{NHEK}=0~,\quad \text{with} \quad   \xi_{\mt{(-1)}}=\frac{4r^2}{4r^2-\tau^2}\p_r+t\,\p_\phi~.
\eea
Such perturbation keeps the temperature invariant and carries no charges. Thus, this is a trivial diffeomorphism. 

When $\delta M\sim \la^0$, the marginal perturbation is divergent in the near horizon limit, but
we can cancel the leading divergence with a diffeomorphism 
\be
\xi[\lambda]=\delta M\le(\lambda^{-1}\xi_{\mt{(-1)}}+\xi_{\mt{(0)}}+\cdots\ri)~,
\ee
where $\xi_{\mt{(-1)}}$ is given by \eqref{ximinus1}, and $\xi_{\mt{(0)}}$ is chosen such that the perturbation surviving the limit 
\be
\delta M\le(h^{\mt{(0)}}+\mathcal L_{\xi_{\mt{(-1)}}}g_{\mt{(1)}}+\mathcal L_{\xi_{\mt{(0)}}}g_\mt{NHEK}\ri)~,
\ee
satisfies the gauge conditions \eqref{gauge1}-\eqref{gauge4}.
As in previous discussions, analyzing the $h_{\theta\theta}$ and $h_{\phi\phi}$ components enables us to identify this mode as 
\be\label{marginalUV}
\chi(\x,\t)={c_{(\phi)}\over4}(1+\cos^2\theta)~, \quad \Phi(\x)={c_{(\phi)}\over2} ~,\qq c_{(\phi)}={4\delta M\over\pert\sqrt{J}} ~.
\ee 
Furthermore preservation of the remaining gauge conditions determines the subleading diffeomorphism to be
\bea
\xi_{\mt{(0)}}=\frac{\tau^2r}{\sqrt{J}(4r^2-\tau^2)}\p_r+\xi^\mt{res}~,
\eea
where $\xi^\mt{res}$ is again given by \eqref{xires}. 

One can easily recognize that \eqref{marginalUV} has angular dependence $\ell=0$ and is just the marginal deformation \eqref{marginal} discussed in Sec.\,\ref{sec:grav}.
This mode carries $\fu(1)$ NHEK charge according to \eqref{jt-energy-ang}, in agreement with the general map between Kerr and NHEK charges \eqref{charge-kerr-nhek}, since the choice \eqref{eq:marg} forces the NHEK mass to vanish. 
Therefore an extremal perturbation with $\delta J=2\sqrt{J}\,\delta M\sim \la^0$ in Kerr corresponds to  a NHEK mode with $\ell=0$ as described by \eqref{marginalUV}. 

This completes our analysis of all smooth and axisymmetric gravitational perturbations interpolating between NHEK and Kerr. A summary of these modes appears in the first two blocks in table~\ref{t:modes}.


\section{Singular perturbations}\label{sec:singularpert}

This last section is devoted to explore further properties of the low-lying modes in cases where we allow for angular singularities. Clearly this takes us to unknown territory, where the rules are less clear and we might encounter more pathologies as we study them.  A fair objection to pursue this direction is that most likely there is no physical process inducing these singular perturbations. Hence, a conservative view might be that if the singularity is not balanced---as we did in Sec.\,\ref{sec:balancing}---one should just discard these configurations as unphysical. 

Nevertheless, there are some formal (theoretical) reasons why it is interesting to consider these singular perturbations. And there are at least three directions that are worth mentioning: 
\begin{enumerate}
\item The potential relation these singular modes might have with superrotations that appear in the asymptotic symmetry analysis of Minkowski space. It has recently been shown in \cite{Strominger:2016wns}, based on earlier work \cite{penrose1972geometry,penrose1992,Podolsky:1999zr,Griffiths:2002hj}, that finite superrotations act on isolated defects on the celestial sphere, and these are closely related to the $C$-metric deformations we described in Sec.\,\ref{sec:marginal}. It would be very interesting to understand the interpretation of other low-lying modes in the context of celestial holography, and if they have a  role in gravitational scattering.  
\item The regularity conditions on low-lying modes arose from the angular dependence of the modes, and not from  physics on the AdS$_2$ portion. These conditions on modes with $\Delta=2,1$ are given by \eqref{eq:JTchiconstraints}. In this context it is interesting to explore if these conditions are universal or only required for specific black holes. More concretely, are there examples of nearly-AdS$_2$ holography  for which $\Delta=2,1$ are not constrained by \eqref{eq:JTchiconstraints} and still well-behaved?  or could one prove that regardless of the gravitational theory and the origin of the AdS$_2$ background these modes are always constrained?  
\item As we will show the matching procedure of the singular perturbations involves non-separable solutions to the Teukolsky's master equations. From a mathematical perspective it is interesting to investigate how the non-separable solutions behave (and contrast to separable solutions).
\end{enumerate}
Answering these questions is outside the scope of this work. Our intention here is to initiate a discussion regarding how one would describe these modes and their properties in the whole Kerr geometry. 

The subsequent discussion will be divided in two parts. We first focus on building the Hertz potential associated with singular perturbations, starting from the perturbations around NHEK. The advantage of transcribing the information of these modes to the Hertz potential is twofold: first, it illustrates some stark differences on the behaviour of $\Psi_{\mt{H}_0}$ for $\ell=0,1$ relative to the propagating modes; second, it allows us to be more systematic as we attempt to extend these modes to the full Kerr geometry via a matching procedure.   This second feature is particularly important for singular modes that have vanishing Weyl scalars on NHEK. The matching procedure is the focus of the second portion of this section, where we highlight places where we can make some progress and also potential obstacles  to reconstruct these modes.

\subsection{Reconstruction of Hertz potential on NHEK}

The task is to build a Hertz potential $\Psi_{\mt{H}_0}$ for the low-lying modes
\be\label{chi-ansatz}
\chi(\x,\theta)=\sin^2\theta \sum_{\ell=0,1}S_\ell(\theta)\chi_\ell(\x)~,  \quad \square_2 \chi_\ell(\x)=\ell(\ell+1)\chi_\ell(\x)~.
\ee
The first steps follow the analysis in Sec.\,\ref{sec:UVpropagating}. In particular we want to use results from that section that do not assume any property of $\chi$ nor $\Psi_{\mt{H}_0}$. This singles out  \eqref{chi-hertz} and \eqref{eqn-chi-hertz}, and from them we want to determine $\Psi_{\mt{H}_0}$. 

 It is important to stress that if we assume a separable ansatz  for $\Psi_{\mt{H}_0}$, such as \eqref{hertz-regular}, it leads to pathologies for $\ell=0,1$. This pathology is clear in \eqref{hertz-chi}. Actually, we will show that Hertz potentials associated to low-lying modes take the form 
\be\label{hertz-low}
\Psi_{\mt{H}_0}=\sum_{\ell =0,1} \le( S_{\ell}(\theta)\mathcal U_{\ell}(\x)-4S_\ell^\mt{inhom}(\theta)\boldsymbol{n}^a\boldsymbol{n}^b\nabla_a\nabla_b\chi_\ell(\x)\ri)~.
\ee
Crucially this Hertz potential is \emph{not} a separable solution in contrast to \eqref{hertz-regular}. Our task is to determine $S_\ell^\mt{inhom}$ and $\mathcal U_{\ell}$ such that  $\Psi_{\mt{H}_0}$ solves \eqref{hertzNHEK}, while being compatible with \eqref{chi-hertz} and \eqref{eqn-chi-hertz}. This will give us a reversible map between $\Psi_{\mt{H}_0}$ and $\chi$.

Let us first determine $S_\ell^\mt{inhom}$. Acting with $\mathcal L_\theta(\mathcal L_\theta+2)$ on \eqref{hertz-low} gives zero on the first term
because
\begin{equation}
  \mathcal L_\theta \, S_0=0\,, \quad \text{and} \quad \le(\mathcal L_\theta +2\ri) S_1=0\,.
\label{eq:ltheta}
\end{equation}
Comparing its action on the second term of \eqref{hertz-low} with \eqref{eqn-chi-hertz} gives us the relation
\be
\mathcal L_\theta(\mathcal L_\theta+2)S_\ell^\mt{inhom}=S_\ell~.
\ee
Using \eqref{eq:ltheta}, the above equation can be integrated to 
\bea
\mathcal L_\theta S^\mt{inhom}_0=\tfrac{1}{2} S_0\,, \quad  \text{and} \quad ( \mathcal L_\theta+2)S^\mt{inhom}_1=-\tfrac{1}{2} S_1~,\label{S01}
\eea
up to an homogenous piece that can be re-absorbed into the second term in \eqref{hertz-low}. Notice we can combine both equations \eqref{S01} as
\be\label{Sin-sol}
(\mathcal L_\theta+\ell(\ell+1))S^\mt{inhom}_\ell=-{\le(\ell-\tfrac{1}{2}\ri)S_\ell}~, \qquad \ell=0,1~.
\ee

Next we require our ansatz \eqref{hertz-low} to solve the Hertz potential master equation \eqref{hertzNHEK}. This gives an equation for $\mathcal U_\ell$
 \be\label{u-hertz}
 \le(\mathcal{L}_\x-2+\ell(\ell+1)\ri)\mathcal U_{\ell}(\x)=2(2\ell-1)\boldsymbol{n}^a\boldsymbol{n}^b\nabla_a\nabla_b\chi_\ell(\x)~,
\ee
where we used the identity
\be\label{u-chi-inhom}
\le(\mathcal{L}_\x-2+\ell(\ell+1)\ri)\boldsymbol{n}^a\boldsymbol{n}^b\nabla_a\nabla_b\chi_\ell(\x)=0~,
\ee
holding for any $\chi_\ell$ satisfying the Klein-Gordon equation \eqref{chi-ansatz}. Furthermore, acting with $\boldsymbol{l}^a\boldsymbol{l}^b\nabla_a\nabla_b$ on \eqref{hertz-low} and plugging it into \eqref{chi-hertz}, we get the further relation 
\be
\boldsymbol{l}^a\boldsymbol{l}^b\nabla_a\nabla_b\,\mathcal U_{\ell}=-\chi_\ell ~,\label{u-inhom}
\ee
where we used \eqref{nnll} and the Klein-Gordon equation for $\chi_\ell$. 

Given the above analysis, our task of mapping any Hertz potential of the form \eqref{hertz-low} to a lower lying $\chi$ mode \eqref{chi-ansatz}, and vice-versa, reduces to solving \eqref{u-hertz} and \eqref{u-inhom} while $\chi_\ell$ satisfying the Klein-Gordon equation. We claim that this reconstruction always has a solution and we provide a detailed derivation for both $\ell=0,1$ in App.\,\ref{app:proof}. The bottom line is that starting from the low-lying  $\chi-$modes there is a corresponding Hertz potential of the form \eqref{hertz-low} which obeys the master equation \eqref{hertzNHEK}. 

Within the singular perturbations there are special cases when the Hertz potential is actually separable. From \eqref{hertz-low}, requiring that the Hertz potential is separable, \emph{i.e.} 
\be\label{hertzjt}
\Psi_{\mt{H}_0}=\sum_{\ell =0,1}S_{\ell}(\theta)\mathcal U_{\ell}(\x)~,
\ee
 implies that 
 \be \label{nnchi} 
 \boldsymbol{n}^a\boldsymbol{n}^b\nabla_a\nabla_b\chi_\ell(\x)=0~.
 \ee
In this situation the master equation becomes 
\be\label{u-sep}
\le(\mathcal{L}_\x-2+\ell(\ell+1)\ri)\mathcal U_\ell(\x)=0~,
\ee
which originates from inserting \eqref{hertzjt} into \eqref{hertzNHEK}. And there is as well the relation \eqref{u-inhom} that relates $\mathcal U_\ell(\x)$ to $\chi_\ell(\x)$---note that \eqref{u-inhom} is compatible with \eqref{nnchi}-\eqref{u-sep}. It is straightforward to construct solutions in this case. One way is to
 first solve for $\mathcal U_\ell(\x)$ from  \eqref{u-sep} and then build the resulting $\chi_\ell$ from \eqref{u-inhom} which can be shown to comply with the constraints \eqref{nnchi} after using \eqref{nnll}.
Another option is to first solve for $\chi_\ell$ from its equation of motion in  \eqref{chi-ansatz} and the constraint \eqref{nnchi}, and then determine $\mathcal U_\ell$  from  \eqref{u-inhom} and \eqref{u-sep}, similarly to the discussion in App.\,\ref{app:proof}. 

One interesting aspect of solutions obeying \eqref{hertzjt}-\eqref{u-sep} is the behaviour of the Weyl scalars evaluated on them. Recalling the relation \eqref{eq:weyl-xp1}, we see that \eqref{nnchi} implies $\Psi_4=0$. If we further set
 \be \label{llchi} 
 \boldsymbol{l}^a\boldsymbol{l}^b\nabla_a\nabla_b\chi_\ell(\x)=0~,
 \ee
 then $\Psi_0=0$. However, the Hertz potential associated to these modes is {\it non-zero}, despite having trivial Weyl scalars. The fact that $\Psi_{\mt{H}_0}$ is separable and non-vanishing establishes clear rules on how we should match this special class of singular modes to the whole geometry. We will discuss this matching procedure in the next subsection.

\subsection{UV behaviour of singular modes}

We have shown that $\Psi_{\mt{H}_0}$ for these low-lying modes  in \eqref{hertz-low} takes the form of sum of two terms, and hence does not comply with the usual basis of solutions that is typically used to describe linearized solutions to the Teukolsky's master equation. We also take this as an indication that the Hertz potential $\tilde\Psi_{\mt{H}_0}$ in Kerr, which reduces to $\Psi_{\mt{H}_0}$ in the decoupling limit \eqref{hertz-glue}, cannot be a single term either. It is not clear to us how to find the most general solutions to the master equation \eqref{Hertzeq2} in the Kerr geometry. Despite this significant obstruction to reconstruct these singular low-lying modes in the entire geometry, we make the following exploratory ansatz
\be\label{eq:aza}
\tilde\Psi_{\mt{H}_0}=e^{-i\tilde \omega\tilde t}\le(\tilde S_{\ell\tilde \omega}^{\mt{(1)}}(\theta)\tilde R^{\mt{(1)}}_{\ell\tilde \omega}(\tilde r)+\tilde S_{\ell\tilde \omega}^{\mt{(2)}}(\theta)\tilde R^{\mt{(2)}}_{\ell\tilde \omega}(\tilde r)\ri)~,
\ee
where the angular functions $\tilde S_{\ell\tilde \omega}^{\mt{(1)}}$ and $\tilde S_{\ell\tilde \omega}^{\mt{(2)}}$ should satisfy
\bea\label{ang-eqn-low}
&&\Big(\mathcal{L}_\theta+K_{\ell\tilde \omega}\Big)  S_{\ell\tilde \omega}^{\mt{(1)}}+\Big(4a\tilde\omega\,\cos\,\theta+a^2\tilde\omega^2\cos^2\theta\Big)\tilde S_{\ell\tilde \omega}^{\mt{(1)}}=-\le(\ell-\tfrac{1}{2}\ri)\tilde S_{\ell\tilde \omega}^{\mt{(2)}}~,\cr
&&\Big(\mathcal{L}_\theta+K_{\ell\tilde \omega}\Big)  S_{\ell\tilde \omega}^{\mt{(2)}}+\Big(4a\tilde\omega\,\cos\,\theta+a^2\tilde\omega^2\cos^2\theta\Big)\tilde S_{\ell\tilde \omega}^{\mt{(2)}}=0~.
\eea
Plugging \eqref{eq:aza} into the master equation \eqref{Hertzeq2}, and using \eqref{ang-eqn-low}, we get two coupled equations for the radial functions
\bea\label{rad-eqn-low}
&&\Big(\mathcal{L}_{\tilde\omega,\tilde r} -2+K_{\ell\tilde \omega}
\Big)\tilde R^{\mt{(2)}}_{\ell\tilde \omega}=-\le(\ell-\tfrac{1}{2}\ri)\tilde R_{\ell\tilde \omega}^{\mt{(1)}}~,\cr
&&\Big(\mathcal{L}_{\tilde\omega,\tilde r} -2+K_{\ell\tilde \omega}
\Big)\tilde R^{\mt{(1)}}_{\ell\tilde \omega}=0~.
\eea
where $\mathcal{L}_{\tilde\omega,\tilde r}$ is given by \eqref{Ltr} by replacing $\p_{\tilde t}\to -i\tilde \omega$.

Without solving this system of equations explicitly, we will show that $\tilde\Psi_{\mt{H}_0}$ would comply with \eqref{hertz-glue} by matching equations in the decoupling limit. Matching of the angular part is straightforward.
In the low frequency limit $\tilde\omega\to0$, it is easy to see that $\tilde S_{\ell\tilde \omega}^{(2)} $ is just the spherical harmonic function $S_\ell$, and $\tilde S_{\ell\tilde \omega}^{(1)}$ satisfies the same differential equation as $S_\ell^\mt{inhom}$ in \eqref{Sin-sol}. Therefore, in the low frequency limit, we have
\be
\lim_{\lambda\to0}\tilde S_{\ell\tilde \omega}^{(1)} = S_\ell^\mt{inhom}\quad  \lim_{\lambda\to0}\tilde S_{\ell\tilde \omega}^{(2)} = S_\ell~.
\ee

Matching of the radial part is also similar to the discussion in Sec.\,\ref{sec:UVpropagating}: the radial equation can be approximated in the near region with ${\frac{\tilde r-r_+}{r_+}} \ll1$ in the parameter regime \eqref{eq:lowfregime}.  
Note that the radial differential operator $\mathcal{L}_{\tilde\omega,\tilde r}$ appearing in \eqref{rad-eqn-low} reduces to $\mathcal{L}_{\omega, r}$
expressed in frequency space, as discussed in \eqref{ltrlimit}.
Using this relation and comparing \eqref{rad-eqn-low} with \eqref{u-hertz} and \eqref{u-chi-inhom}, it is straightforward to see that in the decoupling limit $e^{-i\tilde\omega\tilde t}\tilde R^{(1)}_{\ell\tilde \omega}$ and $e^{-i\tilde\omega\tilde t}\tilde R^{(2)}_{\ell\tilde \omega}$ satisfy the same equation as $\boldsymbol{n}^a\boldsymbol{n}^b\nabla_a\nabla_b\chi_\ell(\x)$ and $\mathcal U_{\ell}$, respectively. 

One can study the behaviour of  $\tilde R^{(1)}_{\ell\tilde \omega}$ and $\tilde R^{(2)}_{\ell\tilde \omega}$  in the very far region of Kerr. From preliminary results, they seem to have a reasonable asymptotic series expansion, but a more detailed analysis that includes the matching region is undoubtedly required. The angular function $S_\ell^\mt{inhom}$ carries logarithmic terms: these do not affect the $h_{\mu\nu}$ in \eqref{hertzmap} for NHEK, but we haven't explored if the logarithmic pieces enter in the metric perturbation for the Kerr geometry. We leave a more systematic study of these non-separable solutions for future work. 

Finally, we discuss how to glue modes of the form \eqref{hertzjt}-\eqref{llchi}, i.e. carrying trivial Weyl scalars on NHEK.  The natural ansatz for the Hertz potential on Kerr is 
\be\label{Psi-ansatz}
\tilde\Psi_{\mt{H}_0}=e^{-i\tilde\omega\tilde t}\tilde S_{\ell\tilde\omega}(\theta)\tilde R_{\ell\tilde\omega}(\tilde r), 
\ee
with $\tilde S_{\ell\tilde\omega}(\theta)$ and $\tilde R_{\ell\tilde\omega}(\tilde r)$ satisfying \eqref{ang-eqn} and \eqref{rad-eqn}. 
Similar to the discussion in Sec.\,\ref{sec:UVpropagating}, the radial equation \eqref{rad-eqn} in Kerr 
can actually be solved in the near and far region as long as 
\be
a|\tilde \omega| \ll 1~, \quad \tilde \tau_H\ll 1~, \quad \hbox{with}\,\,  {\tilde \omega\over \tilde \tau_H}\,\,\hbox{ fixed. }
\ee
A very interesting remark in this case regards the behaviour of the Weyl scalars between NHEK and Kerr. 
While we imposed the vanishing of the Weyl scalars in NHEK, this condition does not have to hold in Kerr. In fact, we can construct explicit solutions where it does not. This is due to the  $\lambda\to 0$ limit in \eqref{weylglue}: the leading contributions vanish as we go near the horizon, but subleading contributions in $\tilde \Psi_{0,4}$ are not necessarily zero. That is, \eqref{weylglue} allows for singular perturbations where $\tilde \Psi_{0,4}$ are non-zero, while as we take the decoupling limit one still gets $\Psi_{0,4}=0$. 
This is a striking feature that we have explored by studying some solutions to  \eqref{hertzjt}-\eqref{llchi} and applying the same matching procedure as in Sec.\,\ref{sec:UVpropagating}. It would be interesting to investigate the properties of these metric perturbations in Kerr more carefully and understand their imprints in the far region. 

\section*{Acknowledgements} 
 AC, WS and BY would like to thank KITP, and in particular the program ``Gravitational Holography,'' for its hospitality during the completion of this work.
The work of AC and VG is  supported by the Delta ITP consortium, a program of the NWO that is funded by the Dutch Ministry of Education, Culture and Science (OCW). VG acknowledges the postdoctoral program at ICTS for funding support through the Department of Atomic Energy, Government of India, under project no. RTI4001.  WS and BY are supported by the NFSC Grant No. 11735001.  This research was supported in part by the National Science Foundation under Grant No. NSF PHY-1748958.

\appendix

\section{Aspects of Teukolsky formalism}\label{app:Teuk}

In this appendix we start by reviewing some of the basic elements of gravitational perturbations, gathering definitions and well known results that are specific to the Kerr background and its near horizon geometry.  Readers can find an excellent review in Appendix C of \cite{Dias:2009ex}, and more recently in \cite{Compere:2018aar}. This discussion includes a summary of Wald's theorem \cite{wald-theorem} characterizing the subset of Kerr perturbations with vanishing Weyl scalars. Also,  appendix \ref{app:proof} presents a set of identities involving differential operators constructed out of the near horizon AdS$_2$ tetrad which are used in Sec.\,\ref{sec:grav} and Sec.\,\ref{sec:UV}.

\subsection{Overview}
Any on-shell metric perturbation $h$ of the Kerr black hole must solve the linearized Einstein's equations
\be
\cE\cdot h = 0~,
\label{eq:linearE}
\ee
where $\cE$ is a self-adjoint linear partial differential operator (PDO). This is a coupled system of partial differential equations typically written in a non-gauge invariant way. Hence, it is hard to solve and to extract the two polarizations carried by the perturbation.

Using the Newman-Penrose formalism, Press and Teukolsky \cite{teukolsky1972rotating,Teukolsky:1973ha,Press:1973zz,Teukolsky:1974yv} used gauge invariant quantities, the Weyl scalars $\Psi$, defined in terms of $h$ by
\be
\Psi = \cT\cdot h~,
\ee
where $\cT$ is a linear PDO, to show that \emph{any} on-shell perturbation satisfied the Teukolsky's master  equation
\be
\cO\cdot\Psi =0\,.
\label{eq:teukolsky}
\ee
A further achievement of the Teukolsky's formalism was that these equations can be solved by separation of variables. Hence, the original problem is mapped to a set of ODEs.

The reconstruction of the gravitational perturbation $h$ from the Weyl scalars is solved using the Cohen-Kegeles formalism \cite{cohen1975space,kegeles-cohen} (see also \cite{chrzanowski1975vector}). This is typically performed in the ingoing (outgoing) radiation gauge IRG (ORG)
\be
l^\mu h_\mn = g^\mn h_\mn=0\quad \text{(IRG)},\qq n^\mu h_\mn = g^\mn h_\mn = 0\quad \text{(ORG)}~,
\ee
and involves the Hertz potential $\Psi_\mt{H}$ whose existence is best understood following the robust mathematical formulation of this reconstruction problem given by Wald \cite{Wald:1978vm}. 

Wald constructed two linear PDO, $\cS$ and $\cO$, satisfying the operator equation \cite{Wald:1978vm} 
\be\label{TeukOpeq}
\cS\cdot \cE = \cO\cdot \cT
\ee
Applying this to $h$ shows that $\cE\cdot h=0$ implies that $\Psi$ must satisfy the master equation \eqref{eq:teukolsky}, in agreement with Teukolsky's work. Furthermore, taking the adjoint (in the sense of operators) of \eqref{TeukOpeq} gives
$\cE\cdot\cS^\dg = \cT^\dg \cdot\cO^\dg$, since $\cE^\dg=\cE$ is self-adjoint. Thus, we can reconstruct the perturbation $h$
\begin{equation}
  h=\cS^\dg \cdot \Psi_\mt{H}
\end{equation}
in terms of a potential $\Psi_\mt{H}$ satisfying
\be
\cO^\dg\cdot\Psi_\mt{H}=0~.
\label{eq:hertz}
\ee
If $\Psi_\mt{H}$ is a Hertz potential, \emph{i.e.} it satisfies $\cO^\dg \cdot \Psi_\mt{H}=0$, we get a solution of Teukolsky equation $\cO\cdot\Psi=0$ given by $\Psi = \cT\cdot\cS^\dg\cdot\Psi_H$. This is simply obtained by reconstructing the metric and computing $\Psi$ from it. This shows that there is a unique Weyl scalar $\Psi$ for a given Hertz potential. However, notice this conclusion doesn't go the other way: different Hertz potentials can give the same $\Psi$.

\subsection{Newman-Penrose formalism \& master equations}

In this Appendix, we collect useful formulas in the Newman-Penrose (NP) formalism \cite{newman1962approach}. 
For book keeping purposes, we introduce the parameter $\iota$ to denote different conventions in the literature : $\iota=1$ corresponds to the mostly positive signature for the metric, namely $(-,+,+,+)$, used in this paper, whereas $\iota=-1$ corresponds to the $(+,-,-,-)$ signature used in, e.g., \cite{Dias:2009ex}.

The Newman-Penrose (NP) formalism \cite{newman1962approach} decomposes the metric components
\begin{equation}
  g_\mn =\iota ( -l_\mu n_\nu -l_\nu n_\mu +m_\mu \bar{m}_\nu + \bar{m}_\mu m_\nu)
\end{equation}
in terms of three complex valued tetrads satisfying
\begin{equation}
\begin{aligned}\label{NPrelations}
& l \cdot m  = l\cdot \bar{m} = n\cdot m = n\cdot\bar{m} = 0~,\-
& l \cdot l  = n\cdot n = m\cdot m  = \bar{m}\cdot \bar{m} = 0~,\-
& l \cdot n  = -\iota, \hspace{2cm} m\cdot \bar{m}  = \iota ~.
\end{aligned}
\end{equation}
We will using the mostly-plus signature conventions, $\iota=1$, which is the opposite of that followed by in \cite{Teukolsky:1973ha,Press:1973zz,Teukolsky:1974yv}. 

There are five inequivalent gauge invariant Weyl scalars built from contractions of the Weyl tensor with these tetrads
\begin{equation}
\begin{aligned}
  \Psi_0 &=\iota\, C_{\mu\nu\alpha\beta}\,l^\mu\,m^\nu\,l^\alpha\,m^\beta\,, \\
  \Psi_1 &=\iota\, C_{\mu\nu\alpha\beta}\,l^\mu\,n^\nu\,l^\alpha\,m^\beta\,, \\
  \Psi_2 &=\iota\, C_{\mu\nu\alpha\beta}\,l^\mu\,m^\nu\,\bar{m}^\alpha\,n^\beta\,, \\
  \Psi_3 &=\iota\, C_{\mu\nu\alpha\beta}\,l^\mu\,n^\nu\,\bar{m}^\alpha\,n^\beta\,, \\
  \Psi_4 &=\iota\, C_{\mu\nu\alpha\beta}\,n^\mu\,\bar{m}^\nu\,n^\alpha\,\bar{m}^\beta\,.
\end{aligned}
\label{eq:weyl-scalars}
\end{equation}
In our manipulations we will denote with lower case, $\psi_i$, the value of the Weyl scalars on the background geometry, and with upper case, $\Psi_i$, the linear contribution due to the metric perturbation. Only $\Psi_0$ and $\Psi_4$ are invariant under tetrad rotations and diffeomorphisms (at linear order). 
Moreover, for the Kerr background, only one, either $\Psi_0$ or $\Psi_4$ is needed to establish the dynamical properties of the linearized solutions. Hence, the remaining Weyl scalars are not needed to describe gravitational waves on Kerr at this order.

Introducing the differential operators\footnote{We have added hats on some of the commonly used symbols to avoid conflict with notation used in the main text.}
\begin{equation}\label{NPdiff}
D = l^\mu \n_\mu\, ,\qq   \hat\D = n^\mu  \n_\mu\,, \qq \d = m^\mu \n_\mu\, , \qq  \bar{\delta}  = \bar{m}^\mu \n_\mu\,,
\end{equation}
and the spin coefficients
\begin{align}
\k & =- \iota m^\mu D l_\mu\,, & \hat\tau & = -\iota m^\mu \hat\D l_\mu\,, \\
\s & = -\iota m^\mu \d l_\mu\,, & \rho &=-\iota  m^\mu \bar{\delta} l_\mu\,, \\
\pi & = \iota \bar{m}^\mu D n_\mu\,, & \nu & = \iota \bar{m}^\mu \hat \D n_\mu\,, \\
\mu & = \iota \bar{m}^\mu \d n_\mu\, , & \hat\la &= \iota \bar{m}^\mu\bar{\delta} n_\nu\,, \\
\ve & = -\iota \tfrac12\le(n^\mu D l_\mu -\bar{m}^\mu Dm_\mu \ri) \,, & \g & =-\iota \tfrac12\le(n^\mu\hat\D l_\mu - \bar{m}^\mu \hat\D m_\mu \ri)\,, \\
\b & =-\iota \tfrac12\le( n^\mu \d l_\mu-\bar{m}^\mu \d m_\mu \ri)\,, & \a &=-\iota \tfrac12\le(n^\mu \bar{\delta} l_\mu - \bar{m}^\mu \bar{\delta} m_\mu \ri)\,.
\end{align}
Teukolsky's master equation \eqref{eq:teukolsky} for the gravitational perturbations $\Psi_0$ and $\Psi_4$ can be written as
\be
\cO_0 \cdot\Psi_0 = 0\,,\qq \cO_4 \cdot\Psi_4= 0\,,
\ee
where
\begin{equation*}
\begin{aligned}
\cO_0 &= ( D-3\ve +\bar{\ve}-4\rho-\bar{\rho})(\hat\D+\mu-4\g) - (\d+\bar{\pi}-\bar{\alpha}-3\b-4\hat\tau)(\bar{\delta}+\pi-4\a)- 3\psi_2 \,,\\
\cO_4 &= (\hat\D+3\g-\bar{\gamma} +4\mu+\bar{\mu})(D+4\ve-\rho) - (\bar{\delta}-\bar{\hat\tau}+\bar{\b}+3\a+4\pi)(\d-\hat\tau+4\b)-3\psi_2\,.
\end{aligned}
\end{equation*}

The adjoint equations \eqref{eq:hertz} satisfied by the Hertz potentials $\Psi_{\mt{H}_0}$ and $\Psi_{\mt{H}_4}$ reconstructing the perturbation $h$ are
\be
\cO_0^\dg\cdot\Psi_{\mt{H}_0}=0~,\qq \cO_4^\dg\cdot\Psi_{\mt{H}_4}=0~,
\ee
with
\bea
\cO_0^\dg\=(\hat\D+3\g -\bar{\g} +\bar{\mu})( D+4\ve+3\rho)-(\bar{\d} +\bar{\b}+3\a - \bar{\tau})(\d+4\b+3\hat\tau) - 3\psi_2~,\\
\cO_4^\dg\=(D-3\ve+\bar{\ve}-\bar{\rho})(\hat\D-4\g-3\mu) - (\d-3\b-\bar{\a}+\bar{\pi})(\bar{\d}-4\a-3\pi)-3\psi_2 ~.
\eea
The reconstructed metric in IRG is then given by
\begin{align}
h^\mt{IRG}_{\mu\nu}\nonumber=&\pert\Big\{l_{(\mu}m_{\nu)}[(D-\rho+\bar\rho)(\delta+4\beta+3\hat\tau)+(\delta+3\beta-\bar\alpha-\hat\tau-\bar\pi)(D+3\rho)] \\&-l_\mu l_\nu(\delta+3\beta+\bar\alpha-{\color{red}{\hat\tau}})(\delta+4\beta+3\hat\tau)-m_\mu m_\nu(D-\rho)(D+3\rho)\Big\}\Psi_{\mt{H}_0}+c.c.~.
\label{eq:metric-recons}
\end{align}
To reconstruct the metric in ORG one would use $\Psi_{\mt{H}_4}$ instead. Notice the existence of a typo (shown in red) in the formula given in \cite{Dias:2009ex}. The one above agrees with, for example, Table 1 in \cite{chrzanowski1975vector}. The relation between the Hertz potential and the defining Weyl scalar is given in \cite{Wald:1978vm,kegeles-cohen,Keidl:2006wk}, 
 which we summarize below \begin{align}
\Psi_0=&\frac{\pert}{2} (D-3\ve +\bar\ve-\bar\rho)(D-2\ve + 2\bar\ve-\bar\rho) (D-\ve+ 3\bar\ve - \bar\rho)(D+4\bar\ve+3\bar\rho)\bar\Psi_{\mt{H}_0}\,,\nonumber\\
\Psi_4=&\frac{\pert}{2}\left\{(\bar\delta+3\alpha+\bar\beta-\bar{\hat\tau})(\bar\delta+2\alpha+2\bar\beta-\bar{\hat\tau})(\bar\delta+\alpha+3\bar\beta-\bar{\hat\tau})(\bar\delta+4\bar\beta+3\bar{\hat\tau})\bar\Psi_{\mt{H}_0}\right.\nonumber\\&\left.+3\Psi_2[\hat\tau(\bar\delta+4\alpha)-\rho(\Delta+4\gamma)-\mu (D+4\ve)+\pi(\hat\delta+4\beta)+2\Psi_2]\Psi_{\mt{H}_0}\right\}\,.
\label{weylhertz}
\end{align}

\subsection{Kerr \& NHEK specifics}

All previous equations hold for any type D vacuum solution. When restricting to the Kerr metric \eqref{eq:kerr}, the Newman-Penrose tetrad usually used is the Kinnersley tetrad \cite{Kinnersley:1969zza} 
\begin{equation}
\begin{aligned}
\tilde{l} &=  {\tilde{r}^2+a^2\/\D} \p_{\tilde{t}} + \p_{\tilde{r}} + {a\/\D}\,\p_{\tilde{\phi}}, \\
\tilde{n} &= {1\/2 \S} \le( (\tilde{r}^2+a^2)\p_{\tilde{t}} - \D\p_r +a\, \p_{\tilde{\phi}}\ri), \\
\tilde{m} &= {1\/\sqrt{2}\G}\le( i\,a\,\r{sin}\,\t\,\p_{\tilde{t}} + \p_\t + {i\/\r{sin}\,\t}\p_{\tilde{\phi}} \ri) ~,
\end{aligned}\label{eq:kerr-tetrad}
\end{equation}
where we defined $\G\equiv \tilde{r}+i a\,\r{cos}\,\t$. The only non-vanishing Weyl scalar is 
$$\tilde\psi_2 = - {M\over \bar\G^3}~, $$
and the spin coefficients are
\begin{align}\nt
\k& =\s=\hat\la=\nu=\e=0 , &&&   \rho & = -{1\/\bar{\G}}~, \\
\b & = {\r{cot}\,\t\/2^{3/2}\G}, & \pi &= {i a\,\r{sin}\,\t\/\sqrt{2}\,\bar\G^2} , & \a & = \pi-\bar\b~,\-
\hat\tau & = -{i a\,\r{sin}\,\t\/\sqrt{2}\,\G^2} ,& \mu & = -{\D\/2\G\bar\G^2} , & \g &=\mu +{\tilde{r}-M\/2 \G\bar\G}~.
\end{align}
Notice we introduced a tilde for these tetrads \eqref{eq:kerr-tetrad} and $\tilde\psi_2$ to identify them as full Kerr quantities, in agreement with the conventions used in the main text.

\paragraph{Near-NHEK tetrad.} In the near-extremal near horizon limit 
\be\label{App:NearNHEKdec}
\tilde r = \sqrt{J} + \lambda\left(r + \frac{\tau^2}{4r}\right),\qq \tilde t = 2 J{{t}\/\la}~,  \qq \tilde\phi = {\phi} + {\sqrt{J}} {t\/\la}~,
\ee
leading to \eqref{eq:near-nhek}, there is a finite NHEK tetrad
\begin{equation}
\begin{aligned}
l &= {1\/(1-{\tau^2\/4 r^2})^2 } \le( {1\/r^2}\p_t + \le(1- {\tau^2\/4r^2}  \ri)\p_r -{1\/r} \le(1+{\tau^2\/4r^2}  \ri) \p_\phi\ri)~, \\
n &= {1\/2 J( 1+\r{cos}^2\t)}\le( \p_t  - r^2\le(1- {\tau^2\/4r^2}  \ri) \p_r -r \le(1+{\tau^2\/4r^2}  \ri)\p_\phi\ri)~,\\
m &= {1\/\sqrt{2} \sqrt{J} (1+i\,\r{cos}\,\t)} \le(\p_\t + i \le( {1\/\r{sin}\,\t} - {\r{sin}\,\t\/2}\ri)\p_\phi \ri)~,
\label{eq:NHEK-tetrad}
\end{aligned}
\end{equation}
related to \eqref{eq:kerr-tetrad} by
\be
l= \lim_{\la\to0} (\la \,\tilde{l}),\qq n  =\lim_{\la\to0} (\la^{-1} \,\tilde{n}),\qq m =\lim_{\la\to0} \tilde{m}~.
\ee
These tetrads satisfy \eqref{NPrelations} and, once more, the only non-vanishing Weyl scalar is 
\be
\psi_2 =  {i\over J(i+\cos\,\theta)^3}~. 
\ee

\paragraph{Master equations (Kerr \& NHEK).} Teukolsky's master equation \eqref{eq:teukolsky} for a spin-$s$ field $\tilde\Psi^{(s)}$ in the full Kerr geometry is \cite{Teukolsky:1973ha}
\begin{multline}
\label{eq:master-Kerr}
  \left[\frac{(\tilde{r}^2+a^2)^2}{\Delta}-a^2\sin^2\theta\right]\,\partial^2_{\tilde{t}}\tilde\Psi^{(s)} + \frac{4Ma\tilde{r}}{\Delta}\,\partial_{\tilde{t}}\p_{\tilde\phi} \tilde\Psi^{(s)} + \le[{a^2\/\D} - {1\/\r{sin}^2\t} \ri]{\p_{\tilde\phi}^2} \tilde\Psi^{(s)} \\
  - \Delta^{-s}\partial_{\tilde{r}}\left(\Delta^{s+1}\partial_{\tilde{r}}\tilde\Psi^{(s)}\right) - \frac{1}{\sin\,\theta}\partial_\theta\left(\sin\,\theta\,\partial_\theta \tilde\Psi^{(s)}\right) -2s\left[\frac{a(\tilde{r}-M)}{\Delta} + \frac{i\,\cos\,\theta}{\sin^2\theta}\right]\partial_{\tilde\phi} \tilde\Psi^{(s)}\\
   -2s\left[\frac{M(\tilde{r}^2-a^2)}{\Delta}-\tilde{r}-i\,a\,\cos\,\theta\right]\,\partial_{\tilde{t}}\tilde\Psi^{(s)} + (s^2\r{cot}^2\theta-s)\,\tilde\Psi^{(s)} = 0\,.
\end{multline}  
Working in ingoing radiation gauge (IRG), the relevant Hertz potential is $\tilde\Psi_{\mt{H}_0}$ and using the adjoint properties of the operator $\mathcal{O}_0$, we learn that it satisfies the master equation for $s=-2$, \emph{i.e.} $ \tilde\Psi_{\mt{H}_0} = \tilde\Psi^{\mt{(-2)}}$.

To solve \eqref{eq:master-Kerr} for normalizable solutions on the 2-sphere, we can use standard separation of variables
\begin{equation}\label{eq:psi17}
  \tilde\Psi^{(s)} = \int d\tilde \omega\sum_{\l, m} e^{-i\tilde{\omega}\tilde{t} + im\tilde{\phi}}\,\tilde{R}_{\l m \tilde\w }(\tilde{r})\,\tilde{S}_{\l m \tilde\w }(\theta)\,,
\end{equation}
to obtain the two decoupled ODEs
\begin{align}
  0 & = \Delta\,\frac{d^2 \tilde{R}_{\l m \tilde\w }}{d\tilde{r}^2} + 2(s+1)(\tilde{r}-M)\,\frac{d \tilde{R}_{\l m \tilde\w }}{d\tilde{r}} + \left(\frac{C^2-2is\,(\tilde{r}-M)\,C}{\Delta} + 4is\tilde{\omega}\tilde{r} - B\right)\,\tilde{R}_{\l m \tilde\w } \,, \cr
 0& =   \frac{1}{\sin\,\theta}\frac{d}{d\theta}\left(\sin\,\theta\, \frac{d\tilde{S}_{\l m\tilde\w}}{d\theta}\right) \-
 & \hspace{1cm} + \left(a^2\tilde{\omega}^2\cos^2\theta - \frac{m^2}{\sin^2\theta} - 2a\tilde{\omega}\,s\,\cos\,\theta - \frac{2ms\,\cos\,\theta}{\sin^2\theta} - \frac{s^2}{\sin^2\theta} - \frac{m^2}{4} + K_{\l m \tilde\w}\right)\tilde{S}_{\l m\tilde{\w}} \,,\nonumber
\end{align}
where we defined
\begin{equation}
  C\equiv  (\tilde{r}^2+a^2)\,\tilde{\omega}^2-am\,, \quad \quad B \equiv K_{\l m \tilde\w} - \frac{m^2}{4} + a^2\tilde{\omega}^2-2am\tilde{\omega} - s(s+1)\,,
\end{equation}
in terms of the separation constant $K_{\l m \tilde\w}$.

\ms 
For \emph{axisymmetric} perturbations, \emph{i.e.} with $m=0$, the above constants become
\be 
C=  (\tilde{r}^2+a^2)\tilde{\omega}^2 ,\qq B=K_{\l\tilde\w} + a^2\tilde{\omega}^2 - s(s+1)~,
\ee
and the angular ODE reduces to
\begin{equation}
  \frac{1}{\sin\,\theta}\frac{d}{d\theta}\left(\sin\,\theta\, \frac{d\tilde{S}_{\l\tilde\w}}{d\theta}\right) + \left(a^2\tilde{\omega}^2\,\cos^2\theta - 2a\tilde{\omega}\,s\,\cos\,\theta - \frac{s^2}{\sin^2\theta} + K_{\l\tilde\w}\right)\,\tilde{S}_{\l\tilde\w} =0\,.
\label{eq:axi-theta-full}
\end{equation}
The radial ODE for axisymetric perturbations can be written in a Schr\"odinger like way
\be
  \Delta^{-s} \frac{d}{d\tilde{r}}\left(\Delta^{s+1}\,\frac{d\tilde{R}_{\l \tilde\w}}{d\tilde{r}}\right) = V(\tilde{r})\,\tilde{R}_{\l \tilde\w}\,, \qq V(\tilde{r}) \equiv B - 4is\,\tilde{\omega}\tilde{r} - \frac{C^2-2is\,(\tilde{r}-M)\,C}{\Delta}\,.
\ee
In the near horizon limit \eqref{App:NearNHEKdec}, the master equation \eqref{eq:master-Kerr} becomes 
\begin{multline}\label{eq:axi-master-NHEK}  
  {16r^2\/ (4r^2-\tau^2)^2}\p_t^2 \Psi^{(s)} - r^2 \p_r^2 \Psi^{(s)} - {2 r(4r^2 + s(4r^2+\tau^2))\/4r^2-\tau^2} \p_r\Psi^{(s)}
\\ -{1\/\r{sin}\,\t} \p_\t(\,\r{sin}\,\t\,\p_\t \Psi^{(s)} ) - { 8 s\, r (4r^2+\tau^2)\/(4r^2-\tau^2)^2} \p_t\Psi^{(s)} + (s^2\,\r{cot}^2\t - s)\Psi^{(s)}=0~.
\end{multline}
Performing some partial separation of variables
\begin{equation}\label{eq:decomppsi1}
  \Psi^{(s)} = \sum_\l U_\l(t,r)\, S_\l(\theta)\,,
\end{equation}
the resulting angular ODE reduces to
\begin{equation}
  S^{\prime\prime}_\l + \cot\theta\,S^\prime_\l - \frac{4}{\sin^2\theta}\,S_\l = -K_\l\,S_\l\,, \quad \quad s=\pm 2~.
\label{eq:theta-NHEK}
\end{equation}
This is the same equation as the one appearing from Einstein's equations in \eqref{eq:stheta} and defines spin-weighted spherical harmonics with $K_\l= \l(\l+1)$. The resulting equation for $U_\l$ is
\begin{multline}\label{eq:radial11}
\p_t^2 U_\l - {\, r^{2s}\/16 (4r^2-\tau^2)^{2s-1}} \p_r\le( r^{-2s} (4r^2-\tau^2)^{2s +1} \p_r U_\l \ri) - {s \/2r}(4r^2+\tau^2) \, \p_t U_\l \\+ {1\/16 r^2} (4r^2-\tau^2)^2 (K-s(s+1))\,U_\l = 0~.
\end{multline}
\subsection{Wald's theorem \cite{wald-theorem}.} 

In this portion we summarise the results of \cite{wald-theorem} regarding gravitational perturbations on Kerr. The content of this work is a theorem which states that  for well-behaved perturbations on a Kerr black hole, $\tilde\Psi_0$ and $\tilde\Psi_4$ uniquely determine each other. To understand this theorem, we need to decode two pieces of it: what it is meant by  ``well-behaved,'' and the implications behind determining $\tilde\Psi_0$ from $\tilde\Psi_4$ (or viceversa). 

In \cite{wald-theorem},  ``well-behaved''  means verbatim: 
\begin{quote}
`` \ldots that at some initial ``time'' (\emph{i.e.}, on an initial spacelike hypersurface which intersects the future event horizon) the perturbation (1) vanishes sufficiently rapidly at infinity, (2) has no angular singularities, and (3) is regular on the future event horizon.'' 
\end{quote}
Our low-lying perturbations in Sec.\,\ref{sec:chi-2} can carry angular singularities, which are interesting for our applications. These cases therefore will lay outside the scope of the theorem. 

Determining  $\tilde\Psi_4$ in terms of $\tilde\Psi_0$, or the reverse, is clearly the important outcome. This implies that the essential information of a gravitational perturbation is encoded in one Weyl scalar, which was particularly useful for the stability analyses of Kerr and astrophysical observations. The key is to show that if  $\tilde\Psi_0=0$ it implies $\tilde\Psi_4=0$, and viceversa.  The proof is relatively straightforward from the Newman-Penrose formalism and uses the well-behaved properties stated above. 

The results in \cite{wald-theorem} also proceed to characterise the solutions to $\tilde\Psi_0=\tilde\Psi_4=0$, and there are four linearly independent solutions:
\begin{enumerate}
\item change in mass, from $M$ to $M+\delta M$,
\item change in angular momentum, from $J$ to $J+\delta J$,
\item a perturbation towards Kerr-NUT,
\item a perturbation towards the rotating $C$-metric. 
\end{enumerate}
The first two are viewed as trivial perturbations, and the last two as physically unacceptable since they are excluded by the boundary conditions deemed as physical in \cite{wald-theorem}.  The analysis  implicitly  treats all diffeomorphisms as trivial transformations.

\subsection{Further identities and proofs}\label{app:proof}

In this subsection we first collect some useful identities regarding differential operators acting on AdS$_2$ that are used in Sec.\,\ref{sec:grav} and Sec.\,\ref{sec:UV}. We then present details of the proofs of the relations between the Hertz potential and the low-lying modes discussed in Sec.\,\ref{sec:singularpert}.

 \paragraph{AdS$_2$ tetrad.} Using the notation of Sec.\,\ref{sec:isometry}, AdS$_2$ tetrads are two-dimensional vectors obeying 
\be\label{eq:nldef}
\boldsymbol{n}\cdot \boldsymbol{n}=\boldsymbol{l} \cdot \boldsymbol{l}=0~,\qquad \boldsymbol{l}\cdot\boldsymbol{n}=-1~.
\ee
In the coordinate system \eqref{eq:2dmetric}, they are explicitly given by
\be
\boldsymbol{l}= {1\/r^2}\p_t + \p_r ~,\qq \boldsymbol{n}={1\over 2}\le( \p_t  - r^2 \p_r\ri) ~.
\ee
For thermal AdS$_2$ with metric \eqref{AdS2thermal}, the corresponding tetrad is
\bea\label{tetradAdS2thermal}
\boldsymbol{l} = {1\/(1-{\tau^2\/4 r^2})^2 } \le( {1\/r^2}\p_t + \le(1- {\tau^2\/4r^2}  \ri)\p_r \ri)~, \qq
\boldsymbol{n}= {1\/2}\le( \p_t  - r^2\le(1- {\tau^2\/4r^2}  \ri) \p_r \ri)~.
\eea
Given an arbitrary scalar function $U(\x)$ on AdS$_2$, the following identities, used in Sec.\,\ref{sec:grav} and Sec.\,\ref{sec:UV}, hold
\begin{equation}
\begin{aligned}
  \Box_2  U(\x) &= -2(\boldsymbol l^a\nabla_a)(\boldsymbol{n}^b\nabla_b)U(\x)\cr
&= -2(\boldsymbol n^a\nabla_a)(\boldsymbol{l}^b\nabla_b)U(\x) +{4} \boldsymbol{n}^b  A_b\, (\boldsymbol{l}^a\nabla_a)U(\x)~, \cr
\boldsymbol{l}^a\boldsymbol{l}^b\nabla_a\nabla_b\, U (\x) &= (\boldsymbol{l}^a\nabla_a)(\boldsymbol{l}^b\nabla_b) U(\x)~,\cr
\boldsymbol{n}^a\boldsymbol{n}^b\nabla_a\nabla_b\,U (\x) &= (\boldsymbol{n}^a\nabla_a)(\boldsymbol{n}^b\nabla_b) U(\x) +  2 \boldsymbol{n}^b  A_b (\boldsymbol{n}^a\nabla_a )U(\x)~,\cr
\Box_2(\Box_2 -2) U(\x) &= 4 \boldsymbol{l}^a\boldsymbol{l}^b\nabla_a\nabla_b\le( \boldsymbol{n}^c\boldsymbol{n}^d\nabla_c\nabla_d\,U (\x)\ri) ~,\cr 
{\mathcal L}_\x({\mathcal L}_\x-2)\,U (\x) &=4\boldsymbol{n}^a\boldsymbol{n}^b\nabla_a\nabla_b\le( \boldsymbol{l}^c\boldsymbol{l}^d\nabla_c\nabla_d\,\,U (\x)  \ri)~,\label{nnll}
\end{aligned}
\end{equation}
where $\mathcal{L}_\x$ is defined in \eqref{hertzNHEK}.

\paragraph{Construction of Hertz potential for low-lying modes.} In this discussion, we give the details of the derivation on how to build $\Psi_{\mt{H}_0}$ from a low-lying $\chi$-mode discussed in Sec.\,\ref{sec:singularpert}. More precisely, given the decompositions $\Psi_{\mt{H}_0}$ in \eqref{hertz-low} and $\chi(\x,\t) = \sin^2\theta\,\chi_\ell(\x)\,S_\ell(\t)$, our matching analysis in Sec.\,\ref{sec:singularpert} showed
 \begin{align}\label{app:u-hertz}
 \le[\mathcal{L}_\x-2+\ell(\ell+1)\ri]\mathcal U_{\ell}(\x)&=2(2\ell-1)\boldsymbol{n}^a\boldsymbol{n}^b\nabla_a\nabla_b\chi_\ell(\x)~,\\
\boldsymbol{l}^a\boldsymbol{l}^b\nabla_a\nabla_b\,\mathcal U_{\ell}&=-\chi_\ell ~.\label{app:u-inhom}
\end{align}
Our task is to show that we can find a $\mathcal U_{\ell}(\x)$ provided that  $\chi_\ell$ satisfies the Klein-Gordon equation 
 \be\label{app:kg}
 \square \chi_\ell=\ell(\ell+1)\chi_\ell~,
\ee
for $\ell=0,1$.

Consider the $\ell=1$ case first.  A solution to \eqref{app:u-inhom} is
\be\label{app:chi-u}
\boldsymbol{l}^a\nabla_a\mathcal U_1(\x)=\boldsymbol n^a\nabla_a\chi_1(\x)~,
\ee
since 
\begin{align}
\boldsymbol{l}^a\boldsymbol{l}^b\nabla_a\nabla_b\,\mathcal U_{1}(\x)&=  (\boldsymbol{l}^a\nabla_a)^2 \,\mathcal U_{1}(\x) \cr
&=  (\boldsymbol l^a\nabla_a)(\boldsymbol{n}^b\nabla_b )\chi_{1} (\x) \cr
&= -{1\over 2}\Box_2 \chi_{1}(\x)\cr
&= - \chi_{1}(\x)~,
\end{align}
where in the last line we used \eqref{app:kg}. Furthermore, it is easy to check that \eqref{app:chi-u} solves \eqref{app:u-hertz}. Hence, any solution to the linear operator equation \eqref{app:chi-u} determines the relation between both modes.

For the case $\ell=0$,  given an on-shell mode $\chi_0$, the general solution to \eqref{app:u-inhom} is
\be\label{u-sol}
\mathcal U_{0}=\mathcal U_{0}^p+\mathcal U^h_0~, \quad \quad \boldsymbol{l}^a\nabla_a  \mathcal U^h_0=0~,
\ee
with $\mathcal U_{0}^p$ a given solution of \eqref{app:u-inhom}, \emph{i.e.} $\boldsymbol{l}^a\boldsymbol{l}^b\nabla_a\nabla_b\,\mathcal U^p_{0}=-\chi_0$. What we show next is that given a solution $\chi_0$ and any particular solution  $\mathcal U^p_0$, we can find a zero mode $\mathcal U^h_0$ such that the Hertz equation \eqref{app:u-hertz} is satisfied. 

To prove this, we eliminate $\chi_0$ in \eqref{app:u-hertz} by plugging \eqref{app:u-inhom} and use \eqref{nnll} to get the differential equation for $\mathcal U_0$,
\be
 (\mathcal{L}_\x-2)^2 \mathcal U_0=0\,.
\label{u-eq}
\ee
Inserting \eqref{u-sol} into \eqref{u-eq} gives
\be\label{sol-uh}
\mathcal  U^h_0=-{1\over4}(\mathcal{L}_\x-2)^2 \mathcal U_{0}^p\,,
\ee
where we used that ${\cal L}_{\x}  \mathcal U^h_0=0$. \eqref{sol-uh} enables us to solve for $\mathcal U^h_0$ for any given $\mathcal U_{0}^p$ determined by a given $\chi_0$. The last step is to show that the right hand side  of \eqref{sol-uh} is necessarily a zero mode of $\boldsymbol{l}^a\nabla_a$ whenever $\square_2 \chi_0=0$. Indeed, applying $\boldsymbol{l}^a\nabla_a$ to \eqref{sol-uh}, we get 
\be 
  (\boldsymbol{l}^a\nabla_a) \mathcal  U^h_0=-\frac{1}{4}(\boldsymbol l^a\nabla_a)(\mathcal L_{\x}-2)^2\,\mathcal U_0^p\,.
\label{derivative-uh} 
\ee
However, since $(\boldsymbol{l}^b\nabla_b ) ({\mathcal L}_\x-2)=2  (\boldsymbol{n}^a\nabla_a)( \boldsymbol{l}^b\nabla_b )(\boldsymbol{l}^c\nabla_c)$, it follows
\be
\begin{aligned}
(\boldsymbol l^b\nabla_b)(\mathcal L_{\x}-2)^2\mathcal U_0^p&=4 (\boldsymbol{n}^a\nabla_a)( \boldsymbol{l}^b \nabla_b)(\boldsymbol{n}^d\nabla_d )(\boldsymbol{l}^c \nabla_c )(\boldsymbol{l}^e \nabla_e)\mathcal U_0^p\\
&=-4 (\boldsymbol{n}^a\nabla_a)( \boldsymbol{l}^b \nabla_b)(\boldsymbol{n}^d\nabla_d)\chi_0=2 (\boldsymbol n^a\nabla_a)\Box_2\chi_0~.
\end{aligned}
\ee
Thus, the right hand side of \eqref{derivative-uh} vanishes as long as $\square_2 \chi_0=0$ and $\mathcal U_0^h$ given by \eqref{sol-uh} is indeed a zero mode of $\boldsymbol l^a\nabla_a$.

\section{Nearly-AdS$_2$ holography}
\label{sec:nearAdS2}

In this appendix we review some aspects of nearly-AdS$_2$ holography;  for a more comprehensive and detailed review we recommend \cite{Sarosi2018}.

The holographic understanding of AdS$_2$ was always more challenging than its higher dimensional cousins, mainly because of the lack of finite energy excitations in a pure gravity theory above the AdS$_2$ vacuum \cite{Fiola:1994ir,Maldacena:1998uz}. However, it was observed in \cite{Almheiri:2014cka} that \emph{nearly} AdS$_2$ was a sensible theory by including the leading corrections away from pure AdS$_2$. 

One natural way to think of these leading corrections is to embed the AdS$_2$ geometry into a different spacetime with potentially different asymptotics. This view appears naturally when discussing near-extremal black hole physics, because their near horizon geometry develops a local AdS$_2$ throat \cite{Kunduri:2007vf}.\footnote{This result can be proved under mild and reasonable assumptions, applies to fairly broad effective actions emerging from string theory dynamics and survives higher order corrections \cite{Kunduri:2007vf} (see also \cite{Kunduri:2013gce} for a review on the allowed such near horizon geometries).} Using the terminology of the AdS/CFT correspondence, such embedding provides a UV completion of the AdS$_2$ physics. It was argued in \cite{Almheiri:2014cka} that these leading gravitational effects are captured by the Jackiw-Teitelboim (JT) theory  \cite{Teitelboim:1983ux,Jackiw:1984je} in AdS$_2$ with action\footnote{The original JT theory corresponds to the second line, where the boundary term was added to have a well defined variational principle. We included the first line to consider this review in light of the more recent perspective on this theory, as we discuss below.}
\begin{align}
  I_{\mt{JT}}[g_{ab},\,\Phi_{\mt{JT}}] &= \frac{\Phi_0}{16\pi G_{\mt{2}}} \int_{\Sigma_2}\dd^2x \sqrt{-g}\,R +  \frac{\Phi_0}{8\pi G_{\mt{2}}} \int_{\partial\Sigma_2}\dd t\sqrt{-\gamma}\, K \cr
  & +\frac{1}{16\pi G_{\mt{2}}} \int_{\Sigma_2}\dd^2x \sqrt{-g}\, \Phi_{\mt{JT}}(R + 2) +  \frac{1}{8\pi G_{\mt{2}}} \int_{\partial\Sigma_2}\dd t\sqrt{-\gamma}\, \Phi_\mt{b}\,(K-1) ~,
\label{eq:JTaction}
\end{align}
where we have set $\ell_{\mt{AdS}_2}=1$ here. This is a particular case of two dimensional dilaton gravity theory with potential $V(\Phi_{\mt{JT}}) = 2\Phi_{\mt{JT}}$, where $\Phi_{\mt{JT}}$ measures the deviations in the size of the extremal black hole horizon $\Phi_0$, \emph{i.e.} $\Phi_0 \gg \Phi_{\mt{JT}}$. The first line is purely topological. It encodes the entropy of the extremal black hole through the size of the extremal horizon captured by $\Phi_0$.\footnote{It equals $-S_0\chi_\mt{Euler}$, where $S_0$ is the extremal entropy and $\chi_\mt{Euler}$ is the Euler number of the manifold $\Sigma_2$.} Variation with respect to the scalar JT-field $\Phi_{\mt{JT}}$ gives rise to $R+2=0$, forcing the two dimensional metric $g_{ab}$ to be locally AdS$_2$. Variation with respect to $g_{ab}$ gives rise to
\be\label{eq:JT1}
  \n_a\n_b \Phi_{\mt{JT}} - g_{ab}\, \Box_{2}\Phi_{\mt{JT}} +g_{ab} \Phi_{\mt{JT}} = 0~.
\ee
We will refer to these as the JT equations and to any scalar field, such as $\Phi_{\mt{JT}}$, satisfying them as JT mode. 

To extract the effective action describing the low energy excitations of nearly-AdS$_2$, we very briefly review the arguments in \cite{Maldacena:2016upp}. Given the UV completion provided by a near-extremal black hole, we want to identify the degrees of freedom responsible for the leading gravitational corrections to pure AdS$_2$. To be definite, describe the latter in Poincar\'e coordinates
\begin{equation}
  ds^2_{\mt{AdS}_2} = \frac{1}{z^2}\left(-\dd t^2 + \dd z^2\right)\,.
\end{equation}
The general solution to the JT equation \eqref{eq:JT} is given by
\begin{equation}
  \Phi_{\mt{JT}} = \frac{1}{z}\left(\alpha + \beta\,t + \gamma(t^2-z^2)\right)\,.
\end{equation} 
Inspired by holography, we glue the JT geometry to the UV spacetime close to the AdS$_2$ boundary $(z=\delta \to 0)$ across a cut-off surface $(t(u),z(u))$, where $u$ stands for the boundary time, by requiring the boundary conditions\footnote{It is important to stress that to keep the near-extremal black hole UV interpretation of the construction, the boundary value $\Phi_{\mt{b}}$ must satisfy $\Phi_{\mt{b}} \propto \delta^{-1} \ll \phi_0$.}
\begin{equation}
  \left. g\right|_{\mt{bdy}} = -\frac{1}{\delta^2}= \frac{-(t^\prime)^2 + (z^\prime)^2}{z^2}\,, \quad \quad \left.\Phi_{\mt{JT}}\right|_{\mt{bdy}} = \Phi_{\mt{b}} = \frac{\Phi_{\mt{r}}}{\delta}
\label{eq:gluing-bc}
\end{equation}
In the absence of the dilaton field (pure AdS$_2$ situation), the first condition can always be solved by an arbitrary $t(u)$, modulo the $\mathrm{SL}(2)$ isometry of pure AdS$_2$, with the choice $z(u)=\delta\,t^\prime$. An alternative way of reaching this conclusion is to consider the asymptotic symmetries of AdS$_2$ \cite{Hotta:1998iq,Cadoni:1999ja,NavarroSalas:1999up}. These are generated by
\begin{equation}
  \zeta^t = f(t),\, \quad \zeta^z = z\,f^\prime(t)\,,
\end{equation}
and would map the cut-off $t(u)=u$ to $t(u)=u+f(u)$. Either way, the reparameterisation symmetry is spontaneously broken by pure AdS$_2$ giving rise to an infinite number of pseudo-Goldstone modes parameterised by $t(u)$. In the presence of the dilaton field, the second boundary condition correlates the shape of the cut-off surface $t(u)$ with the source $\Phi_{\mt{r}}(u)$
\begin{equation}
  \frac{\alpha + \beta\,t(u) + \gamma\,t^2(u)}{(t^\prime(u))^2} = \Phi_{\mt{r}}(u)\,.
\end{equation}
If we interpret the latter as an equation of motion for a dynamical field $t(u)$, it was noticed in \cite{Maldacena:2016upp} that it can be obtained from varying the effective action
\begin{equation}
  I_{\mt{Schwarzian}} = -\frac{1}{8\pi G_{\mt{2}}} \int_{\partial\Sigma_2} du\,\Phi_{\mt{r}}(u) \{t(u),u\}\,,
\label{eq:schwarzian}
\end{equation}
where 
\begin{equation}
  \{t(u),u\} \equiv \left(\frac{t^{\prime\prime}}{t^\prime}\right)^\prime - {1\/2} \le({t''\/t'} \ri)^2\,.
\label{def:schw}
\end{equation}
In fact, if we ignore the topological terms, the action \eqref{eq:schwarzian} originates from the boundary term in \eqref{eq:JTaction} by using the $\Phi_{\mt{JT}}$ equation of motion, \emph{i.e.} $R=-2$, and evaluating the extrinsic curvature $K$ using the boundary conditions \eqref{eq:gluing-bc}  \cite{Maldacena:2016upp}. We conclude that the zero modes get a non-vanishing action, proportional to the Schwarzian \eqref{def:schw}, whenever there is a source, \emph{i.e.} a non-trivial boundary condition for the dilaton field which is also responsible for explicitly breaking the AdS$_2$ asymptotic symmetry group. From purely kinematic considerations, the effective action \eqref{eq:schwarzian} is the simplest local boundary action linear in the source $\Phi_{\mt{r}}(u)$  and invariant under the global $\mathrm{SL}(2)$ acting on the space of pseudo-Goldstone modes $t(u)$.

Before closing this review, we would like to stress two further points. The first one is concerned with the energy of the excitations described by the JT action \eqref{eq:JTaction}
\begin{equation}
  M(u) = -\frac{\Phi_\mt{r}}{8\pi G_{\mt{2}}}\,\{t(u),u\}\,.
\end{equation}
In the absence of a source $\Phi_\mt{r}$, it vanishes, as it should for pure AdS$_2$. In the presence of a source, it can be finite. This is a consequence of appropriately embedding the leading gravitational corrections to pure AdS$_2$ in a UV complete scheme, as originally envisioned in \cite{Almheiri:2014cka}. In fact, in the absence of matter, conservation of energy is equivalent to the equation of motion for $t(u)$. The second one has to do with the expected universality of the physics just reviewed. The Schwarzian action \eqref{eq:schwarzian}, together with the addition of relevant matter degrees of freedom, captures the thermodynamics of near-extremal black holes at low temperatures and is expected to arise as a universal low energy sector in near-extremal black hole physics. 



\section{Pleba\'{n}ski–Demia\'{n}ski type D solutions}\label{App:Delta1}

The complete family of type D spacetimes in Einstein-Maxwell theory was given by Pleba\'{n}ski and Demia\'{n}ski \cite{Plebanski:1976gy}. We refer to  \cite{Griffiths:2005qp} for a review and reinterpretation of these geometries. In this Appendix, we will focus on the solutions of pure Einstein gravity with  $\L=0$, which are summarized in Fig.~1 of \cite{Griffiths:2005qp}. In addition to the mass and angular momentum, the line element also contains a NUT parameter $ n $ and an acceleration parameter $\a$. In the notations of \cite{Griffiths:2005qp}, the line element is
\bea\dd s^2 \= {1\/\Om^2 } \le( {Q\/\rho^2} \le[\dd t - \le( a\,\r{sin}^2\t + 4n\,\r{sin}^2(\tfrac\t2) \ri)\dd\phi \ri]^2 - {\rho^2\/Q} \dd r^2 - {\rho^2\/\wt{P}}\,\r{sin}^2\t\,\dd\t^2 \ri.\-
 &&  \hspace{6cm}\le. -{\wt{P}\/\rho^2} \le[  a\,\dd t - (r^2+ (a+n)^2)\dd\phi\ri]^2  \ri)~,
\eea
where we have\footnote{This formula is obtained from (17) of \cite{Griffiths:2005qp} with $e=g=\L=0$ and with $\w=1$. }
\bea\nt
\rho^2 \= r^2 + (n+a\,\r{cos}\,\t)^2~, \\
\Om \= 1- \a r (n+a\,\r{cos}\,\t) ~,\-
Q \= {\a^2(a-n)^2 r^2-1\/1+3 \a^2 n^2 (a^2-n^2)}( 2r (M_0+\a n (M_0\a n -1)(a^2-n^2))-(1+2 M_0\a n)(r^2+a^2 -n^2))~,\-
\wt{P}\= (1-2 a \a M_0 \,\r{cos}\,\t)\,\r{sin}^2\t + {\a^2 a (a^2-n^2)(1+2 \a n M_0)\/1+3 \a^2 n^2(a^2-n^2)}(4 n+ a\,\r{cos}\,\t)\,\r{cos}\,\t\,\r{sin}^2\t~.
\eea

\subsection{Changing extremal mass}

The Kerr metric is obtained after setting $\a= n =0$ and extremality is achieved with $a=M_0$. The near horizon geometry is obtained taking the limit  \eqref{eq:near-horizon} which leads to the NHEK metric \eqref{NHEK}. We can consider the perturbation which changes the extremal mass $M_0$
\be
J\ra J+ \pert \,\d J+ O(\pert ^2)~,
\ee
 This gives the perturbation  \eqref{NHEKfullansatz} with 
\be
\chi(\x,\t) = { \d J\/J} (1+\r{cos}^2\t)~,\qq\Phi(\x) = { 2\d J\/J}~.
\ee
which corresponds to the even $\D=1$ mode. We see that the constant $\Phi$ mode is turned on and cancels the two conical singularities at $\t=0$ and $\t=\pi$ carried by $\chi$.

\subsection{Adding NUT charge}

We obtain the Kerr-NUT metric after setting the acceleration parameter $\a$ to zero. The extremal limit is achieved for
\be
M^2_0 \equiv M^2= a^2-n^2~.
\ee
The near horizon geometry can then be obtained using 
\be
t \ra 2 a (a+n)\,{ t\/\la}\,,\qq r \ra M_0+\la r\,,\qq \phi \ra \phi + a\,{t\/\la}~,
\ee
in the limit $\la\to 0$. We obtain the NHEK-NUT metric 
\bea
ds^2 \= \le(a^2(1+\r{cos}^2\t) + {2a  n \,\r{cos}\,\t} \ri) \le( -r^2 \dd t^2 + {\dd r^2\/r^2 } + \dd\t^2 \ri) \-
&& \hspace{2cm}+ {4 a \, \r{sin}^2\t\,\/a(1+\r{cos}^2\t )+ {2  n \,\r{cos}\,\t}} \le((a+ n )\dd\phi+ M r \dd t \ri)^2~,
\eea
which was derived in \cite{Ghezelbash:2009gy}.\footnote{Our formula differs from their by a rescaling of the angle $\phi\ra {a\/a+n}\phi$.} The perturbation corresponding to adding NUT charge to the NHEK is obtained by writing
\be
 n  = \pert \,\d n  + O(\pert ^2)~,
\ee 
which gives the perturbation  \eqref{NHEKfullansatz} with 
\be
\chi(\x,\t) = {2 \d n \/M_0} \,\r{cos}\,\t~,\qq\Phi(\x) = {2\d n \/ M_0}~,
\ee
corresponding to an $\l=0$ mode.

\subsection{Accelerated NHEK}

After setting the NUT parameter $n$ to zero, we obtain the spinning $C$-metric \cite{Kinnersley:1970zw,Pravda:2000vh}. For the extremal case $M=a$, we can take the near horizon limit using 
\be
t \ra {2 J\/1-\a^2 J^2}{ t\/\la}\,,\qq r \ra r_++\la r\,,\qq \phi \ra \phi + {\sqrt{J}\/1-\a^2 J^2}{t\/\la}~,
\ee
with $\la\to0$. This leads to the accelerated NHEK geometry:
\bea\nt
ds^2 \= {J(1+\r{cos}^2\t)\/(1-\a J\,\r{cos}\,\t)^2}\le[ {1\/1-\a^2 J^2} \le(- r^2 \dd t^2 + {\dd r^2\/r^2}\ri) + {\dd \t^2\/(1-\a J\,\r{cos}\,\t)^2} \ri] \\
&& \hspace{5cm}+ {4J\,\r{sin}^2\t\/1+\r{cos}^2\t} \le(\dd \phi+ {r\/1-\a^2 J^2} \dd t\ri)^2~,
\eea
which was studied in \cite{Astorino:2016xiy}. We can then consider the perturbation of NHEK corresponding to adding acceleration:
\be
\a =  \pert\, \d\a + O(\pert ^2)~.
\ee
We have to  redefine $\t\ra \t-\pert \,\d\a \,\r{sin}\,\t$ to fit this perturbation into the ansatz  \eqref{NHEKfullansatz} and we obtain an $\l=0$ mode:
\be
\chi(\x,\t) = 4 J \d\a\,\r{cos}\,\t~,\qq\Phi(\x) = 0~.
\ee

\section{Isometries of (near-)NHEK}\label{app:isometries}

Starting from the $\fsl(2)$ Killing vectors of the NHEK geometry given in \eqref{NHEK:iso}, the $\fsl(2)$ Killing vectors of the near-NHEK geometry at temperature $\tau/(2\pi)$ are obtained using the diffeomorphism \eqref{Sdiffeo} with $f(t)$ satisfying
\be
\{f(t),t\} = -{\tau^2\/2}~. 
\ee
The metric of the AdS$_2$ factor takes the form
\be\label{AdS2thermal}
ds^2 = -r^2\le( 1-{\tau^2\/4r^2}\ri)^2 \dd t^2 + {\dd r^2\/r^2}~.
\ee
The general solution is given by
\be
f(t) = {a \, e^{\tau t}+ b\/c \,e^{\tau t}+ d}~,\qq \le(\begin{smallmatrix} a & b\\ c & d\end{smallmatrix}\ri) \in \r{GL}(2,\R)~.
\ee
In the main text, we use the choice $f(t) = e^{\tau t}$ which is the only choice for which $\z_0$ is proportional to $\p_t$. This leads to the Killing vectors
\be\label{App:SL2expbasis}
\zeta_0=\frac{1}{\tau}\p_t,\qq \zeta_\pm=\le(\frac{4 r^2+\tau^2}{\tau(4r^2-\tau^2)}\p_t\mp r\p_r-\frac{4\tau r}{4 r^2-\tau^2}\p_\phi\ri)e^{\pm\tau t}~,
\ee
satisfying the $\fsl(2)$ algebra
\be
[\z_0,\z_\pm ] = \pm \z_\pm,\qq [\z_-,\z_+] = 2\z_0~,
\ee
and corresponding to the dual scalars
\be
\Phi_{\zeta_0}={r\/\tau} \le(1+{\tau^2\/4 r^2}\ri),\qq \Phi_{\zeta_\pm}={r\/\tau} \le(1-{\tau^2\/4 r^2}\ri)e^{\pm\tau t}~.
\ee
Another natural choice is 
\be
f(t)  =\tfrac{2}{\tau} \, \r{tanh}\le(\tfrac{\tau t}{2} \ri)~,
\ee
which has a smooth limit $\lim_{\tau\to 0}f(t) = t$ corresponding to the Poincaré basis of NHEK. This leads to the Killing vectors
\bea\nt
\xi_0 \= {4r^2 + \tau^2\/\tau(4r^2-\tau^2)} \,\r{sinh}(\tau t)\,\p_t - r\,\r{cosh}(\tau t)\,\p_r - {4 \tau r\/4r^2 -\tau^2}\,\r{sinh}(\tau t) \,\p_\phi~,\-
\xi_{-} \= {1\/2}\le( 1+{4r^2 +\tau^2\/4r^2-\tau^2} \,\r{cosh}(\tau t)\ri) \p_t - {\tau r\/2}\,\r{sinh}(\tau t)\,\p_r - {2 r \tau^2\/4r^2-\tau^2}\,\r{cosh}(\tau t)\,\p_\phi~,\-
\xi_{+} \= -{2\/\tau^2}\le( 1-{4r^2 +\tau^2\/4r^2-\tau^2} \,\r{cosh}(\tau t)\ri) \p_t - {2 r\/\tau}\,\r{sinh}(\tau t)\,\p_r - {8 r \/4r^2-\tau^2}\,\r{cosh}(\tau t)\,\p_\phi~,
\eea
corresponding to the dual scalars
\begin{align}
\Phi_{\xi_-}&=\frac{r}{2}\le(1+{\tau^2\over 4 r^2}\ri)+\frac{r}{2}\le(1-{\tau^2\over 4 r^2}\ri)\cosh(\tau t)~,\cr
\Phi_{\xi_{0}}&={r\over \tau}\le(1-{\tau^2\over 4 r^2}\ri) \sinh(\tau t)~,\\
\Phi_{\xi_+}&=-\frac{2r}{\tau^2}\le(1+{\tau^2\over 4 r^2}\ri)+\frac{2r}{\tau^2}\le(1-{\tau^2\over 4 r^2}\ri)\cosh(\tau t)~.
\nonumber
\end{align}
We have the following decomposition in the basis \eqref{App:SL2expbasis}:
\bea
\xi_{-} \= {\tau\/4} \z_{-} + {\tau\/2} \z_0 + {\tau\/4}\z_{+}~, \-
\xi_{0} \= -{1\/2} \z_{-} + {1\/2}\z_{+}~, \-
\xi_{+} \= {1\/\tau} \z_{-} -{2\/\tau} \z_0 + {1\/\tau}\z_{+}~,
\eea
with the same decomposition for the dual scalars.  In the limit $\tau\to 0$, we obtain the Poincaré Killing vectors with their associated dual scalars:
\begin{align}
& \xi_-  = \p_t,& & \Phi_{\xi_-} = r~,\\ 
& \xi_0 = t\p_t - r \p_r, & & \Phi_{\xi_0} = r t~,\\
& \xi_+ = \le(t^2 +{1\/r^2} \ri) \p_t - 2 t r \p_r -{2\/r}\p_\phi ,&  &\Phi_{\xi_+} = rt^2 -{1\/r}~.
\end{align}

\bibliographystyle{JHEP-2}
\bibliography{all}

\end{document}